\documentclass[a4paper,11pt]{article}
\pdfoutput=1 

\usepackage{jheppub} 
\usepackage[T1]{fontenc}
\usepackage[italian,english]{babel}
\usepackage{hyperref}
\usepackage{ifpdf}
\usepackage{subfigure}
\usepackage{amssymb}
\usepackage{amsfonts}
\usepackage{epsf}
\usepackage{rotating}
\usepackage{graphicx}
\usepackage{amsmath}
\usepackage{fancyhdr}
\usepackage{lineno}
\usepackage{babel}
\usepackage{graphics}
\usepackage{pstricks}
\usepackage{color}
\usepackage{multirow}
\usepackage{nicefrac}
\usepackage{bm}
\usepackage{slashed}
\usepackage{float}
\usepackage{array}
\usepackage{boldline}

\usepackage{mathrsfs}
\usepackage{stackengine}

\newcolumntype{?}{!{\vrule width 3pt}}

\title{Investigating the Production of Leptoquarks by Means of Zeros of Amplitude at Photon Electron Collider}

\author[a]{Priyotosh Bandyopadhyay,}
\author[a]{Saunak Dutta,}
\author[a]{Anirban Karan.}
\affiliation[a]{Indian Institute of Technology Hyderabad, Kandi,  Sangareddy-502285, Telengana, India}

\emailAdd{bpriyo@phy.iith.ac.in}
\emailAdd{ph17resch11002@iith.ac.in}
\emailAdd{kanirban@iith.ac.in}

\preprint{IITH-PH-0003/20 }

\begin{document}

	\abstract{Leptoquarks are one of the possible candidates for explaining various anomalies in flavour physics. Nonetheless, their existence is yet to be confirmed from experimental side. In this paper we have shown how zeros of single photon tree-level amplitude can be used to extract information about leptoquarks in case of $e$-$\gamma$ colliders. Small number of standard model backgrounds keep the signal clean in this kind of colliders. Unlike other colliders, the zeros of single photon amplitude  here depend on $\sqrt{s}$ as well as the mass of leptoquark  along with its electric charge. We perform a PYTHIA based simulation for reconstructing the leptoquark from its decay products of first generation and estimating the background with luminosity of 100 fb$^{-1}$. Our analysis is done for all the leptoquarks that can be seen at $e$-$\gamma$ collider with three different masses (70 GeV, 650 GeV and 1 TeV) and three different centre of momentum energy (200 GeV, 2 TeV and 3 TeV). The effects of non-monochromatic photons on the zeros of amplitude under laser backscattering and equivalent photon approximation have also been addressed.}

	\maketitle
	\flushbottom
	
	\section{Introduction}
	\label{sec:intro}
	
	Leptoquarks  are proposed particles that 
	couple to quarks and leptons simultaneously, and hence carry non-zero baryon number as well as lepton number. They emerge naturally in various extensions of the Standard Model (SM), such as Pati-Salam model \cite{pati-salam}, GUT based on $SU(5)$ or $SO(10)$ \cite{GUT1,GUT2,GUT3}, extended technicolor models \cite{technicolor1,technicolor2}, etc. These colour-triplet electromagnetically charged bosons (spin zero or one) could be singlet, doublet or triplet under $SU(2)_L$ group \cite{Rev1a,Rev2a,Rev3a,Rev4a,Rev5a,Rev6a}. Detection of leptoquark would be a signal for the unification of matter fields. Anomalies observed in the lepton flavour universality ratios $R_K$, $R_{K*}$, $R_D$, $R_{D*}$ related to rare B decays \cite{exptRdRk1,exptRdRk2,exptRdRk3,exptRdRk4,exptRdRk5} and the deviations in the measurements of angular observables from their theoretical estimates can be addressed using several leptoquark models. Some of these models can explain observed discrepancy in muon $g-2$ \cite{g-21,g-22} and also accommodate the excess of $2.4\sigma$ in a Higgs decay branching fraction to $\mu\tau$  at 8 TeV with 19.7 fb$^{-1}$ luminosity \cite{htomutau}. Because of their great importance in elucidating several issues of flavour physics \cite{Motiv1a,Motiv1b,Motiv1c,Motiv1d,Motiv1e,Motiv1f,Motiv1g,Motiv2a,Motiv3a,Motiv4a,Motiv4b,Motiv4c,Motiv4d,Motiv4e,Motiv4g,Motiv5a,Motiv6a,Motiv7a,Motiv8a,Motiv9a, Motiv10a,Motive11,Motive12a,Motive12b,Motive13a,Motive13b,Motive13c,Motive13d,Motive13e,Motive13f,Motive14,Motive15,Motive16a}, leptoquarks have been studied in literature in gory details during  last few decades \cite{Rev1a,Rev2a,Rev3a,Rev4a,Rev5a,Rev6a,ModS1a,ModS2a,ModS2b,ModS2c,ModS3a,ModS4a,ModS5a,ModS6a,ModS6b,ModS7,ModS8a,ModS8b,Exp1a,Exp10a,Exp13a,Exp13b}. In parallel, numerous searches for leptoquarks have been performed in different colliders \cite{Exp2a,Exp2b,Exp3a,Exp5a,Exp6a,Exp7a,Exp8a,Exp9a,Exp11a,Exp12a,Exp14a,Exp15a,LEP1,LEP2}.

	On the other hand, the phenomenon of RAZ (\textit{radiation amplitude zero}) was first discussed for $q_i \bar q_j\to W^\pm\gamma$ process at $pp$ or $p\bar p$ collider in order to probe the magnetic property of $W$-boson \cite{RAZ1}. This phenomenon has been studied extensively in literature for various BSM models like supersymmetry, leptoquarks, other gauge theories, etc and physics behind its occurrence has also been scrutinized \cite{RAZ2,RAZ2p,RAZ3,RAZ4,RAZ5,RAZ7,RAZ8,RAZ9,RAZ10,RAZ11,RAZ12,RAZ13,RAZ14,RAZ15,RAZ16,RAZ17,RAZ18,RAZ19,RAZ19a,RAZ20,RAZ21,RAZ22,RAZ23,RAZ24,RAZ25,RAZ26,RAZ27,RAZ27p,RAZ28,RAZ29,RAZ30,RAZ31,RAZ32}. In non-Abelian theories the tree-level amplitudes\footnote{The word ``amplitude" in this context is synonymous to $|\mathcal M|^2$ where $\mathcal M$ is the matrix element for a given process.} for single photon emission processes, which is the sum generated by attaching photon to the internal and external particles in all possible ways, can be factorized into two parts: a) the first part contains the combination of generators of the gauge group, various kinematic invariants, charges and other internal symmetry indices, whereas b) the second part corresponds to the actual amplitudes of the Abelian fields containing the dependence on the spin or polarization indices \cite{RAZ2, RAZ2p}. The first factor goes to zero in certain kinematical zones depending on the charge and four momenta of external particles and  forces the single-photon tree amplitudes to vanish \cite{RAZ3}. The general criterion for tree-level single photon amplitude to vanish is that $\displaystyle\Big(\frac{p_j\cdot k}{Q_j}\Big)$ must be same for all the external particles (other than photon) involved in the process \cite{RAZ3} where, $p_j^\mu$ and $Q_j$ are the four momentum and charge of $j$th external particle and $k^\mu$ is the four momenta of photon. For $2\to 2$ scattering processes with photon in final state, this condition reduces to:
	\begin{equation}
	\label{eq:costheta}
	\cos\theta^*=\frac{Q_{f_2}^{}-Q_{f_1}^{}}{Q_{f_2}^{}+Q_{f_1}^{}}
	\end{equation} 
	where, 	$Q_{f_1}^{}$ and $Q_{f_2}^{}$ are the charges for the incoming particles $f_1$ and $f_2$ and $\theta^*$ is the angle between photon and $f_1$ in the centre of momentum (CM) frame at which RAZ occurs provided that the masses of colliding particles are negligible with respect to total energy of the system, i.e. $\sqrt{s}$. 
	
	
	Linear colliders in the range of a few hundred GeV to 1.5 TeV are going to be build in near future. These colliders can provide the possibility for studying electron-photon interactions at very high energy \cite{EPhC1a,EPhC2a,EPhC2b,EPhC2c,EPhC2d,EPhC2e,EPhC3a,EPhC3b,EPhC3c,EPhC4a}. Using modern laser technology, high energetic photons with large luminosity  can be prepared through laser backscattering for this kind of studies. Since very few SM processes contribute to the background for these electron-photon colliders, they can reveal clean signals of leptoquarks through zeros of tree-level single photon amplitude \cite{EphLQ1,EphLQ2,EphLQ3}. In this paper we have studied this possibility in detail. The phenomena of RAZ in various leptoquark models has already been described in literature in context of $e$-$p$ colliders where leptoquark is expected to be produced associated with a photon or the it undergoes radiative decays \cite{RAZLQ1,RAZLQ2}. Though our scenario looks quite similar to it, there arises great difference between these two colliders while considering the position of zero amplitude in the phase space. It is evident from Eq.~\eqref{eq:costheta} that RAZ for $e$-$p$ colliders occurs at some particular angle between the photon and the quark which depends only on the electric charge of electron and the quark; however, we show that the same angle for zero amplitude at $e$-$\gamma$ colliders depends on the mass of the leptoquark as well as $\sqrt{s}$ along with the electric charge \cite{RAZLQ3}. Nevertheless, the general condition for tree-level single photon amplitude being zero \cite{RAZ3} still remains valid.
	
	In this paper we have analysed all kinds of leptoquarks that are going to be  produced at $e$-$\gamma$ colliders for three different masses (70 GeV, 650 GeV and 1 TeV) with three different centre of momentum energy (200 GeV, 2 TeV and 3 TeV). Though leptoquark with light mass seems to be ruled out, most of these analysis assumes coupling of leptoquark to single generation of quark and lepton, whereas, the results from UA2 and CDF collaboration show that there is still room for low mass leptoquark with sufficiently small couplings and appropriate branching fractions to different generations of quarks and leptons. On the other hand, the bounds on couplings and branching fractions of higher mass leptoquarks are more relaxed. The leptoquark will eventually decay to a lepton and a quark, and hence we it will produce a mono-lepton plus di-jet signal at detector. In a PYTHIA based analysis, we reconstruct the leptoquark from the invariant mass of the lepton and one jet. Then we look for the angle between the reconstructed leptoquark and electron and construct the angular distribution which should match with the theoretical estimates. Observation of the zeros of this distribution at the theoretically predicted portion of phase space would indicate the presence of some leptoquarks. Furthermore, we have studied the effects of non-monochromatic photons on the zeros of angular distribution under laser backscattering and equivalent photon approximation schemes considering the current experimental limitations of electron-photon colliders.  
	
	The paper is disposed in the following way. In the next section (sec. \ref{sec:theory}) we describe the theoretical approach to the production of scalar as well as vector leptoquarks at $e$-$\gamma$ collider and find the conditions for the zeros of angular distribution. The experimental constrains on the mass, coupling and branching fractions of the leptoquarks have been summarised in sec. \ref{sec:mass-coupling}. In sec. \ref{sec:model-simulation}, we describe  the simulation set up, choice of benchmark points and centre of momentum energies, production cross sections and branching fractions of the leptoquarks and PYTHIA based simulation for different types of leptoquarks produced at the electron-photon collider. Sec. \ref{sec:non-mono} deals with effects of non-monochromatic photons on the zeros of differential distribution. Finally, we conclude in sec. \ref{sec:concl}.
	
	\section{Theoretical formalism}
	\label{sec:theory}
	
\begin{figure}[H]
\includegraphics[scale=0.45]{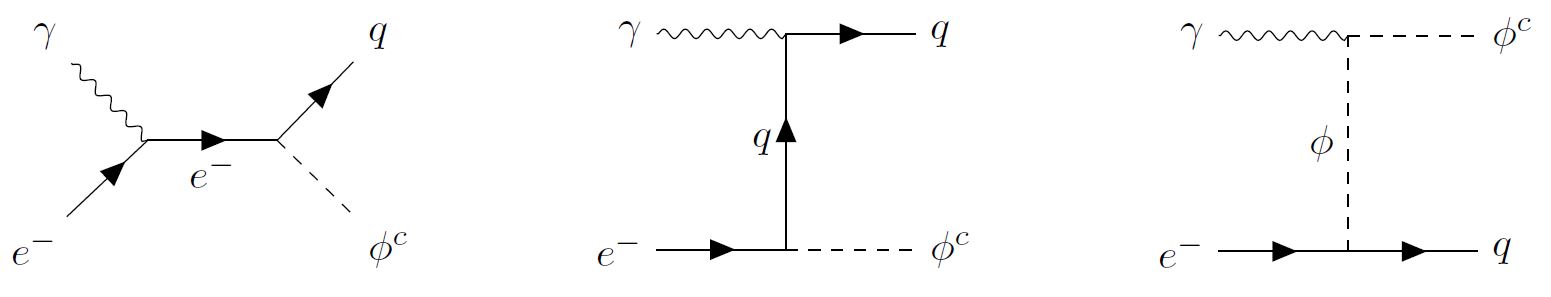}
		\caption{Feynman diagrams for $e^-\,\gamma\to q\,\phi$}
		\label{fig:egdp}
		\hspace*{0.5cm}
	\end{figure}

In this section, we develop the theoretical formalism for the production of a leptoquark	(more precisely anti-leptoquark) associated with a quark or an anti-quark at the electron-photon collider to get the mathematical expression for the differential distribution of this process. We consider the process $e^-\gamma\to q \,\phi^c$ where $q$ is a quark and $\phi$ is a leptoquark (the sign `$c$' indicates charge conjugate), for which there are three possible tree-level Feynman diagram, as shown in fig. \ref{fig:egdp}.

	\subsection{Scalar Leptoquark}
	\label{sub:scalar}	
 If the leptoquark $\phi$ is a scalar one, the matrix elements for the respective diagrams are as follows:
 	\begin{align}
	\label{eq:M1}
	&\mathcal{M}_1^S=\bar u(p_q)\,(-iY_L^{eq} P_L-iY_R^{eq} P_R)\,\frac{i}{(\slashed p_e+\slashed p_\gamma)}\,(ie\gamma^\mu)\,u(p_e)\,\epsilon_\mu^\gamma~,\\
	\label{eq:M2}
	&\mathcal{M}_2^S=\bar u(p_q)\,(-ieQ_q\gamma^\mu)\,\frac{i}{(\slashed p_q-\slashed p_\gamma)}\,(-iY_L^{eq} P_L-iY_R^{eq} P_R)\,u(p_e)\,\epsilon_\mu^\gamma~,\\
	\label{eq:M3}
	&\mathcal{M}_3^S=\bar u(p_q)\,(-iY_L^{eq} P_L-iY_R^{eq} P_R)\,u(p_e)\,\frac{i}{(p_q-p_e)^2-M_\phi^2}\,[ie(1+Q_q)(2p_e^\mu-2p_q^\mu+p_\gamma^\mu)]\,\epsilon_\mu^\gamma~.
	\end{align}
	where $p_e^\mu$, $p_\gamma^\mu$ and $p_q^\mu$ are the four momenta of the particles electron, photon and the produced quark respectively, $Y_{L,R}$  are $3\times 3$ matrices describing the couplings of leptoquark with left-handed and right-handed leptons and quarks respectively, $e$ denotes the charge of positron, $Q_q$ signifies the charge of $q$ quark in the unit $e$, $M_\phi$ indicates the mass of leptoquark, $\epsilon_\mu^\gamma$ is the polarization of the photon and $P_{L,R}\equiv(1\mp\gamma^5)/2$. Here, we have deliberately neglected the masses of electron and the quark since they would have insignificant effects in determining the zero of amplitude involving production of very heavy leptoquark for all practical purposes unless the produced quark is top. Therefore, after taking the spin and polarization sum of initial and final state particles, the modulus squared matrix element for this mode becomes:
	\begin{equation}
	\begin{split}
	\label{eq:modmsq}
	\sum_{spin}|\mathcal{M}^S|^2=&\frac{e^2\,[(Y_L^{eq})^2+(Y_R^{eq})^2]\Big[(s-M_\phi^2)(1-\cos\theta)+2sQ_q\Big]^2}{s\,(s-M_\phi^2)\,(1-\cos\theta)\,\Big[s(1+\cos\theta)+M_\phi^2(1-\cos\theta)\Big]^2}\\
	&\qquad\qquad\qquad\qquad\qquad\qquad\qquad\times\Big[(s-M_\phi^2)^2(1+\cos\theta)^2+4M_\phi^4\Big]
	\end{split}
	\end{equation}
	where, $s=(p_e+p_\gamma)^2$ and $\theta$ is the angle between electron and the leptoquark or equivalently photon and the quark  $q$. 
	
	\subsection{Vector Leptoquark}
	\label{sub:vector}
	
	Now, if the leptoquark $\phi$ be a vector particle, the matrix elements will get modified in the following way:
	\begin{align}
	\label{eq:Mv1}
	&\mathcal{M}_1^V=\epsilon^\gamma_\nu\,\epsilon^{\phi}_\mu\,\bar u(p_q)\,(-iY_L^{eq}\gamma^\mu P_L-iY_R^{eq}\gamma^\mu P_R)\,\frac{i}{(\slashed p_e+\slashed p_\gamma)}\,(ie\gamma^\nu)\,u(p_e)~,\\
	\label{eq:Mv2}
	&\mathcal{M}_2^V=\epsilon^\gamma_\nu\,\epsilon^{\phi}_\mu\,\bar u(p_q)\,(-ieQ_q\gamma^\nu)\,\frac{i}{(\slashed p_q-\slashed p_\gamma)}\,(-iY_L^{eq}\gamma^\mu P_L-iY_R^{eq}\gamma^\mu P_R)\,u(p_e)~,\\
	\label{eq:Mv3}
	&\mathcal{M}_3^V=\epsilon^\gamma_\nu\,\epsilon^{\phi}_\mu\,\bar u(p_q)\,(-iY_L^{eq}\gamma^\rho P_L-iY_R^{eq}\gamma^\rho P_R)\,u(p_e)\,\frac{i}{(p_q-p_e)^2-M_\phi^2}\,\nonumber\\
	&\qquad\;\;[ie(1+Q_q)\{(2p_e^\nu-2p_q^\nu+p_\gamma^\nu)g_{\mu\rho}+(p_q^\rho-p_e^\rho-2p_\gamma^\rho)g_{\mu\nu}+(p_\gamma^\mu-p_e^\mu+p_q^\mu)g_{\nu\rho}\}]~.
	\end{align}
	Here, $\epsilon^{\phi}_\mu$ is polarization vectors for the vector leptoquark. After taking the spin and polarization sum of initial and final state particles\footnote{It should be noted that: $\displaystyle\sum_{polarization}\epsilon_\mu^{\phi*}\epsilon_\nu^{\phi}=\Big(-g_{\mu\nu}+\frac{p_{\phi\mu}p_{\phi\nu}}{M_\phi^2}\Big)$ where $p_\phi^\mu$ is the four-momentum of the leptoquark.}, the modulus squared matrix element becomes:
	\begin{equation}
	\begin{split}
	\label{eq:modmsqv}
	\sum_{spin}|\mathcal{M}^V|^2=&\frac{2e^2\,[(Y_L^{eq})^2+(Y_R^{eq})^2]\Big[(s-M_\phi^2)(1-\cos\theta)+2sQ_q\Big]^2\,}{s\,(s-M_\phi^2)\,(1-\cos\theta)\,\Big[s(1+\cos\theta)+M_\phi^2(1-\cos\theta)\Big]^2}\\
	&\qquad\qquad\qquad\qquad\times\Big[\{s(1-\cos\theta)+M_\phi^2(1+\cos\theta)\}^2+4(s-M_\phi^2)^2\Big]
	\end{split}
	\end{equation}

	The differential cross-section for this process turns out to be:
	\begin{equation}
	\label{eq:diffdist}
	\frac{d\sigma}{d\cos\theta}=\frac{s-M_\phi^2}{32 \pi s^2}\bigg(\frac{3}{4}\sum_{spin}|\mathcal M^{(S,V)}|^2\bigg)
	\end{equation}
	Here, the one fourth factor comes because of the average over initial state spins and polarizations; on the other hand, the factor three indicates the number of colour combinations available in the final state.

\begin{figure}[h!]
		\includegraphics[scale=0.35]{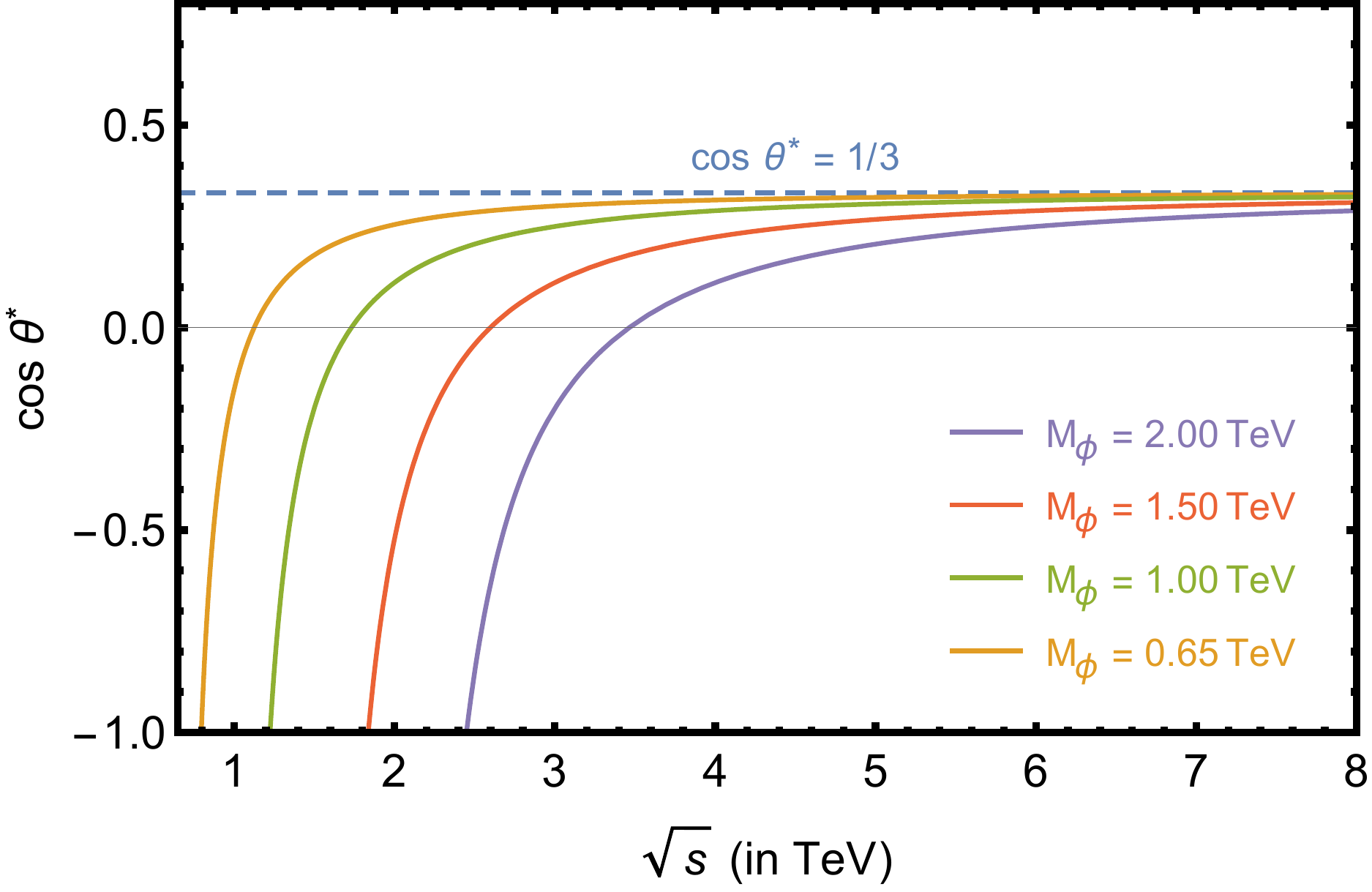}
		\hfill
		\includegraphics[scale=0.35]{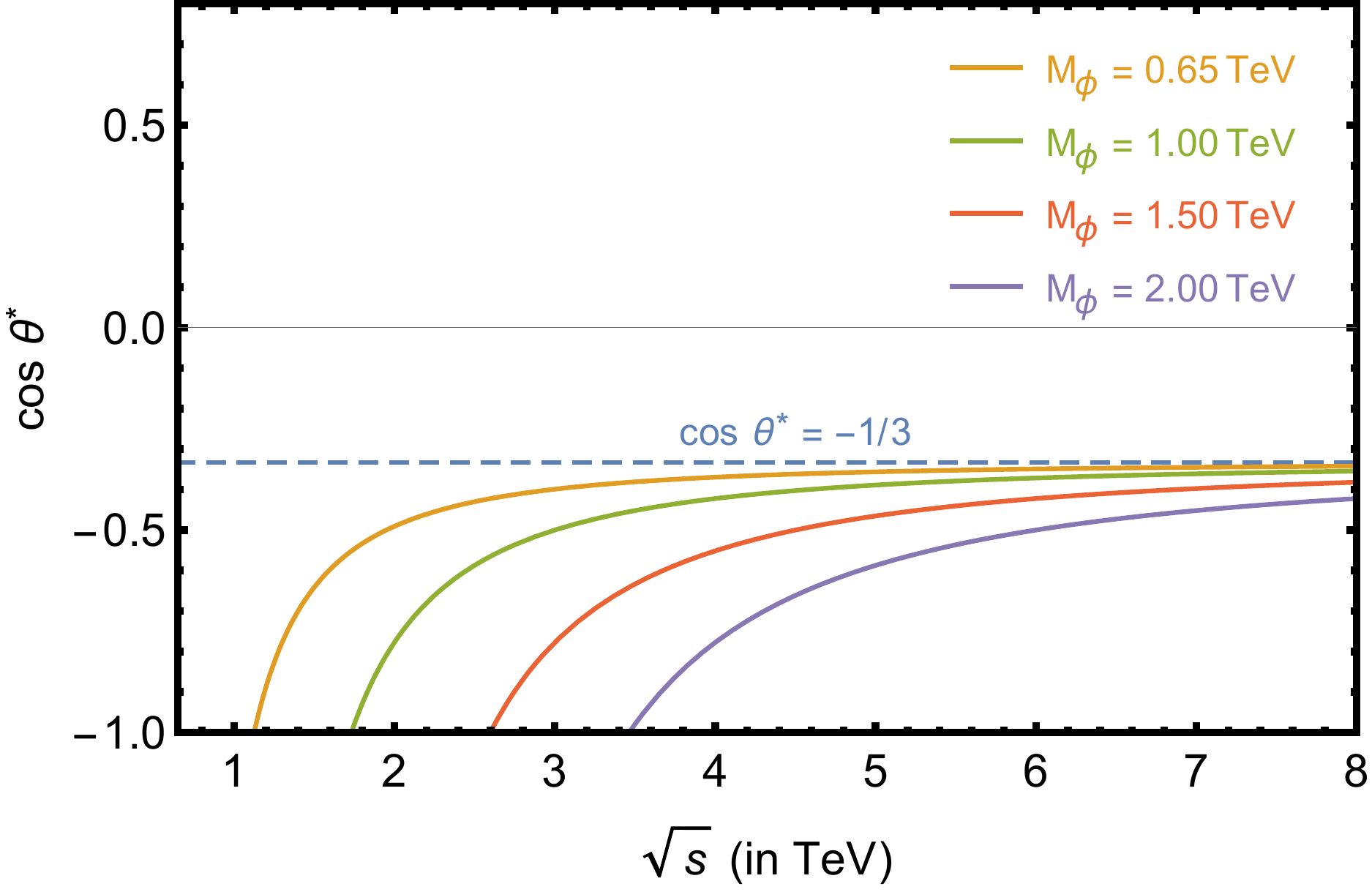}
		\caption{Variation of $\cos\theta^*$ with respect to $\sqrt{s}$ for $Q_q=-1/3$ and $Q_{\bar q}=-2/3$, respectively for different masses of leptoquark.}
		\label{fig:costhst_rts}
	\end{figure}

	Now, it is evident from the Eqs. \eqref{eq:modmsq} and \eqref{eq:modmsqv} that the differential cross-section vanishes iff:
	\begin{equation}
	\label{eq:costhst}
	(s-M_\phi^2)(1-\cos\theta^*)+2sQ_q=0\quad \implies\quad
	\cos\theta^*=1+\frac{2\,Q_q}{[1-(M_\phi^2/s)]}=f(Q_q,M_\phi^2/s)~,
	\end{equation}
	since all the other terms are positive quantities.
	This  also follows from the general condition for tree-level single photon amplitude to vanish \cite{RAZ3} :
	\begin{equation}
	\label{eq:nullcond}
	\frac{p_e.p_\gamma}{-1}=\frac{p_q.p_\gamma}{Q_q}=\frac{p_\phi.p_\gamma}{Q_\phi}
	\end{equation} 
	where $Q_\phi$ is the charge of leptoquark in unit of $e$ and can be expressed as: $Q_\phi=-(1+Q_q)$. However, the striking difference between single photon emission with two body final state and this process is that $\cos\theta^*$ in the former case does not depend on the mass of fourth particle as well as the centre of momentum energy  (as shown in Eq. \eqref{eq:costheta}) after neglecting the masses of fermions, whereas $\cos\theta^*$ in the later scenario does depend on the mass of leptoquark and $\sqrt s$ (as can be seen from Eq. \eqref{eq:costhst}). The variation of $\cos\theta^*$ with increasing centre of momentum energy ($\sqrt s$) for different masses of leptoquark has been shown in fig. \ref{fig:costhst_rts}; the left panel depicts the variation for production of a leptoquark associated with a down-type quark, while the right panel describes the same with a up-type anti-quark. It can also be observed from Eq. \eqref{eq:costhst} that $\cos\theta^*$ approaches $(1+2Q_q)=(Q_q^{}-Q_{\phi}^{})$ amyptotically when $\sqrt s>>M_{\phi}$.
	For the vanishing amplitude to be inside the physical region, the condition that must satisfy is:
	\begin{equation}
	\label{eq:physreg}
	Q_q<0 \quad\text{ and }\quad\frac{M_{\phi}}{\sqrt s}\leq\sqrt {-Q_{\phi}}
	\end{equation}
	which in turn would imply that 
	\begin{equation}
	\label{eq:condition}
	-1<Q_\phi<0~.
	\end{equation}

	It should be noted that instead of quark $q$, if the leptoquark is produced with an anti-quark $\bar q$, all the expressions for that process can be achieved by replacing $\bar u(p_q^{})$ with $\bar v(p_{\bar q}^{})$ and $Q_q$ with $Q_{\bar q}$ in the equations from Eq. \eqref{eq:M1} to Eq. \eqref{eq:physreg} where $Q_{\bar q}$ is the charge of $\bar q$ in unit of $e$.
	
All of the leptoquarks \cite{Rev1a,Rev2a,Rev3a,Rev4a,Rev5a,Rev6a}, that can be produced at $e$-$\gamma$ collider, have been listed in table \ref{tab:LQ}. Here, $\Psi_q$, $\Psi_l$ are quark and lepton doublets whereas $q_u$, $q_d$ and $l_e$ are fields for $u$-quark, $d$-quark and electron respectively. The transpose $T$ acts on $SU(2)$ indices only. $S_3^{ad}$ and $U_3^{ad}$ denote scalar and vector triplet respectively in the adjoint representation of SU(2); they are defined as:
\begin{equation*}
S_3^{ad}=\begin{pmatrix}
\frac{S_3^{+1/3}}{\sqrt 2}&S_3^{+4/3}\\S_3^{-2/3}&-\frac{S_3^{+1/3}}{\sqrt 2}
\end{pmatrix}
\quad \text{and}\quad 
U_3^{ad}=\begin{pmatrix}
\frac{U_3^{+2/3}}{\sqrt 2}&U_3^{+5/3}\\U_3^{-1/3}&-\frac{U_3^{+2/3}}{\sqrt 2}
\end{pmatrix}.
\end{equation*}

		\setlength{\tabcolsep}{13pt}
\renewcommand{\arraystretch}{1.5}
\begin{table}[H]
	\begin{tabular}{?>{\centering}m{0.5cm}||>{\centering}m{0.5cm}||>{\centering}m{0.5cm}||c||c||c?}
		\hlineB{5}
		LQ & Y & $Q_{em}$ & Interaction &  Process & $\cos\theta^*$ \\
		\hlineB{5}
		\multicolumn{6}{?c?}{Scalar Leptoquarks}\\
		\hline\hline
		\multirow{2}{*}{$S_1$}  & \multirow{2}{*}{\nicefrac{2}{3}} & \multirow{2}{*}{\nicefrac{1}{3}} &$\overline\Psi_{q}^c\,P_L\,i \sigma_2\Psi_l^{}\,S_1$,   & \multirow{2}{*}{$\bar u\, \Big({S_1^{\nicefrac{+1}{3}}}\Big)^c$} & \multirow{2}{*}{$f(\nicefrac{-2}{3},\nicefrac{M_\phi^2}{s})$} \\
		& &  & $\bar q_u^c\,P_R\,l_e^{}\,S_1$ & &   \\[2mm]
		\cline{1-6}
		$\widetilde S_1$ &\nicefrac{8}{3} & \nicefrac{4}{3}& $\bar q_d^c\,P_R\,l_e\,\widetilde S_1$ & $\bar d \, \Big({\widetilde S_1^{\nicefrac{+4}{3}}}\Big)^c$ & ------ \\[2mm]
		\cline{1-6}
		\multirow{3}{*}{$\vec{S}_3$} & \multirow{3}{*}{\nicefrac{2}{3}} &\nicefrac{4}{3}  &  & $\bar d \,  \Big({S_3^{\nicefrac{+4}{3}}}\Big)^c$ & ------ \\
		&  &\nicefrac{1}{3}& $\overline\Psi_{q}^c\,P_L\,(i\sigma_2\, S_3^{ad})\,\Psi_l^{}$ & $\bar u\, \Big({S_3^{\nicefrac{+1}{3}}}\Big)^c$ & $f(\nicefrac{-2}{3},\nicefrac{M_\phi^2}{s})$\\
		&  &$\nicefrac{-2}{3}$&   & ------ & ------\\[2mm]
		\cline{1-6}
		\multirow{2}{*}{$R_2$} & \multirow{2}{*}{\nicefrac{7}{3}} &\nicefrac{5}{3}&$\overline\Psi_q\,P_R\,R_2\,l_e$, & $u \,  \Big({R_2^{\nicefrac{+5}{3}}}\Big)^c$ & ------ \\
		&  &$\nicefrac{2}{3}$ & $\bar q_u^{}\,P_L\,(R_2^T\,i\sigma_2)\,\Psi_l$ & $d\, \Big({R_2^{\nicefrac{+2}{3}}}\Big)^c$ & $f(\nicefrac{-1}{3},\nicefrac{M_\phi^2}{s})$\\[2mm]
		\cline{1-6}
		\multirow{2}{*}{$\widetilde{R}_2$} & \multirow{2}{*}{\nicefrac{1}{3}} &\nicefrac{2}{3} &\multirow{2}{*}{$\bar q_d^{}\, P_L\,(\widetilde R_2^T\,i\sigma_2)\,\Psi_l$} & $d \,  \Big({\widetilde R_2^{\nicefrac{+2}{3}}}\Big)^c$ & $f(\nicefrac{-1}{3},\nicefrac{M_\phi^2}{s})$ \\
		&  &$\nicefrac{-1}{3}$ &  & --- & ---\\[2mm]
		\hlineB{5}
		\multicolumn{6}{?c?}{Vector Leptoquarks}\\
		\hline\hline
		\multirow{2}{*}{$V_{2\mu}$}&  \multirow{2}{*}{\nicefrac{5}{3}} &\nicefrac{4}{3} & $\overline\Psi_{q}^c\,\gamma^\mu P_R\,(i\sigma_2\,V_{2\mu})l_e^{}$ & $\bar d \, \Big({ V_{2\mu}^{\nicefrac{+4}{3}}}\Big)^c$ & ------ \\
		& &$\nicefrac{1}{3}$ & $\bar q_d^{c}\,\gamma^\mu P_L\,(V_{2\mu}^T\, i\sigma_2)\,\Psi_l^{}$& $\bar u\, \Big({V_{2\mu}^{\nicefrac{+1}{3}}}\Big)^c$ & $f(\nicefrac{-2}{3},\nicefrac{M_\phi^2}{s})$\\[2mm]
		\cline{1-6}
		\multirow{2}{*}{$\widetilde V_{2\mu}$} & \multirow{2}{*}{$\nicefrac{-1}{3}$} &\nicefrac{1}{3} &  \multirow{2}{*}{$\bar q_u^{c}\,\gamma^\mu P_L\,(\widetilde V_{2\mu}^T\, i\sigma_2)\,\Psi_l^{}$} & $\bar u \, \Big({\widetilde V_{2\mu}^{\nicefrac{+1}{3}}}\Big)^c$ &$f(\nicefrac{-2}{3},\nicefrac{M_\phi^2}{s})$  \\
		&  &$\nicefrac{-2}{3}$& &--- & ------\\[2mm]
		\cline{1-6}
		\multirow{2}{*}{$U_{1\mu}$} & \multirow{2}{*}{\nicefrac{4}{3}} & \multirow{2}{*}{\nicefrac{2}{3}}& $\overline\Psi_{q}\,\gamma^\mu P_L\,\Psi_l^{}\,U_{1\mu}$ & \multirow{2}{*}{$d\,\Big({ U_{1\mu}^{\nicefrac{+2}{3}}}\Big)^c$} & \multirow{2}{*}{$f(\nicefrac{-1}{3},\nicefrac{M_\phi^2}{s})$}\\
		&&&$\bar q_d^{}\,\gamma^\mu P_R\,l_e^{}\,U_{1\mu}$&& \\[2mm]
		\cline{1-6}
		$\widetilde U_{1\mu}$ &\nicefrac{10}{3} & \nicefrac{5}{3} & $\bar q_u^{}\,\gamma^\mu P_R\,l_e^{}\,\widetilde U_{1\mu}$ & $u \,\Big({ \widetilde U_{1\mu}^{\nicefrac{+5}{3}}}\Big)^c$ & ------ \\[2mm]
		\cline{1-6}
		\multirow{3}{*}{$\vec{U}_{3\mu}$}& \multirow{3}{*}{\nicefrac{4}{3}} &\nicefrac{5}{3} & \multirow{3}{*}{$\overline\Psi_{q}\,\gamma^\mu P_L\,U_{3\mu}^{ad}\,\Psi_l^{}$} & $u \,  \Big({U_{3\mu}^{\nicefrac{+5}{3}}}\Big)^c$ & ------ \\
		&  &\nicefrac{2}{3}  &  & $ d\,\Big({ U_{3\mu}^{\nicefrac{+2}{3}}}\Big)^c$ & $f(\nicefrac{-1}{3},\nicefrac{M_\phi^2}{s})$\\
		&  &$\nicefrac{-1}{3}$ & & --- & ---\\[2mm]
		\hlineB{5}
	\end{tabular}
	\caption{The values of $\cos\theta^*$ for production of different leptoquarks at $e^-\gamma$ collider.}
	\label{tab:LQ}
\end{table}

\section{Mass and coupling}
\label{sec:mass-coupling}

\begin{figure}[H]
	\includegraphics[scale=0.35]{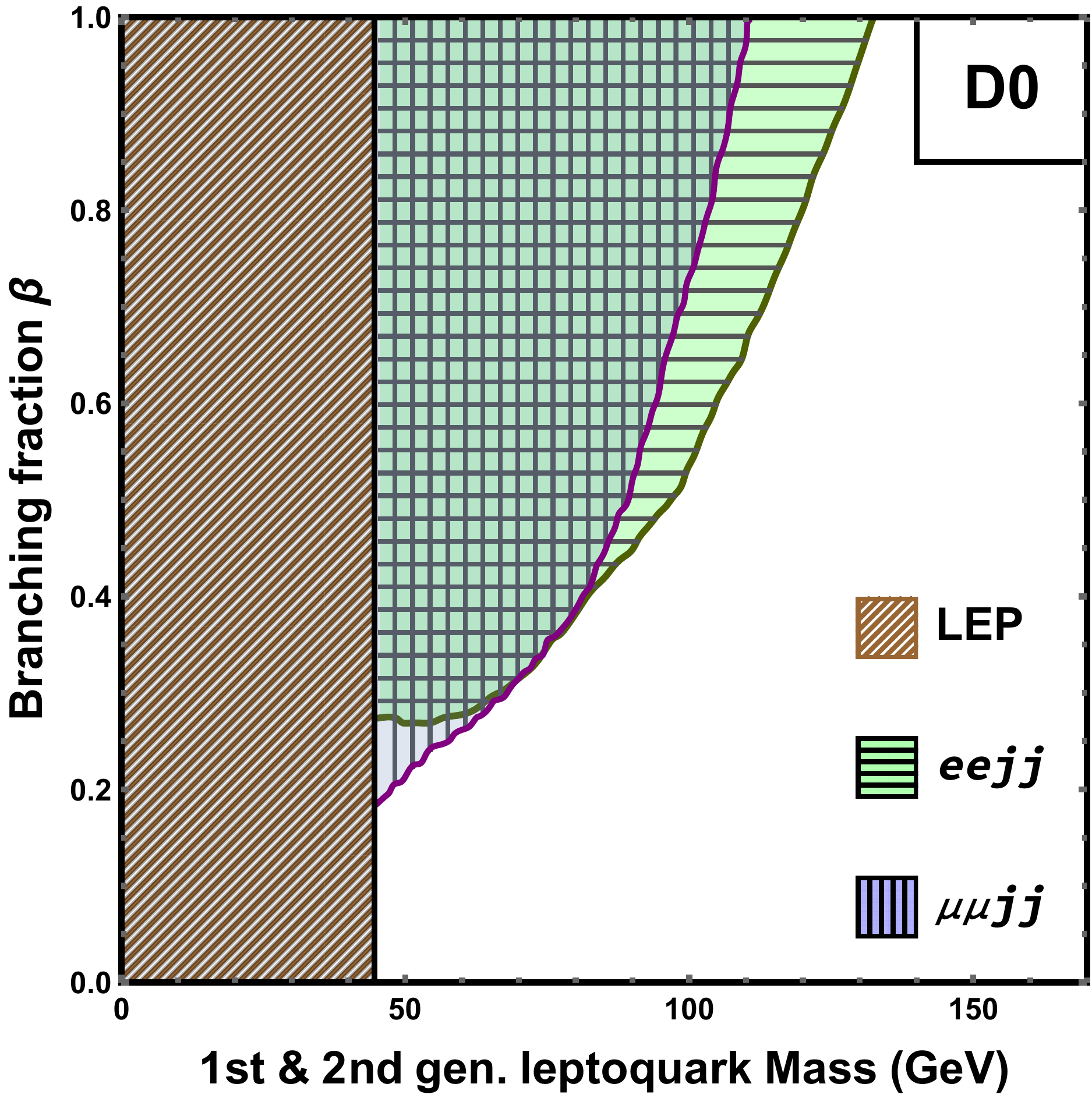}\hfill
	\includegraphics[scale=0.35]{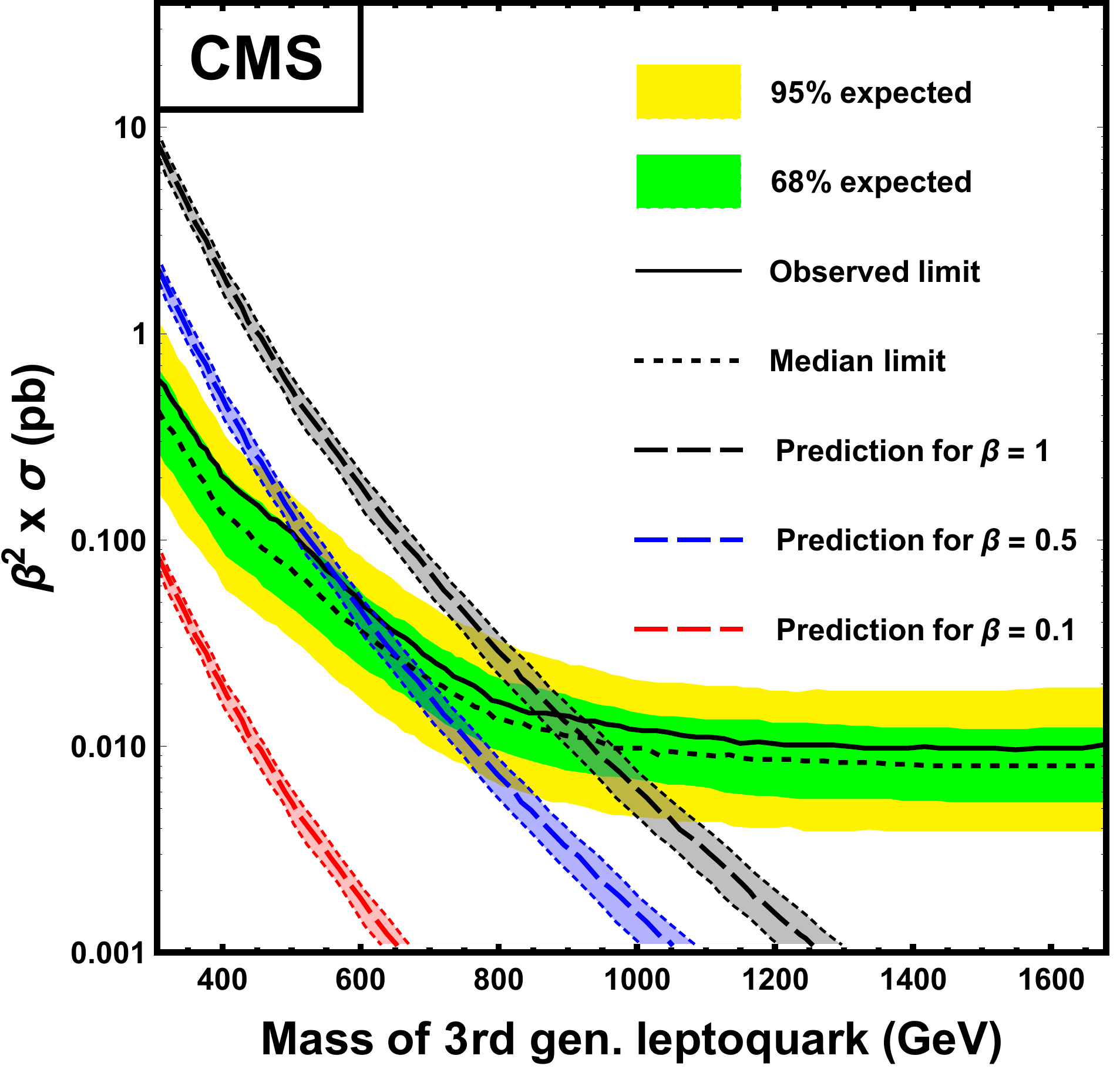}\\
	
	\includegraphics[scale=0.35]{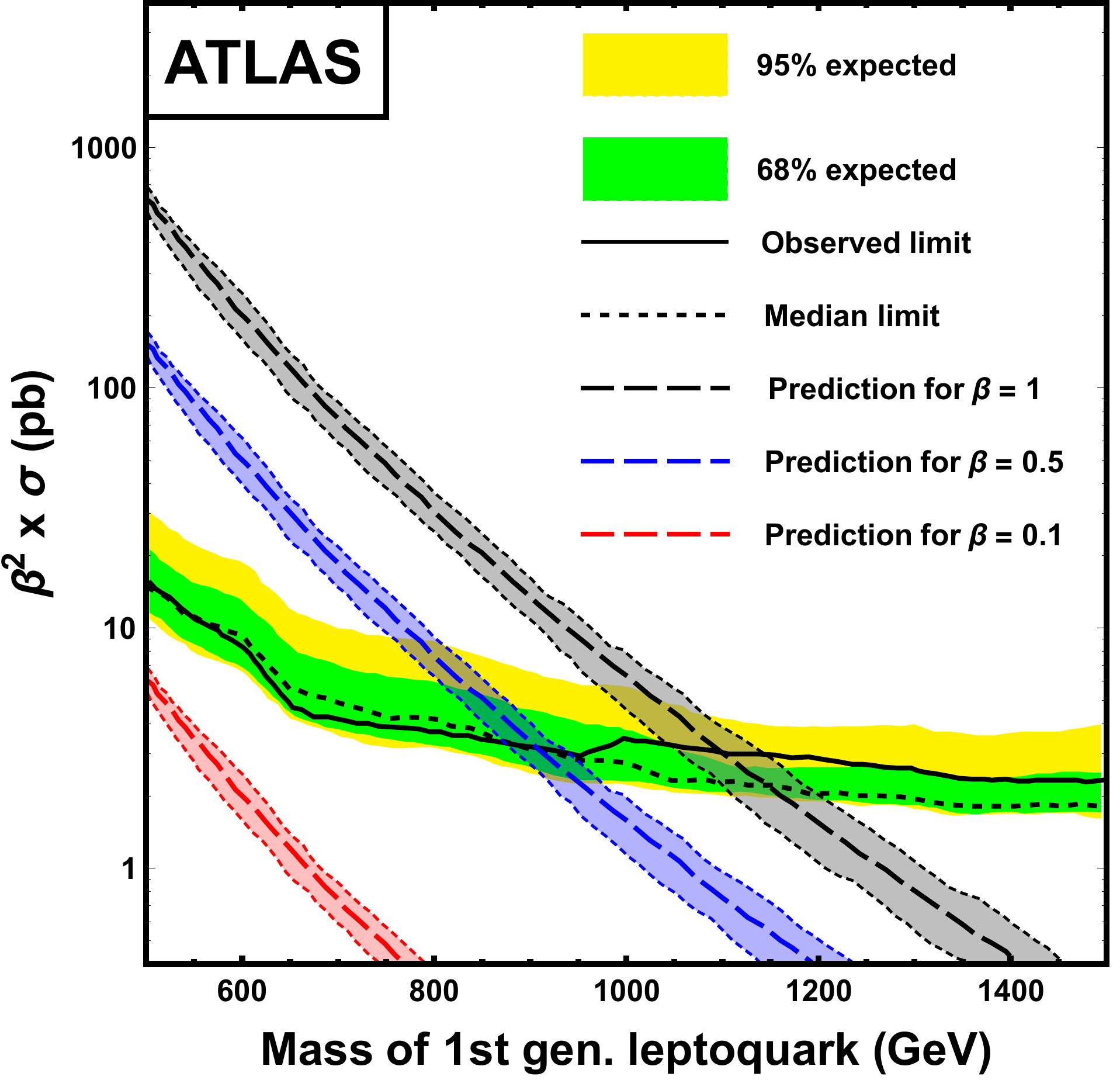}\hfill
	\includegraphics[scale=0.35]{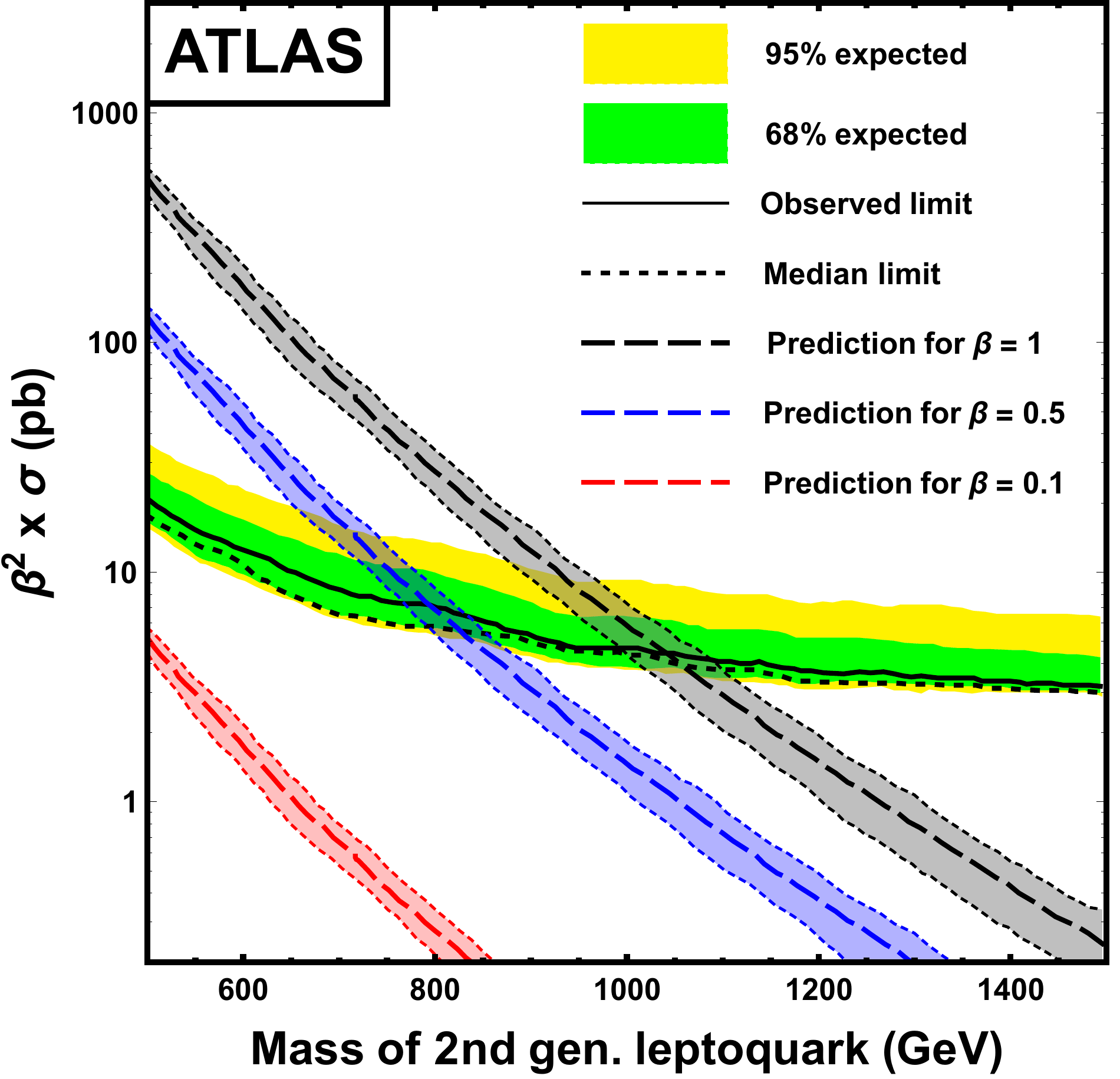}\\
	\caption{Data from D0, CMS and ATLAS for the branching fraction against the allowed mass range for different generations of leptoquarks.}
	\label{fig:expt}
\end{figure}

 The measurement of $R-$ratio from PEP and PETRA constrains the scalar leptoquarks to have $M_{\phi}\gtrsim15-20$ GeV \cite{Exp10a} in a model-independent way depending on the charges of them only where they are assumed to be pair-produced in the decay of a virtual photon.  Measurement from AMY \cite{Exp9a} provides $M_{\phi}\geq22.6$ GeV for scalar leptoquarks and similar bound for vector ones too. The LEP constrains $M_{\phi}\geq44$ GeV \cite{LEP1,LEP2} with the coupling to $Z^0$ to be $\nicefrac{1}{3}\sin^2\theta_w$  assuming the pair-production of leptoquarks from $Z^0$ and further decay of them into jets and two leptons. For decay into first two generations of quarks and leptons, this lower bound is almost independent of branching fraction; however for third generation it depends slightly. UA2 provides the relation between lowest allowed mass and the branching ratio of the leptoquark \cite{Exp6a}. Assuming $50\%$ branching to first generation, di-electron+ di-jet channel gives $M_{\phi}\geq58$ GeV, electron+$\slashed{p}_T$+di-jet channel shows $M_{\phi}\geq60$ GeV and combination of them provides $M_{\phi}\geq67$ GeV. However, $100\%$ branching to first generation will exclude the mass lower than 74 GeV. DELPHI concludes  $M_{\phi}\geq77$ GeV \cite{Exp7a}, but their analysis assumes large coupling for leptoquark-lepton-quark $(\lambda\geq e)$. CDF and D0 suggest the mass of leptoquarks to be greater than 113 GeV and 126 GeV \cite{Exp8a} respectively, on first and second generation of leptoquarks. Several bounds from meson decays, meson-antimeson mixing, lepton flavour violating decays, lepton-quark universality, $g-2$ of muon and electron, neutrino oscillation and other rare processes have been presented in Ref. \cite{Rev6a,Motiv5a,LQbound1,LQbound2,pdg}. If the leptoquark couples to left handed quarks and leptons of first generation only, then according to pdg \cite{pdg} $\lambda^2\leq0.07\times\widetilde M_{\phi}^2$ for scalar leptoquark and  $\lambda^2\leq0.4\times\widetilde M_{\phi}^2$ for the vector one where $\widetilde M_{\phi}\equiv(\frac{M_{\phi}}{1TeV})$; however, the constraints change for the second generation as $\lambda^2\leq0.7\times\widetilde M_{\phi}^2$ (scalar) and  $\lambda^2\leq0.5\times\widetilde M_{\phi}^2$ (vector). This analysis is done for leptoquark induced four-fermion interaction. Results from ATLAS and CMS \cite{Exp2a,Exp2b,Exp3a} rule out leptoquarks with mass upto 1500 GeV for first and second generation leptoquarks with 100\% branching and 1280 GeV for 50\% branching.

  In the fig. \ref{fig:expt}, we show the plots for branching fraction against the mass of leptoquark from Tevatron and LHC. In the top left panel, data from D0 has been presented, where the brown (obliquely meshed) region represents the disallowed mass range for leptoquark from LEP experiment and the  greenish (horizontally meshed) and bluish (vertically meshed) areas indicate the excluded portions for the mass of first and second generation leptoquarks from two-electron plus two-jet and two-muon plus two-jet channels at D0. The rest three plots are from LHC for three generations of leptoquarks. The continuous black line signify the observed limit whereas the green and yellow areas indicate $1\sigma$ and $2\sigma$ regions. The black, blue and red portions with dashed line inside show theoretical predictions with branching $(\beta)$ to be 100\%, 50\% and 10\% respectively.  Nevertheless, all these analyses have been done assuming that one leptoquark couples to quark and lepton from one generation only. The scenario changes drastically if branching for a leptoquark to quarks and leptons of all the generations are kept open.

\section{Leptoquark models and simulation}
\label{sec:model-simulation}

For our purpose, we choose four leptoquarks of different charges from scalar sector and same from the vector sector separately. For every leptoquark scenario, we have studied three different benchmark points (with mass 70 GeV, 650 GeV and 1.5 TeV respectively and different couplings), each of which has been scrutinised at three distinct energy scale (200 GeV, 2
TeV, 3 TeV). The couplings have been picked out in such a way that they lie inside the allowed region, as shown in fig. \ref{fig:expt}. For low mass leptoquark we use the data from D0, which allows around 25\% branching to first and second generations of quarks and leptons at $M_\phi=70$ GeV. For the heavy leptoquark scenarios, one should look at the graphs from ATLAS and CMS. There is no data for ATLAS beyond the mass range $500 \text{ GeV}>M_\phi> 1.5\text{ TeV}$; similarly CMS probes the mass range for leptoquark to be $300 \text{ GeV}>M_\phi> 1.7\text{ TeV}$.
\begin{table}[H]
	\begin{center}
		\renewcommand{\arraystretch}{1.4}
		\begin{tabular}{?>{\centering}m{1.2cm}|>{\centering}m{1.2cm}||c||>{\centering}m{0.6cm}|>{\centering}m{0.6cm}|>{\centering}m{0.6cm}||>{\centering}m{0.6cm}|>{\centering}m{0.6cm}|m{0.6cm}?}\hlineB{5}
			Lepto-&{Bench-} &	$M_{\phi}$ & \multirow{3}{*}{$Y_L^{11}$}&\multirow{3}{*}{$Y_L^{22}$}&\multirow{3}{*}{$Y_L^{33}$}&\multirow{3}{*}{$Y_R^{11}$}&\multirow{3}{*}{$Y_R^{22}$}&\multirow{3}{*}{$Y_R^{33}$}\\ 
			quarks	&	mark& in	&&&&&&\\
			&	points& GeV	&&&&&&\\\hline\hline
			$(S_1^{\nicefrac{+1}{3}})^c$,&	BP1& 70	& 0.035 & 0.04 & 0.035 & 0.03 & 0.03 & 0.03  \\ \cline{2-9}
			$(R_2^{\nicefrac{+5}{3}})^c$,&	BP2& 650	& 0.1 & 0.1 & 0.1 & 0.1 & 0.1 & 0.1  \\  \cline{2-9}
			$(U_{1\mu}^{\nicefrac{+2}{3}})^c$	&	BP3& 1500	& 0.1 & 0.1 & 0.1 & 0.1 & 0.1 & 0.1  \\ \hline\hline
			
			$(\widetilde R_2^{\nicefrac{+2}{3}})^c$,	&BP1  & 70	& 0.07 & 0.07 & 0.1  &---&---&---\\ \cline{2-9}
			$(S_3^{\nicefrac{+4}{3}})^c$,	&BP2  &650	& 0.07 & 0.07 & 0.1 & --- &---&---\\ \cline{2-9}
			$(\widetilde V_{2\mu}^{\nicefrac{+1}{3}})^c$,	& \multirow{2}{*}{BP3}  &\multirow{2}{*}{1500}	& \multirow{2}{*}{0.07} & \multirow{2}{*}{0.07} & \multirow{2}{*}{0.1} & \multirow{2}{*}{---} &\multirow{2}{*}{---}&\multirow{2}{*}{---}\\
			$(U_{3\mu}^{\nicefrac{+5}{3}})^c$&&&&&&&&\\
			\hline\hline
			
			\multirow{3}{*}{$(V_{2\mu}^{\nicefrac{+4}{3}})^c$}&	BP1& 70	& 0.05 & 0.05 & 0.1 & 0.1 & 0.1 & 0.1  \\ \cline{2-9}
			&	BP2& 650	& 0.05 & 0.05 & 0.1 & 0.1 & 0.1 & 0.1  \\  \cline{2-9}
			&	BP3& 1500	& 0.05 & 0.05 & 0.1 & 0.1 & 0.1 & 0.1  \\ \hlineB{5}
		\end{tabular}
		\caption{Benchmark points for different leptoquark scenarios.}
		\label{tab:bp}
	\end{center}
\end{table}

 \begin{figure}[H]
 	\begin{center}	
     \includegraphics[scale=0.45]{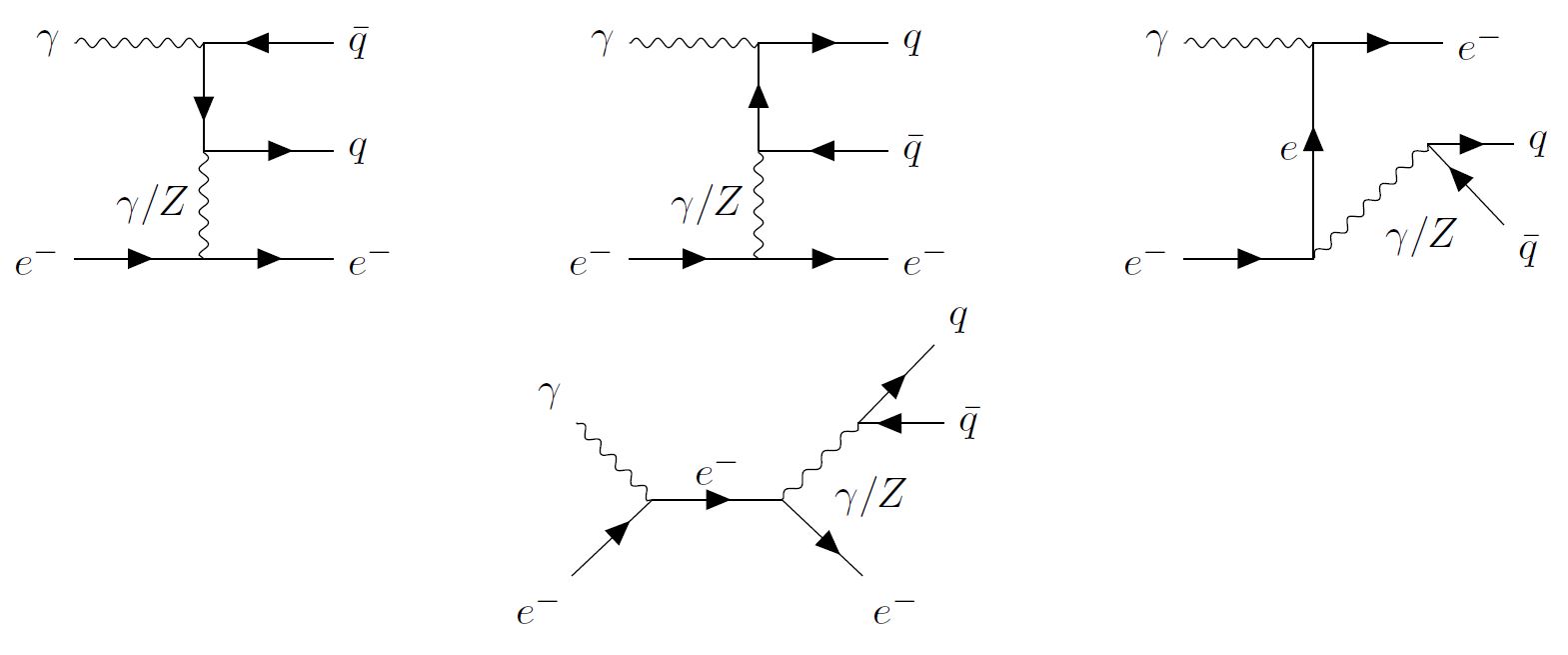}
 	\end{center}
 	\caption{Feynman diagrams for the SM background of the process $e^-\,\gamma\to e^- +2 jets$}
 	\label{fig:background}
 	\hspace*{0.5cm}
 \end{figure}

\begin{table}[H]
	\begin{center}
		\renewcommand{\arraystretch}{1.4}
		\begin{tabular}{?>{\centering}m{0.6cm}||c|c|c?>{\centering}m{0.6cm}||c|c|c?}\hlineB{5}
			$\sqrt{s}$ in &\multicolumn{3}{c?}{Cross-section in fb}&$\sqrt{s}$ in &\multicolumn{3}{c?}{Cross-section in fb}\\ \cline{2-4}\cline{6-8}
			TeV&BP1&BP2&BP3 &TeV&BP1&BP2&BP3\\ \hlineB{5}
			\multicolumn{4}{?c?}{Leptoquark $(S_1^{\nicefrac{+1}{3}})^c$}&\multicolumn{4}{c?}{Leptoquark $({U}_{1\mu}^{+2/3})^c$}\\\hline\hline
			
			0.2&430.24&---&---&0.2&482.41&---&---\\ \hline
			2.0&6.61&50.65& 31.95 &2.0&803.82&58.95&14.84 \\ \hline
			3.0&3.30&26.03&17.98&3.0&812.59&68.04&10.55\\ \hlineB{5}
			\multicolumn{4}{?c?}{Leptoquark $(R_2^{\nicefrac{+5}{3}})^c$}&\multicolumn{4}{c?}{Leptoquark $({V}_{2\mu}^{+4/3})^c$}\\\hline\hline
			0.2&517.5&---&---&0.2&12343.51&---&---\\ \hline
			2.0&8.10&59.30&35.96 &2.0&19110.75&152.70&15.38\\ \hline
			3.0&3.70&30.79&20.70&3.0&19214.64&181.61&21.40\\ \hlineB{5}
			
			\multicolumn{4}{?c?}{Leptoquark $(\widetilde R_2^{\nicefrac{+2}{3}})^c$}&\multicolumn{4}{c?}{Leptoquark $(\widetilde{V}_{2\mu}^{+1/3})^c$}\\\hline\hline

			0.2&226.83&---&---&0.2&2127.02&---&---\\ \hline
			2.0&3.61&2.89& 1.78 &2.0&485.34&26.58&16.38\\ \hline
			3.0&1.66&1.49&1.02&3.0&477.98&15.46&9.18\\ \hlineB{5}
			
			\multicolumn{4}{?c?}{Leptoquark $(S_3^{\nicefrac{+4}{3}})^c$}&\multicolumn{4}{c?}{Leptoquark $(U_{3\mu}^{+5/3})^c$}\\\hline\hline
			0.2&327.44&---&---&0.2&9579.55&---&---\\ \hline
			2.0&5.33&3.95& 2.27 &2.0&11769.27&117.41&21.17\\ \hline
			3.0&2.43&2.08&1.36&3.0&11783.95&124.50&20.50\\  \hlineB{5}
		\end{tabular}
		\caption{Production cross-sections for the chosen leptoquarks at $e$-$\gamma$ collider for the benchmark points listed in table \ref{tab:bp} at centre of momentum energies to be 200 GeV, 2 TeV and 3 TeV.}
		\label{tab:cross-section}
	\end{center}
\end{table}

The benchmark points used in our analysis for different leptoquarks are described in table \ref{tab:bp}. It should be kept in mind that $\widetilde{R}_2$, $\vec S_3$, $\widetilde V_{2\mu}$ and $U_{1\mu}$
do not have any coupling to right-handed leptons. The production cross-sections and branching fractions for all the leptoquarks under consideration have been put together at table \ref{tab:cross-section} and \ref{tab:branching} respectively.  The tree-level cross-sections and branching fractions have been calculated using {\tt CalcHEP 3.7.5}~\cite{Belyaev:2012qa}. It should be noticed that the mass of the leptoquark being higher than the centre of momentum energy, the scenarios BP2 and BP3 can not be explored at $\sqrt s=$200 GeV. On the other hand, top being heavy than the  leptoquarks of BP1 case, it will not get produced by decay of the later one. The production cross-sections for the vector modes are in general higher than that of the scalar modes which happens mainly because of two reasons. Firstly, vector leptoquarks couple to the vector currents giving rise to very different distribution from the scalar case. Secondly, any vector leptoquark has three states of polarizations which enhance the production cross-section.

\begin{table}[H]
	\begin{center}
		\renewcommand{\arraystretch}{1.4}
		\begin{tabular}{?>{\centering}m{1cm}||c|c|c?>{\centering}m{1cm}||c|c|c?}\hlineB{5}
		
			 \multirow{2}{*}{Modes}	& \multicolumn{3}{c?}{Branching fraction}&\multirow{2}{*}{Modes}&\multicolumn{3}{c?}{Branching fraction}\\ \cline{2-4}
			\cline{6-8}
			&  BP1  &	BP2 &BP3&&BP1  &	BP2 &BP3\\ \hlineB{5}
				\multicolumn{4}{?c?}{Leptoquark $(S_1^{\nicefrac{+1}{3}})^c$}&\multicolumn{4}{c?}{Leptoquark $({U}_{1\mu}^{+2/3})^c$}\\\hline\hline
			$u e$	 & 0.245 & 0.229 & 0.223 &$\bar d e$& 0.222&0.225&0.223\\ \hline
			$c \mu$  & 0.288 & 0.229 & 0.223 &	$\bar s \mu$&0.261&0.225&0.223\\ \hline
			$t \tau$ &    ---    & 0.199 & 0.218&$\bar b \tau$ &0.222&0.225&0.223\\ \hline
			$d \nu_e$ & 0.141 & 0.114 & 0.112 &$\bar u \nu_e$&0.128&0.112&0.111\\ \hline
			$s \nu_\mu$ & 0.185 & 0.114 & 0.112 &$\bar c \nu_\mu$&0.167&0.112&0.111\\ \hline
			$b \nu_\tau$ & 0.140 & 0.114 & 0.112 &$\bar t \nu_\tau$&---&0.101&0.109\\ \hlineB{5}
			
			\multicolumn{4}{?c?}{Leptoquark $(R_2^{\nicefrac{+5}{3}})^c$}&\multicolumn{4}{c?}{Leptoquark $({V}_{2\mu}^{+4/3})^c$}\\\hline\hline
		$\bar u e$	&0.458&0.349&0.336& $de$&0.278&0.278&0.278\\ \hline
		$\bar c \mu$	&0.542&0.349&0.336&$s\mu$&0.278&0.278&0.278\\ \hline
		$\bar t \tau$	&---&0.302&0.327&$b\tau$&0.444&0.444&0.444\\ \hlineB{5}
		
		\multicolumn{4}{?c?}{Leptoquark $(\widetilde R_2^{\nicefrac{+2}{3}})^c$}&\multicolumn{4}{c?}{Leptoquark $(\widetilde{V}_{2\mu}^{+1/3})^c$}\\\hline\hline
		$\bar d e$&0.248&0.247&0.247&$u e$& 0.500&0.261&0.250\\ \hline
		$\bar s\mu$&0.248&0.247&0.247&$c \mu$&0.500&0.261&0.250\\  \hline
		$\bar b \tau$&0.503&0.505&0.505&$t \tau$ &---&0.478&0.500\\ \hlineB{5}
		
		\multicolumn{4}{?c?}{Leptoquark $(S_3^{\nicefrac{+4}{3}})^c$}&\multicolumn{4}{c?}{Leptoquark $(U_{3\mu}^{+5/3})^c$}\\\hline\hline
		$\overline{d} e^+$    & 0.248 & 0.247 & 0.247 &$u e^+$	 & 0.5 & 0.261 & 0.25 \\ \hline
		$\overline{s} \mu^+$  & 0.248 & 0.247 & 0.247 &	$c \mu^+$  & 0.5 & 0.261 & 0.25 \\ \hline
		$\overline{b} \tau^+$ & 0.503 & 0.505 & 0.505 &$\overline{b} \tau^+$ & 0.503 & 0.505 & 0.505 \\ \hlineB{5}
		\end{tabular}
		\caption{Branching fractions of the leptoquarks for the given benchmark points.}
		\label{tab:branching}
	\end{center}
\end{table}

The zeros of amplitude shows up for the leptoquarks having charges $\nicefrac{-1}{3}$ and $\nicefrac{-2}{3}$ only since the other ones fail to satisfy Eq. \eqref{eq:condition}. The zeros for all these scenarios have been merged in table \ref{tab:zeros}. It should be noted that unlike BP2 and BP3 at $\sqrt s=$ 200 GeV, leptoquark of 1.5 TeV mass (BP3) and charge $\nicefrac{-1}{3}$ gets produced at $\sqrt s=$ 2 TeV; but it does not show the zero in distribution since the ratio of its mass squared to $s$ is larger than its charge violating the condition in Eq. \eqref{eq:physreg}.  It should also be noticed that due to low mass of letoquark in BP1, $\cos\theta^*$ reaches the asymptotic value of $\nicefrac{\pm1}{3}$ at $\sqrt{s}=$ 2 TeV and 3 TeV in both the cases of $Q_\phi$ being $\nicefrac{-1}{3}$ and $\nicefrac{-2}{3}$. In the next few sections, we discuss the kinematical distributions leading to appropriate cuts and final states. Later, we present the signal and background number for those final states for different centre of momentum energies at the integrated luminosity of 100 fb$^{-1}$. 

\begin{table}[H]
	\begin{center}
		\renewcommand{\arraystretch}{1.4}
		\begin{tabular}{?c||c|c|c||c|c|c?}\hlineB{5}
			Benchmark 	& 	\multicolumn{6}{c?}{Values of $\cos\theta^*$ for zeros of $(d\sigma/d\cos\theta)$ at different $\sqrt{s} $ }\\ \cline{2-7}
			\noalign{\vskip\doublerulesep\vskip-\arrayrulewidth} \cline{2-7}
			points  &\multicolumn{3}{c||}{For $Q_{\bar q}=\nicefrac{-2}{3}$ or $Q_\phi=\nicefrac{-1}{3}$  }&\multicolumn{3}{c?}{For $Q_{q}=\nicefrac{-1}{3}$ or $Q_\phi=\nicefrac{-2}{3}$ }\\ \cline{2-7}
			& 0.2 TeV & 2 TeV & 3 TeV&0.2 TeV & 2 TeV & 3 TeV \\\hline
			\hline
			BP1 & $-\;0.52$ & $-\;0.33$ & $-\;0.33$& $0.24$ & $0.33$ & $0.33$ \\ \hline
			BP2 & --- & $-\;0.49$ & $-\;0.40$&--- & $0.25$ & $0.30$  \\ \hline
			BP3 & --- & --- & $-\;0.78$&--- & $-0.52$ & $0.11$ \\ \hlineB{5}
		\end{tabular}
		\caption{Values of $\cos\theta^*$ corresponding to zeros of differential cross-section for production of leptoquark at different centre of momentum energy for various benchmark points.}
		\label{tab:zeros}
	\end{center}
\end{table}

\subsection{Simulation set up}\label{sub:simul}
For the simulation in electron-photon collider we implement the scenarios in  {\tt SARAH 4.13.0} \cite{Staub:2013tta}. Later models files are generated for  {\tt CalcHEP 3.7.5} which is used for signal and background event generation. The generated events have then been simulated with {\tt PYTHIA 6.4}  \cite{pythia}. The simulation at hadronic level has been performed using the Fastjet-3.2.3 \cite{Fastjet} with with the {\tt CAMBRIDGE AACHEN} algorithm. For this, the jet size have been selected to be R = 0.5, with the following criteria:
\begin{itemize}
	\item Calorimeter coverage: $|\eta| < 4.5$.
	\item Minimum transeverse momentum of each jet: $p_{T,min}^{jet} = 20.0$ GeV; jets are ordered in $p_{T}$.
   \item Leptons ($\ell=e,\mu$)are selected with $p_T\geq 10$ GeV and $| \eta| \leq 2.5$.
	\item No jet should be accompanied by a hard lepton in the event.
	\item Jet-lepton isolation $\Delta R_{lj} > 0.4$ and lepton-lepton isolation$\Delta R_{ll} > 0.2$ 
	\item Selected leptons are hadronically clean, \textit{i.e,} hadronic activity within a cone of $\Delta R < 0.3$ around each lepton should be less than $15\%$ of the leptonic transeverse momentum, \textit{i.e.} $ p_{T}^{\mathrm{had}}< 0.15 p_{T}^{\text{lep}}$ within the cone.
\end{itemize}

Prepared with this set up, we analyse different leptoquark scenarios and plot the required invariant mass for jet and lepton and their angular correlations. This would guide us to choose the kinematical cuts appropriately.

The leptoquark will eventually decay into a quark (or antiquark) and a lepton providing mono-lepton plus di-jets signal at the electron photon collider. The SM background for this process, shown in fig.~\ref{fig:background}, is governed by eight Feynman diagrams for each generation of quark-antiquark pair mediated by photon and $Z$-boson (neglecting the one with Higgs boson propagator since its coupling with electron is very small).While plotting against the invariant mass of lepton-jet pair $(M_{\ell j})$, the background gives a continuum, whereas the signal shows a peak at $M_\phi$. So, to reconstruct the leptoquark, we first put a cut constraining  $(M_{\ell j})$ to deviate from $M_\phi$ by 10 GeV at most, which is denoted as ``cut1'' in all the signal background analysis table. Next, to distinguish the daughter jet produced by the decay of leptoquark, we apply an angular cut on the angle between the lepton and each of the jets depending on the boost of the leptoquark. If the three momentum of the leptoquark becomes small, the path of the daughter jet will make an obtuse angle with the final state lepton providing negative values of $\cos \theta_{\ell j}$, whereas for a highly boosted leptoquark, it makes an acute angle with the lepton giving positive valued  $\cos \theta_{\ell j}$. To enhance the significance, we choose the angular cut in such a way that the background reduces conspicuously without much change in the signal event.

\subsection{Scalar leptoquarks}
\label{sub:sig-bg-scalar}

\subsubsection{Leptoquark $(S_1^{\nicefrac{+1}{3}})^c$}

In table \ref{tab:S1_recons}, we summarise the signal background analysis for the scalar leptoquark $(S_1^{\nicefrac{+2}{3}})^c$. In case of BP1, all the three values of $\sqrt{s}$ (i.e. 200 GeV, 2 TeV and 3 TeV) are allowed for the production of 70 GeV leptoquark associated with a light jet. As discussed in last paragraph, the leptoquark produced at   $\sqrt{s}=$ 200 GeV will not be boosted highly and hence, we apply the angular cut as $-0.2\leq \cos \theta_{\ell j}\leq1$, which increases the significance from $47.5\sigma$ to $50.5\sigma$. But for $\sqrt{s}$ equal to 2 TeV and 3 TeV the leptoruark will be very highly boosted; so we put an angular cut of   $0.9\leq \cos \theta_{\ell j}\leq1$ that changes the significance from $6.8\sigma$ to $6.4\sigma$ and $3.7\sigma$ to $3.9\sigma$, respectively. In case of BP2, centre of momentum energy of 200 GeV  is forbidden for the production of 650 GeV leptoquark. For the rest of two values of $\sqrt{s}$, the leptoquark will be moderately boosted. So, an angular cut of  $0\leq \cos \theta_{\ell j}\leq1$ has been employed for both the cases. It 	elevates the significance from $8.1\sigma$ to $14.5\sigma$ and $4.1\sigma$ to $8.8\sigma$ for $\sqrt{s}$ to be 2 TeV and 3 TeV, respectively. On the other hand, for BP3 also, real leptoquark gets produced at 2 TeV and 3 TeV centre of momentum energy. At $\sqrt{s}=$ 2 TeV, the produced leptoquark of mass 1.5 TeV moves very slowly and hence an angular cut of   $-0.9\leq \cos \theta_{\ell j}\leq1$ has been implemented which enhances the significance to $7.7\sigma$ from $8.2\sigma$. Similarly, at $\sqrt{s}=$ 3 TeV, also a slow leptoquark gets produced for BP3. So, we  put an angular cut of   $-0.8\leq \cos \theta_{\ell j}\leq1$ which enhances the significance to $5.4\sigma$ from $3.5\sigma$.

\begin{table}[H]
	\begin{center}
		\renewcommand{\arraystretch}{1.4}
		\begin{tabular}{?>{\centering}m{1cm}||>{\centering}m{1cm}||c||>{\centering}m{1cm}|>{\centering}m{1cm}||m{0.8cm}?}\hlineB{5}
			Bench-mark points&$\sqrt s$ in TeV&Cut& Signal&Back-ground&Signi-ficance\\ \hline\hline
			\multirow{6}{*}{BP1}&	\multirow{2}{*}{0.2}&$|M_{lj}-M_{\phi}|\leq10$ GeV&11133.6&43725.0&47.5\\ 
			\cline{3-6}
			&&cut1+$(-0.2)\leq \cos \theta_{\ell j}\leq1$&10537.8&32989.8&50.5\\	\cline{2-6}
			&	\multirow{2}{*}{2}&$|M_{lj}-M_{\phi}|\leq10$ GeV&147.5&319.4&6.8\\ 	\cline{3-6}
				&&cut1+$(0.9)\leq \cos \theta_{\ell j}\leq1$&91.5&114.2&6.4\\	
			\cline{2-6}
			&	\multirow{2}{*}{3}&$|M_{lj}-M_{\phi}|\leq10$ GeV&61.2&219.8&3.7\\ 	\cline{3-6}
				&&cut1+$(0.9)\leq \cos \theta_{\ell j}\leq1$&34.5&44.2&3.9\\	
\hline\hline
		
				\multirow{4}{*}{BP2}&	\multirow{2}{*}{2}&$|M_{lj}-M_{\phi}|\leq10$ GeV&394.4&2003.6&8.1\\ 
				\cline{3-6}
			&&cut1+$0\leq \cos \theta_{\ell j}\leq1$&299.5&129.1&14.5\\	\cline{2-6}
			&	\multirow{2}{*}{3}&$|M_{lj}-M_{\phi}|\leq10$ GeV&176.5&1660.7&4.1\\ 
			\cline{3-6}
			&&cut1+$0\leq \cos \theta_{\ell j}\leq1$&159.0&167.5&8.8\\ \hline\hline
						\multirow{4}{*}{BP3}&
							\multirow{2}{*}{2}&$|M_{lj}-M_{\phi}|\leq10$ GeV& 280.8&1061.6&7.7\\ 
							\cline{3-6}
							&&cut1+$(-0.9)\leq \cos \theta_{\ell j}\leq1$&199.8&391.5&8.2\\	
							\cline{2-6}
							&\multirow{2}{*}{3}&$|M_{lj}-M_{\phi}|\leq10$ GeV&106.2&815.0&3.5\\ 
					\cline{3-6}
					&&cut1+$(-0.8)\leq \cos \theta_{\ell j}\leq1$&101.6&254.7&5.4\\	\hlineB{5}
		\end{tabular}
		\caption{Signal background analysis for  leptoquark $(S_1^{\nicefrac{+1}{3}})^c$   with luminosity 100 fb$^{-1}$ at $e$-$\gamma$ collider.}
		\label{tab:S1_recons}
	\end{center}
\end{table}

\begin{figure}[H]

\belowbaseline[-2mm]{\includegraphics[scale=0.30,angle=-90]{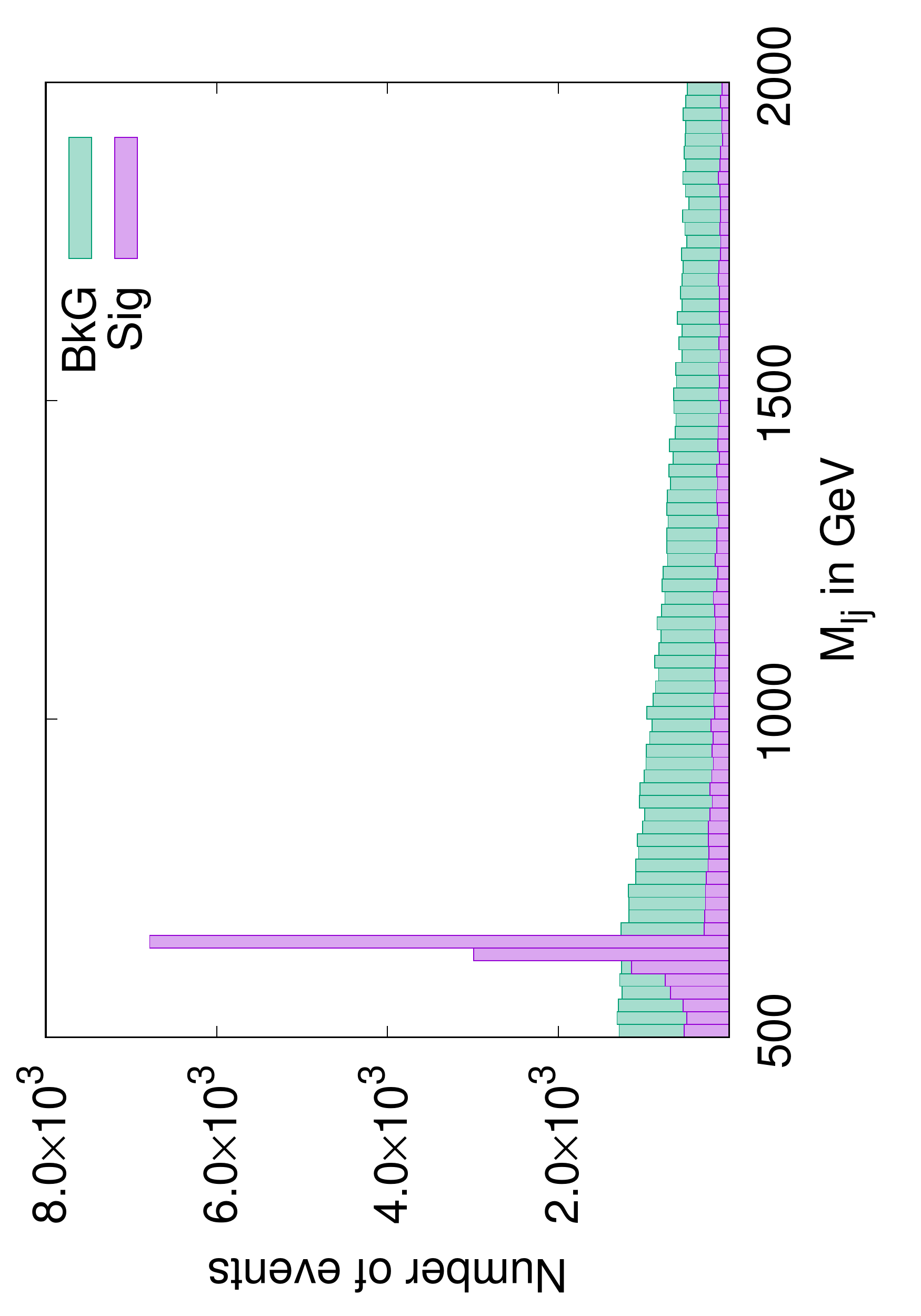}}~
\belowbaseline[0pt]{\includegraphics[scale=0.41]{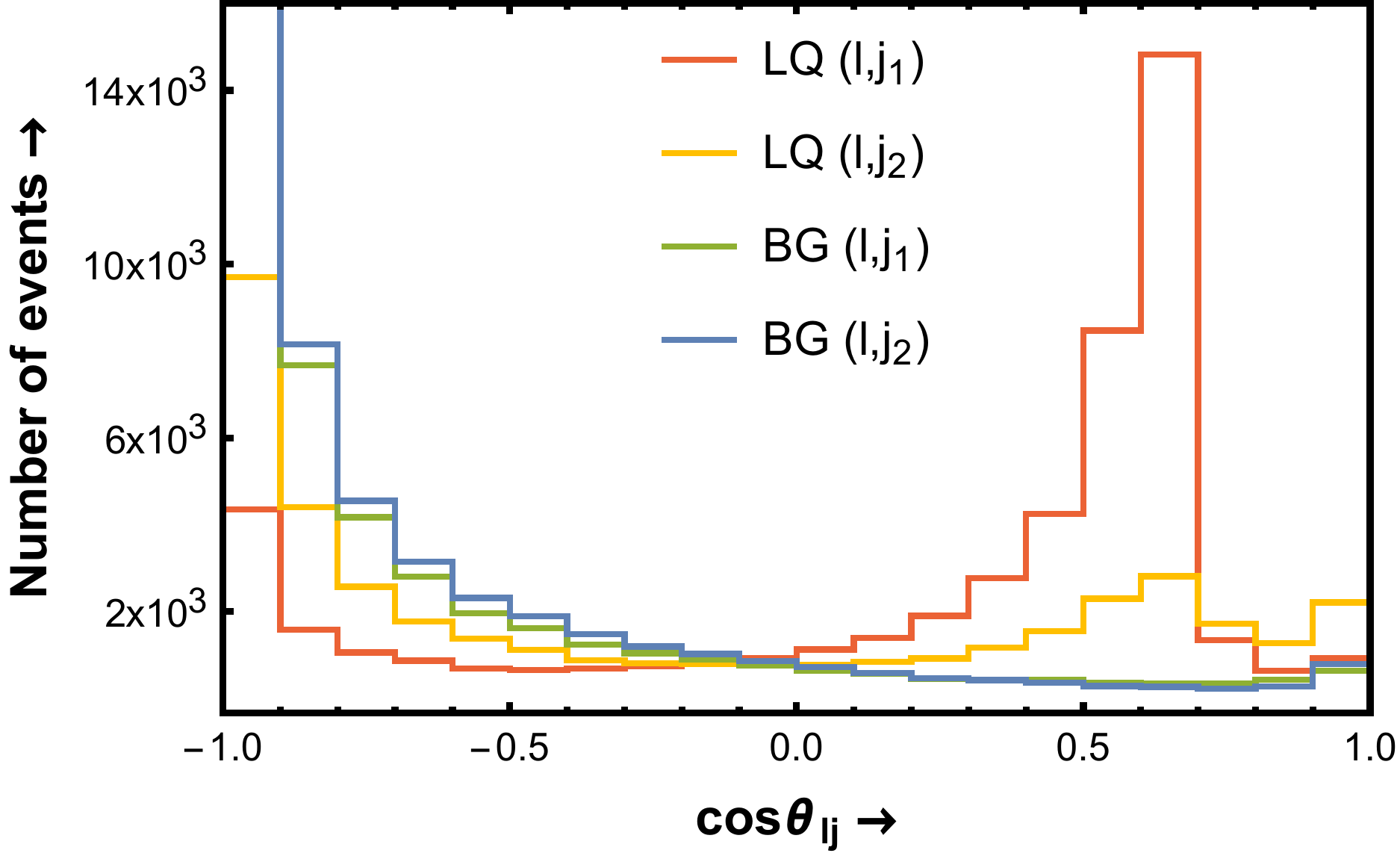}}\\[-5mm]

\belowbaseline[-2mm]{\includegraphics[scale=0.30,angle=-90]{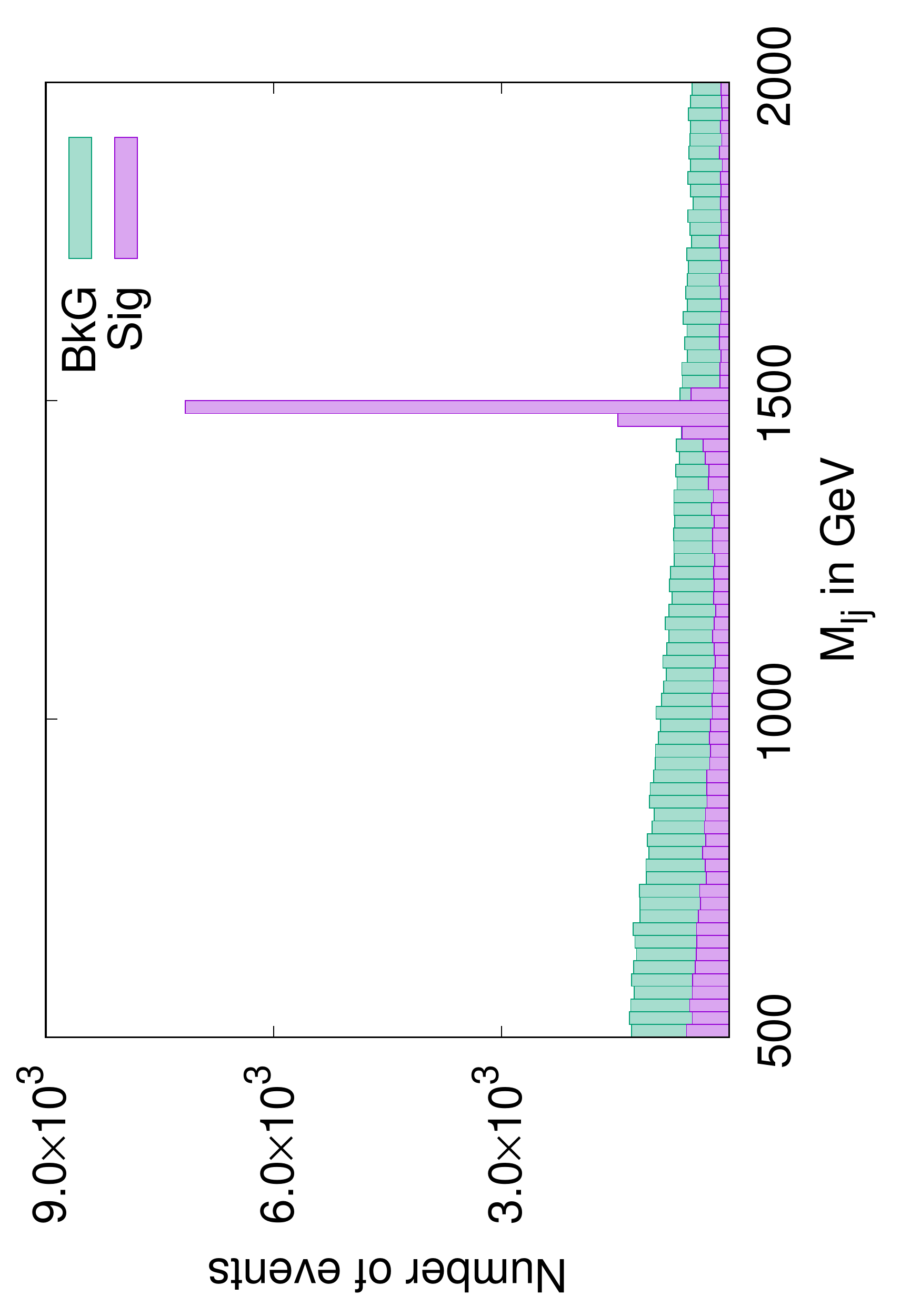}}~
\belowbaseline[0pt]{\includegraphics[scale=0.41]{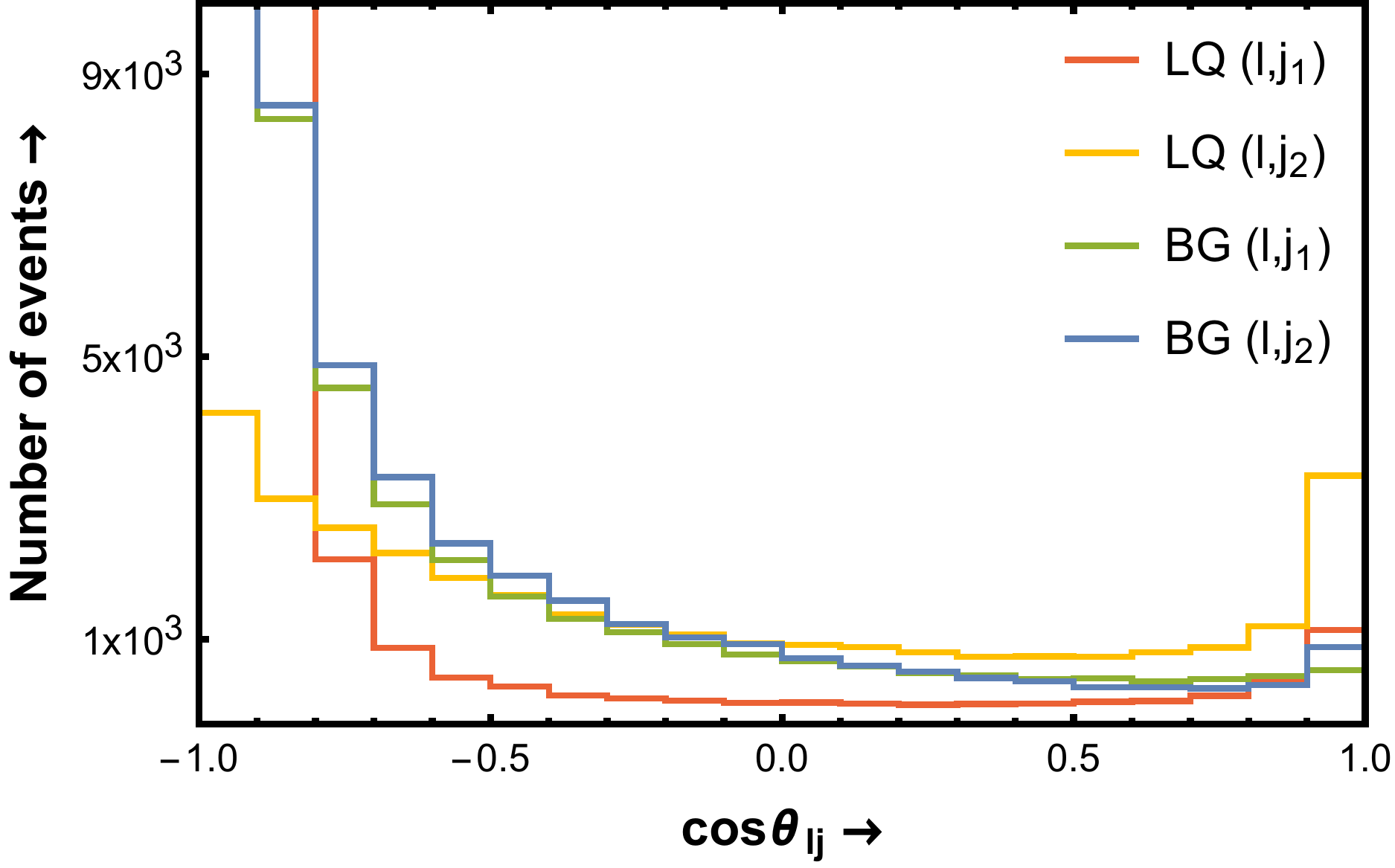}}\\[-5mm]

\belowbaseline[-2mm]{\includegraphics[scale=0.30,angle=-90]{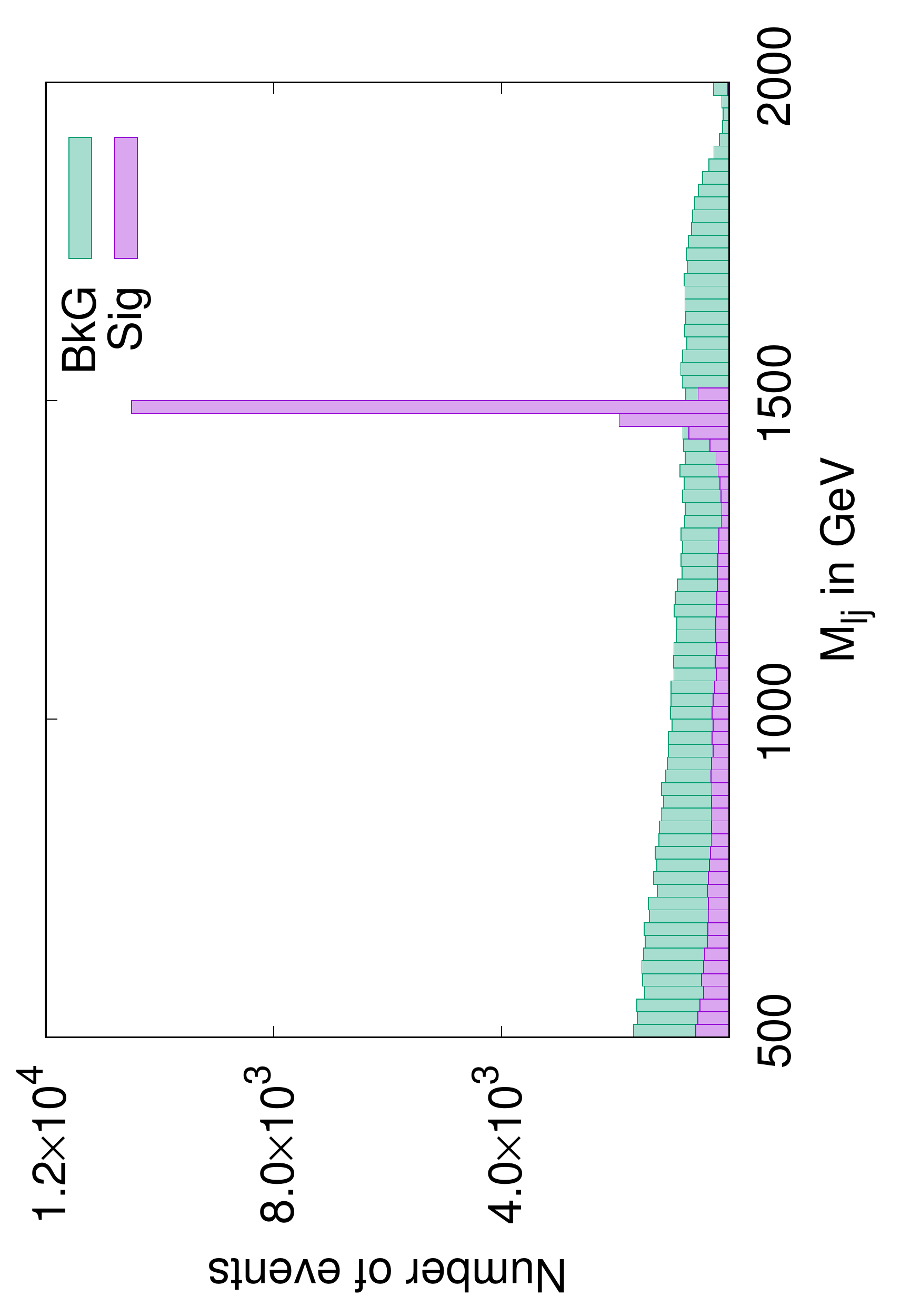}}~
\belowbaseline[0pt]{\includegraphics[scale=0.41]{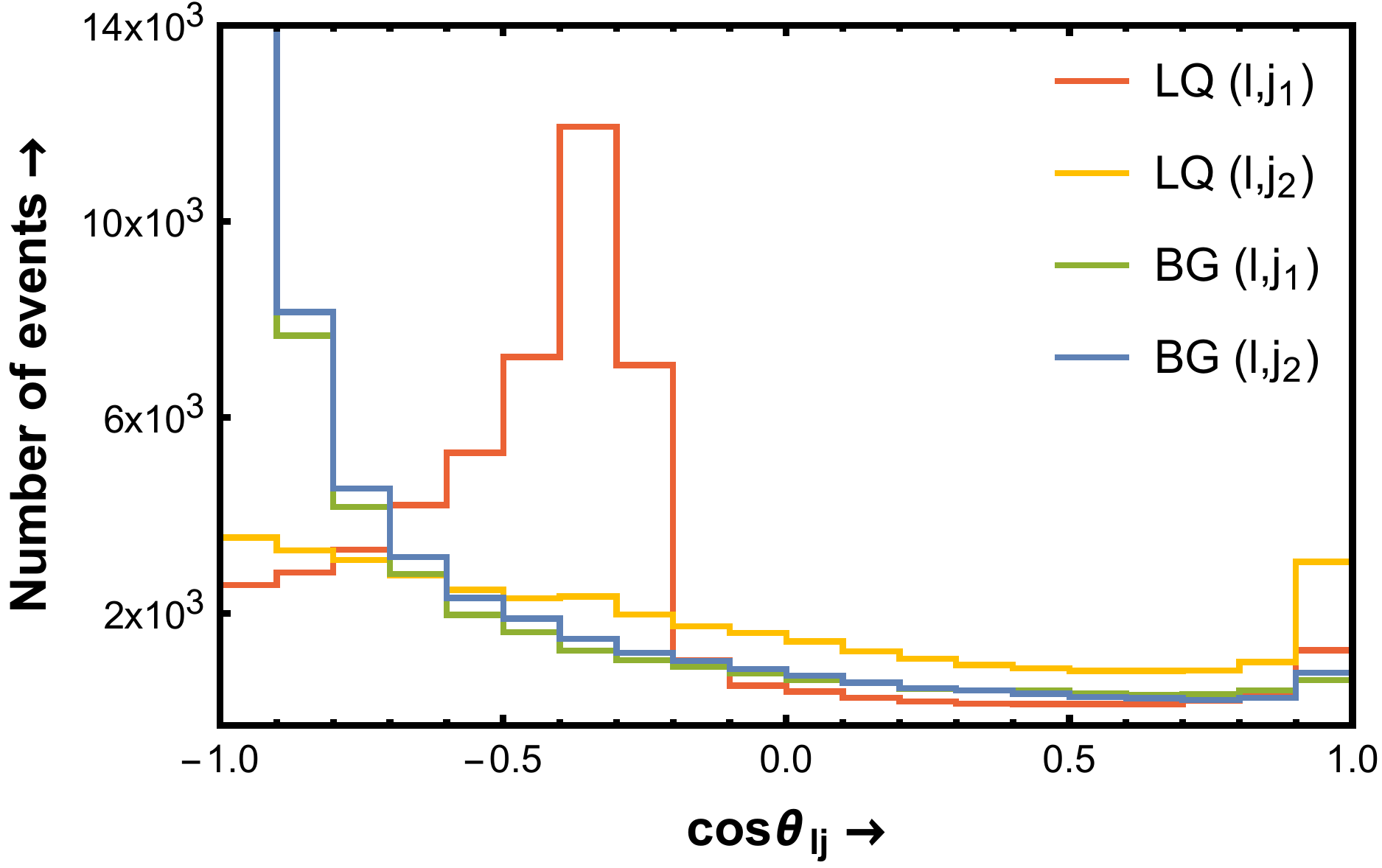}}\\[-5mm]
\caption{Signal background simulation for  leptoquark $(S_1^{\nicefrac{+1}{3}})^c$ for $10^5$ number of events. The plots are organized in  the same order as in table \ref{tab:S1_recons}. In the left panel, we show the number of events against the invariant mass of electron-jet pair for both signal (purple) and background (aqua). In the right panel we present the number of events for signal and backgrounds against the cosine of angle between final state electron and jet. The red and yellow symbolizes the signal events for electron with first and second jet respectively, whereas the green and blue indicate the background events for the same.}
\label{fig:sc_sing}
\end{figure}

In fig. \ref{fig:sc_sing}, we present the detailed pictorial description of our PYTHIA simulation  with $10^5$ number of events and luminosity of 100 fb$^{-1}$. The graphs are arranged in the same order like in table \ref{tab:S1_recons}. In the left panel, the number of events has been plotted against the invariant mass of electron and jet for both signal and background at different centre of mass energies for the three benchmark points. The greenish (aqua) regions indicate the SM background whereas, the purple regions signify the signal events. As expected, the signal events peak around the masses of leptoquarks. On the other hand, the number of events against the cosine of angle between the final state electron and the two jets has been plotted in the right panel for same benchmark points with same $\sqrt s$. While the blue and green lines represent the background events, the yellow and red lines depict the signal events. These plots justify our choice of cuts for invariant mass and the angle between final state lepton and the two jets. If any of the two jets passes those two cuts, we identify that as signal event.

\begin{figure}[H]
	\belowbaseline[-1mm]{\includegraphics[scale=0.3,angle=-90]{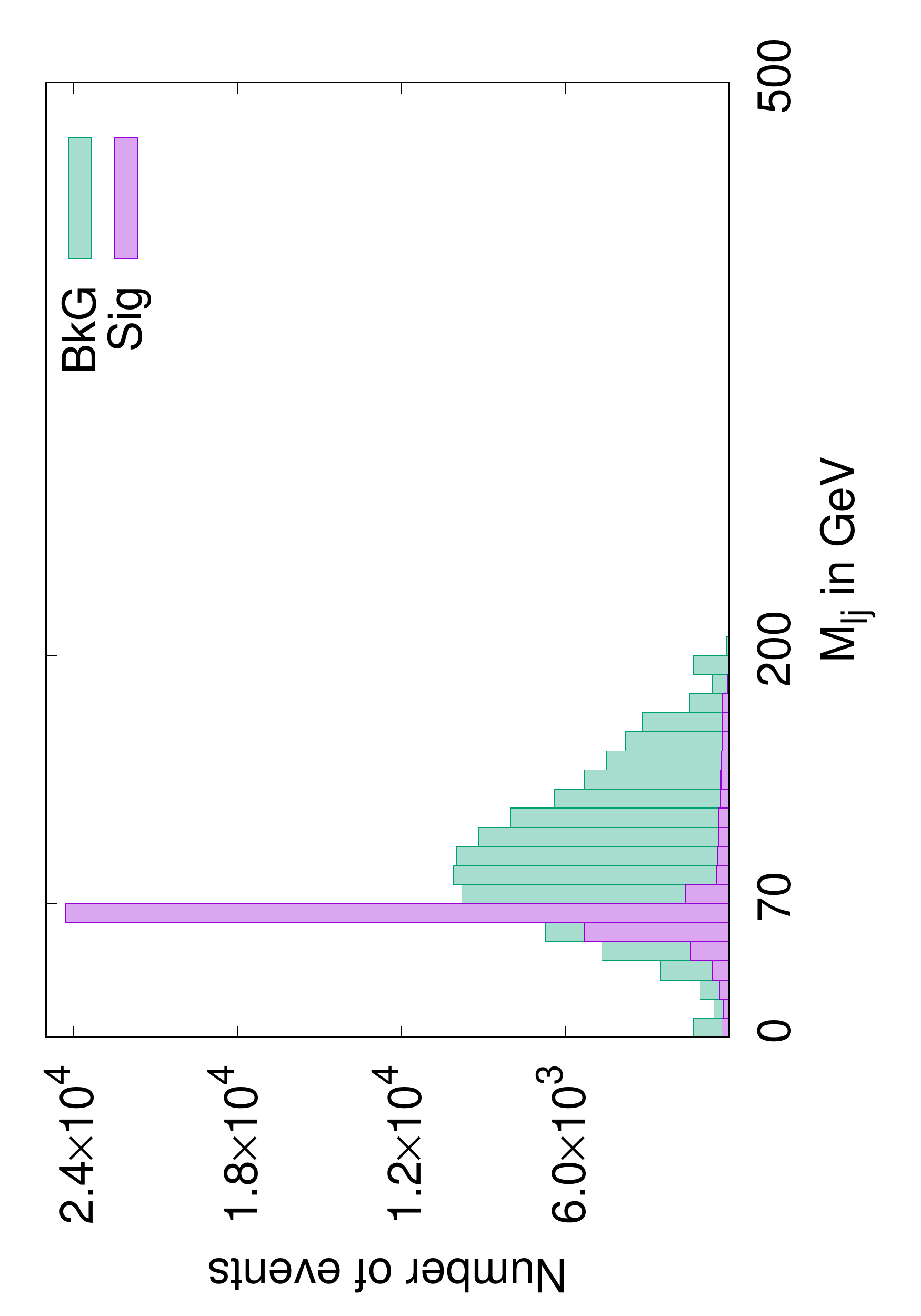}}~ \belowbaseline[0pt]{\includegraphics[scale=0.41]{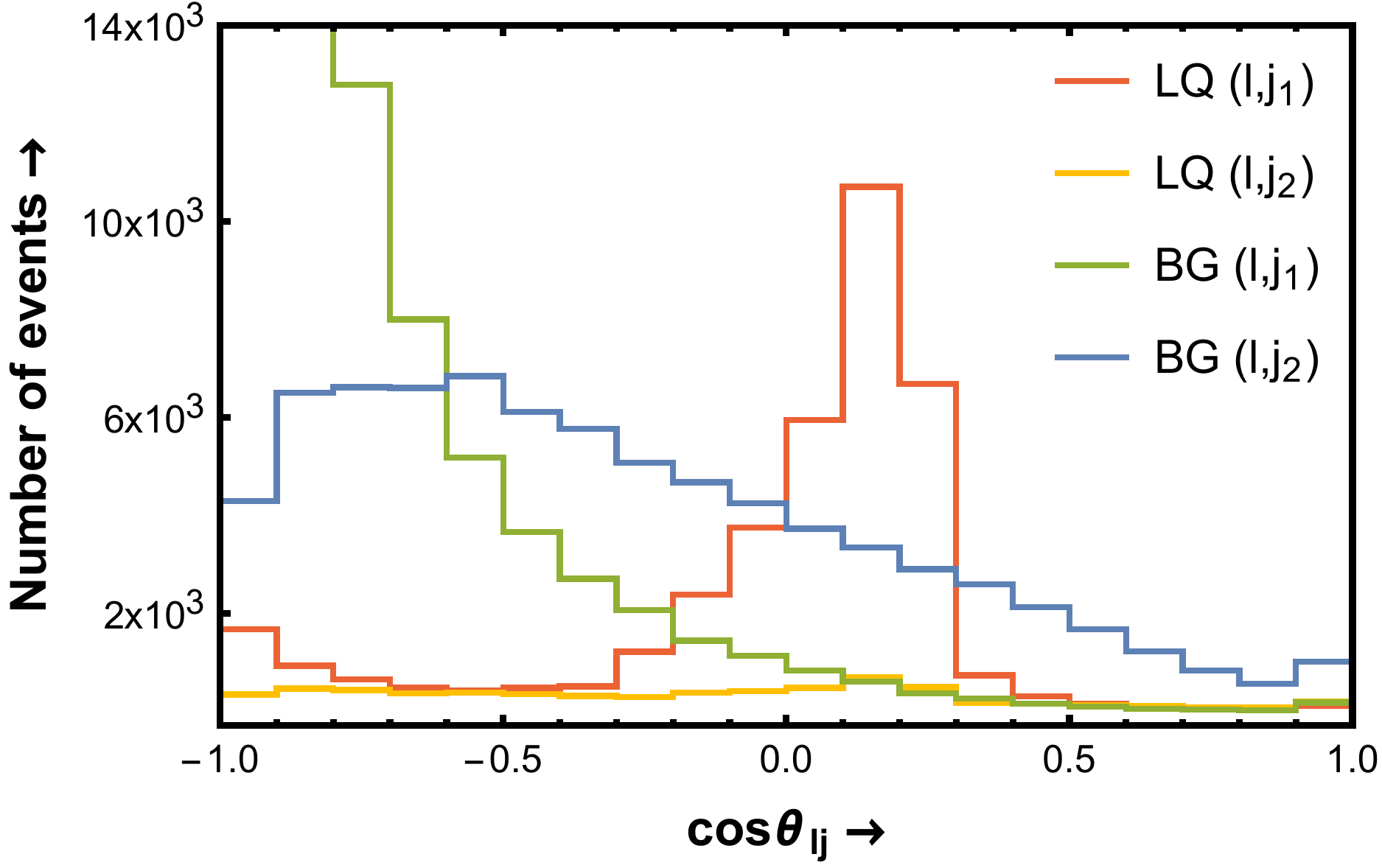}}\\
	
	\belowbaseline[-2mm]{\includegraphics[scale=0.30,angle=-90]{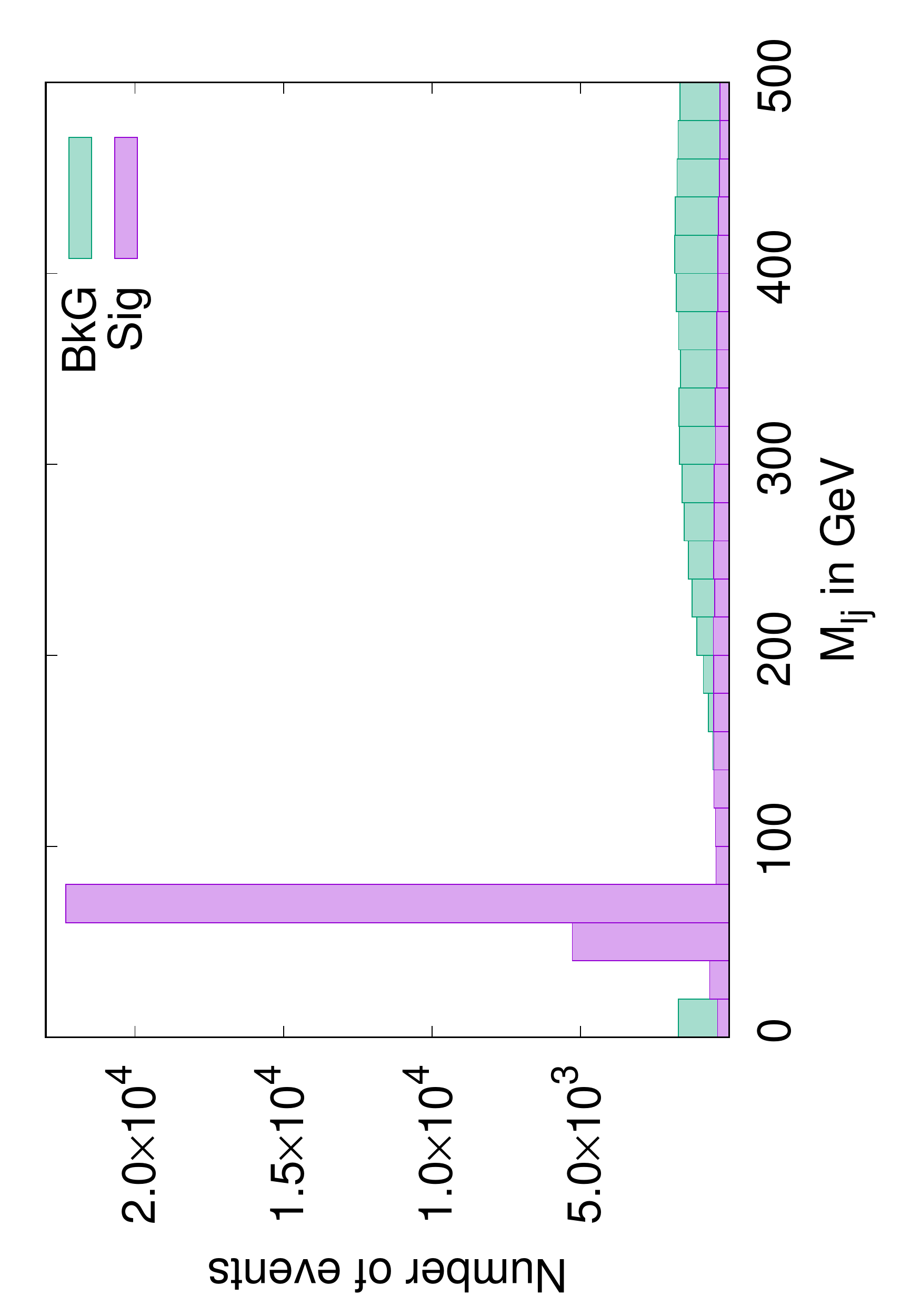}}~
    \belowbaseline[0pt]{\includegraphics[scale=0.41]{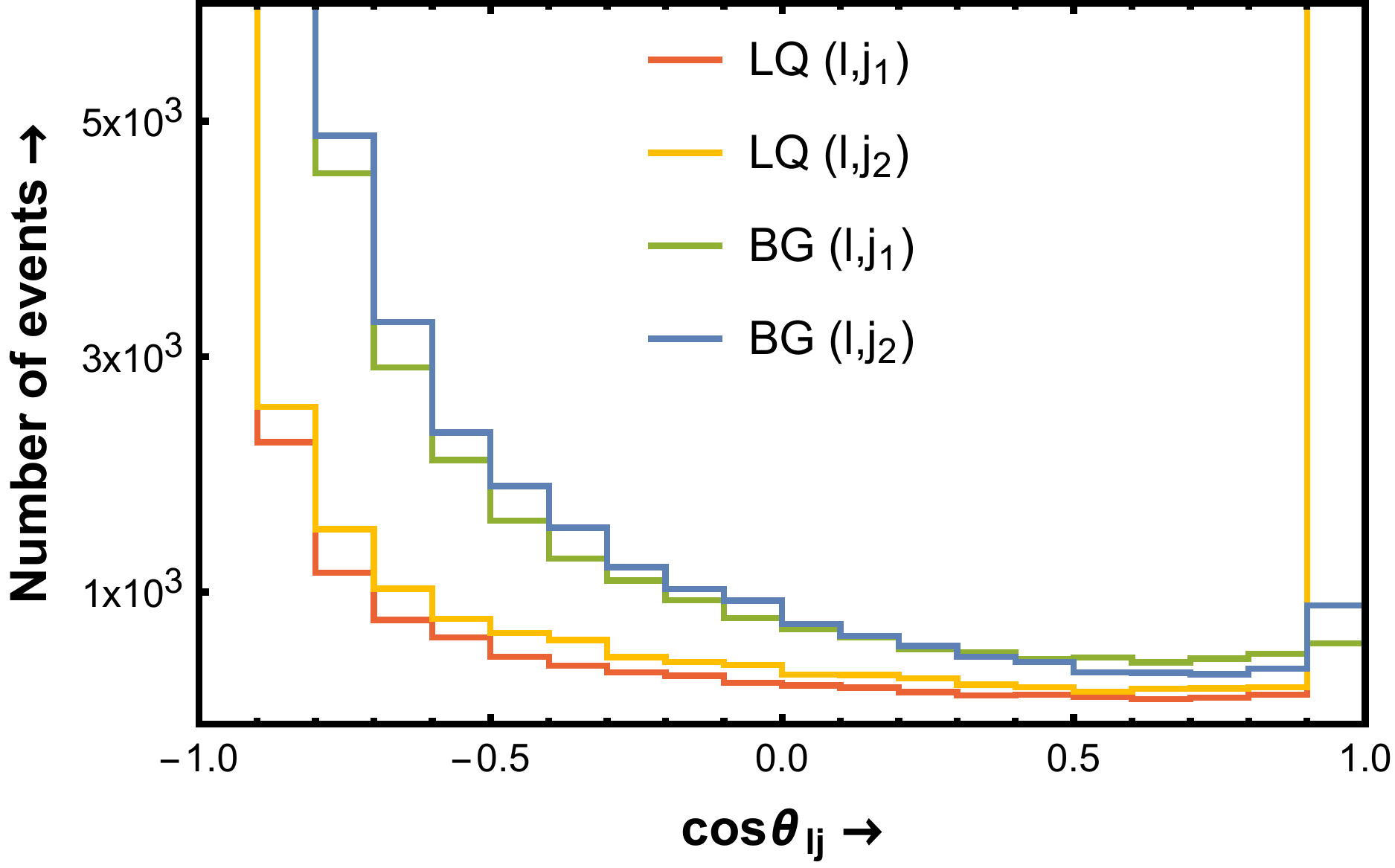}}\\
    
    \belowbaseline[-2mm]{\includegraphics[scale=0.30,angle=-90]{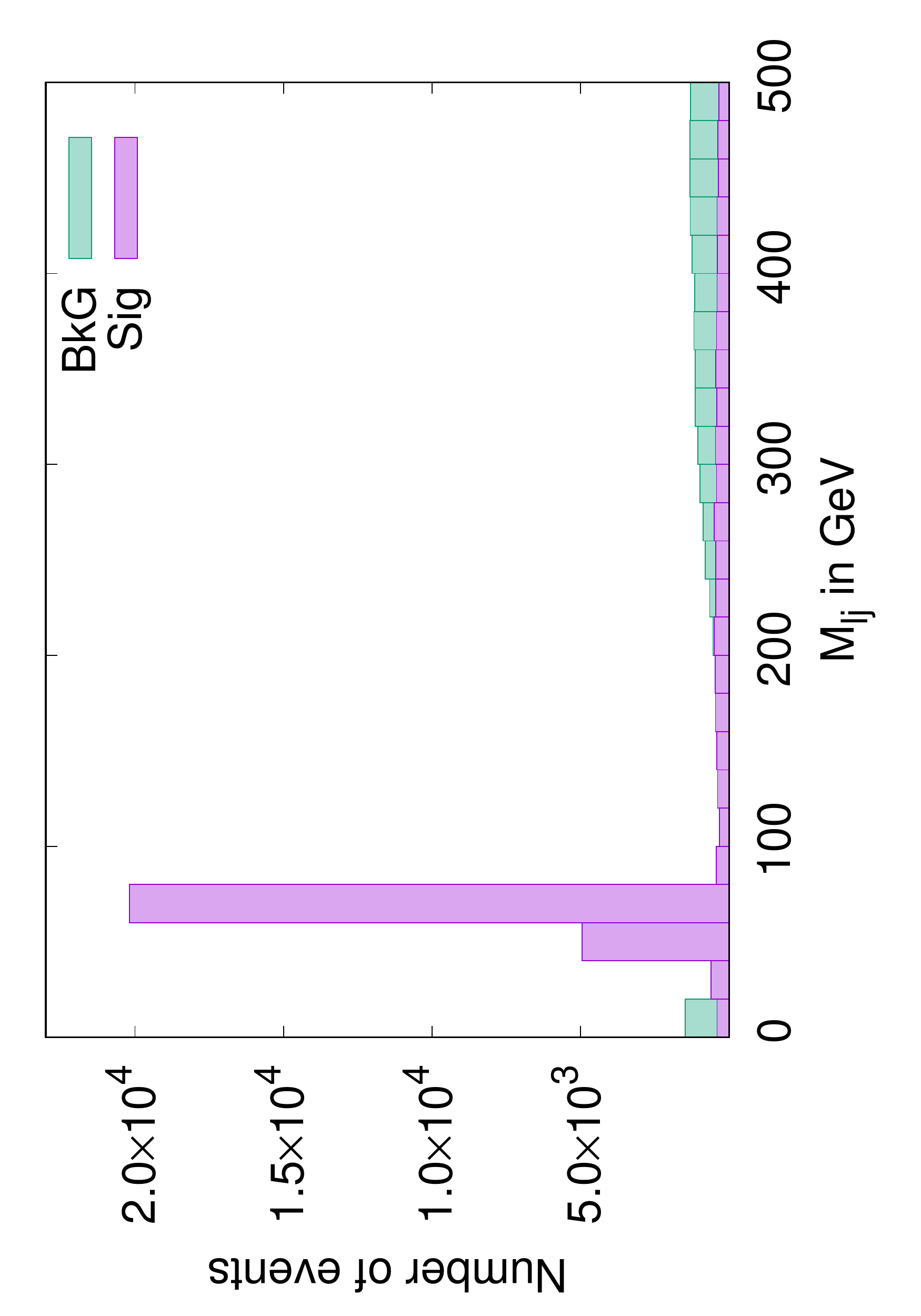}}~
    \belowbaseline[0pt]{\includegraphics[scale=0.41]{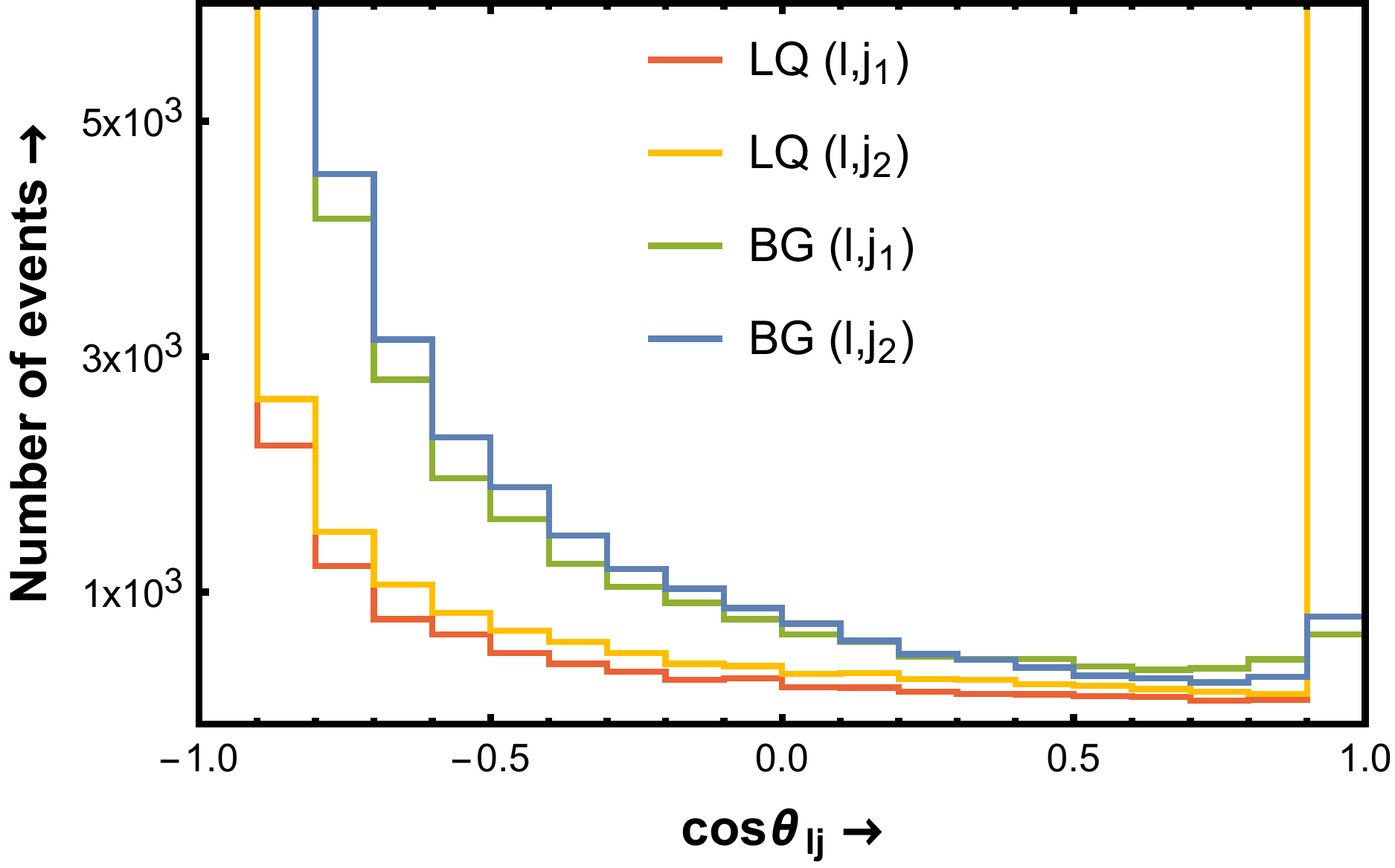}}\\
    
    \belowbaseline[-2mm]{\includegraphics[scale=0.30,angle=-90]{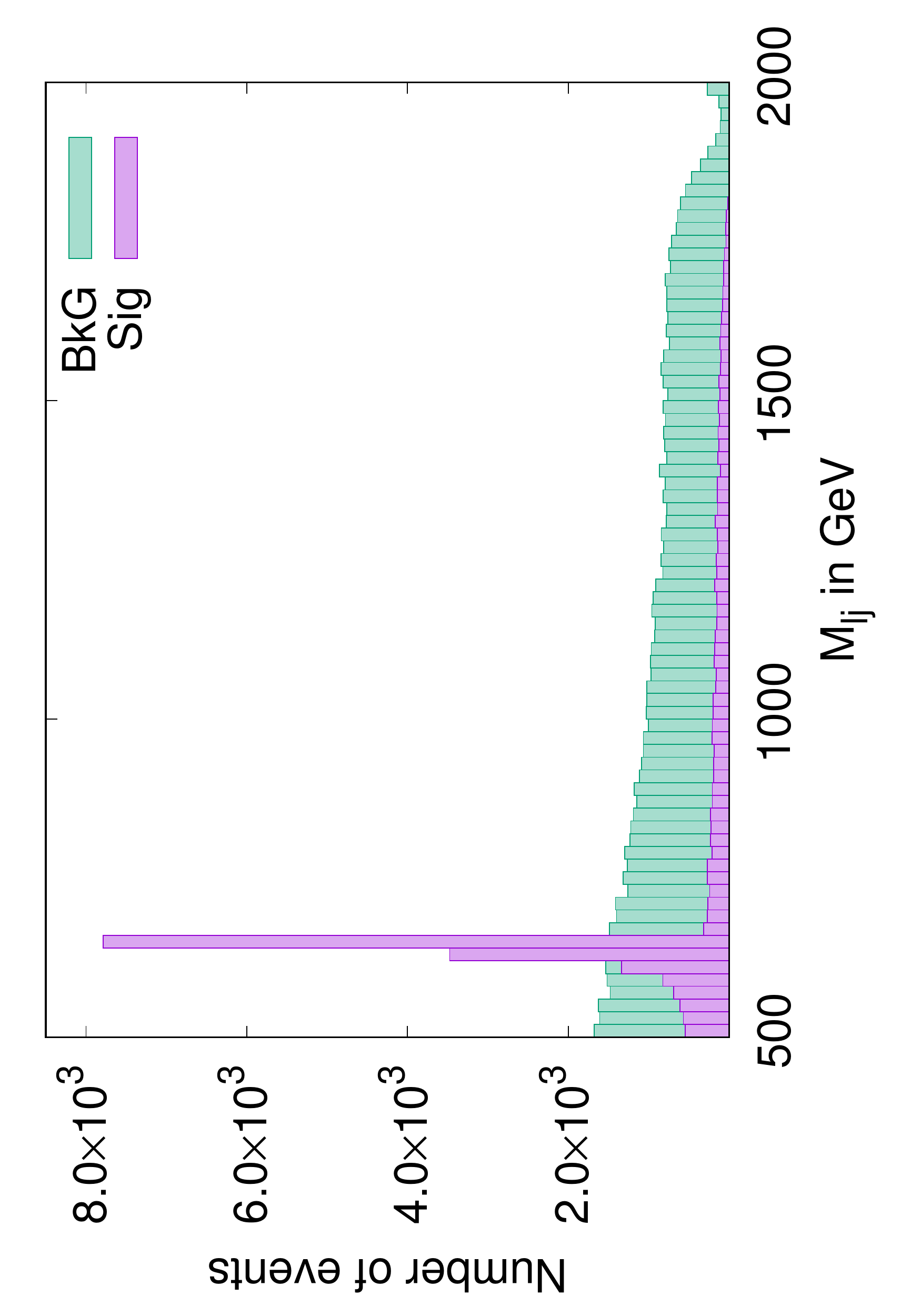}}~
    \belowbaseline[0pt]{\includegraphics[scale=0.41]{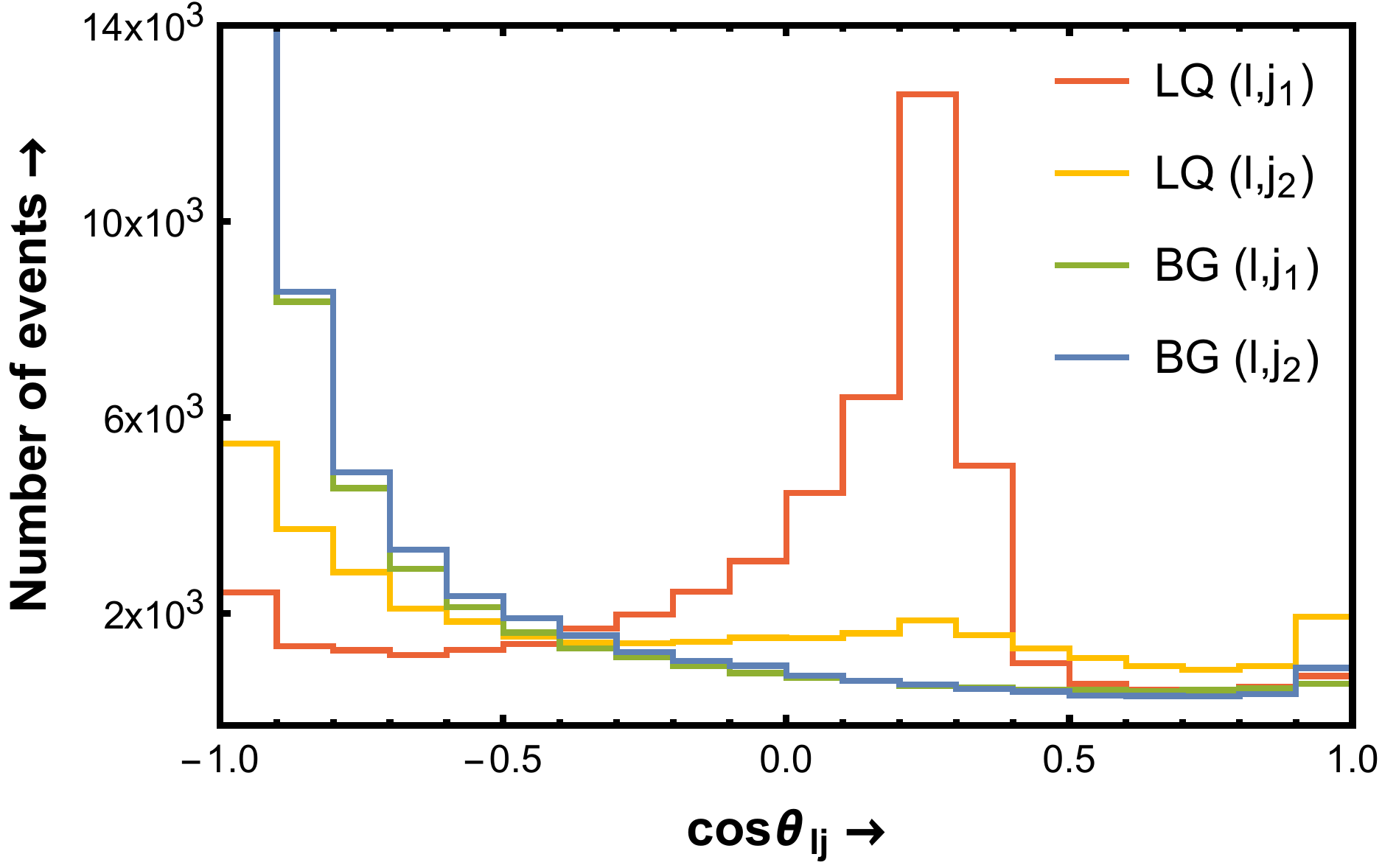}}\\
	
\end{figure}

\begin{figure}[H]
 	\begin{center}	
 		\includegraphics[scale=0.4]{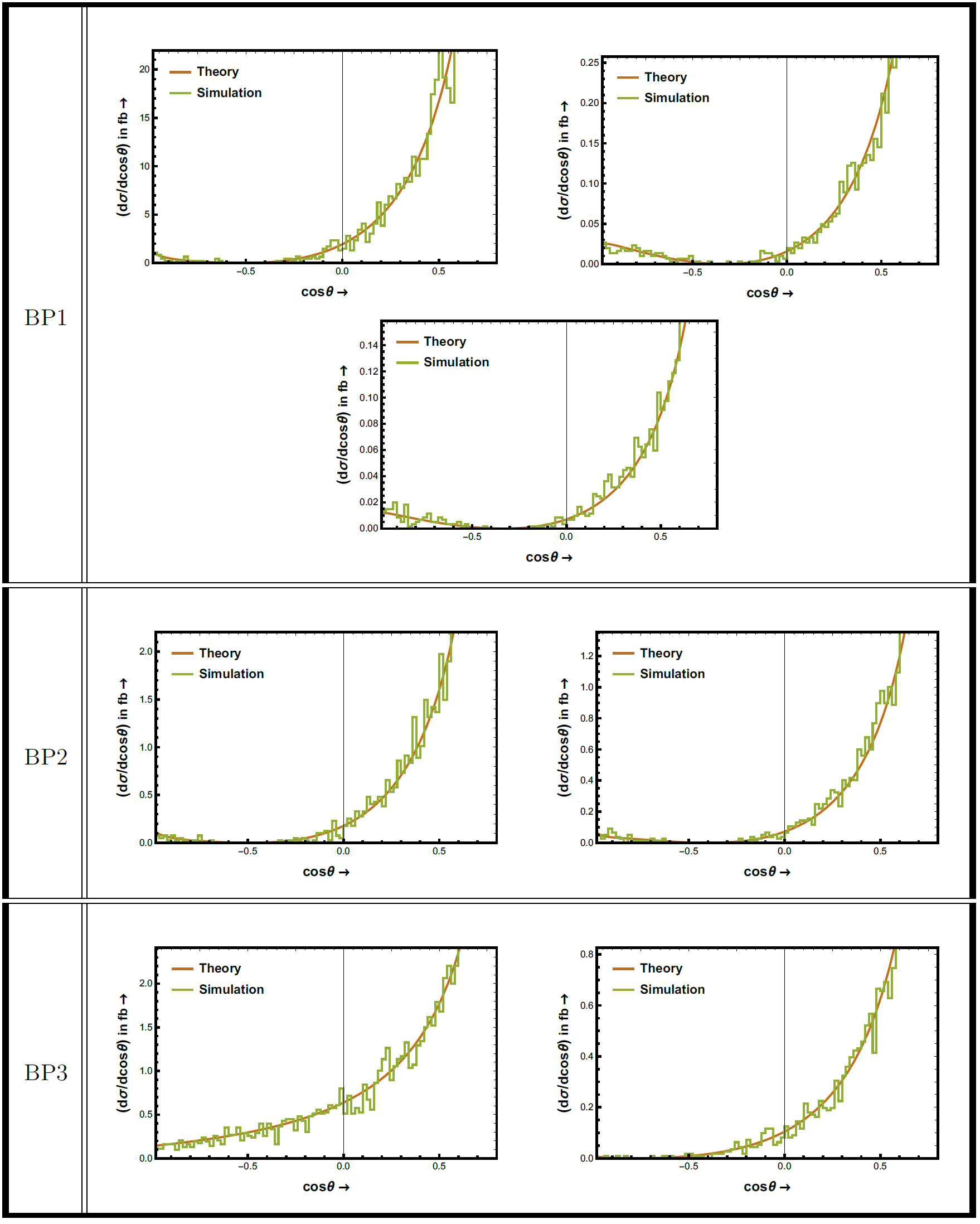}
 		
 	\end{center}
 	\caption{Angular distribution for the production of  $(S_1^{\nicefrac{+1}{3}})^c$ at various centre of momentum energies for different benchmark points, arranged in the order of table \ref{tab:S1_recons}. The brown (smooth) curves indicate the theoretical expectations whereas the green (jagged) lines signify the PYTHIA simulated data.}
 	\label{fig:S1}
 \end{figure}

 In the fig. \ref{fig:S1}, the differential cross-section has been delineated against the cosine of the angle between initial state electron and the leptoquark (or equivalently, the angle between photon and the quark that is produced associated with the leptoquark) at different centre of momentum energies for various benchmark points. The green (ragged) lines portray the simulated data with hundred bins within the range $-1<\cos\theta<1$ whereas, the brown (smooth)  lines  represent  the  theoretical  predictions given by Eq. \eqref{eq:diffdist}. The plots are\hfill\\
 \noindent arranged in the order of table \ref{tab:S1_recons}. The left and right plots at the top in BP1 row are for 200 GeV and 2 TeV centre of momentum energies respectively while the third one is for 3 TeV. In BP2 row, the first and second plots are done for 2 TeV and 3 TeV centre of momentum energies respectively. Likewise, for BP3 also the plots for $\sqrt s$ valued 2 TeV and 3 TeV are presented in the left and right panel of the third row. As can be seen, the angular distribution in each graph vanishes at some point except the first one in third row  which fails to satisfy the condition described by Eq. \ref{eq:physreg}. The positions of zeros can be verified from the left column ( titled ``$Q_{\bar q}=\nicefrac{-2}{3}$ or $Q_\phi=\nicefrac{-1}{3}$'') of table \ref{tab:zeros}.

\subsubsection{Leptoquark $(\widetilde {R}_2^{\nicefrac{+2}{3}})^c$}

\begin{table}[H]
	\begin{center}
		\renewcommand{\arraystretch}{1.4}
		\begin{tabular}{?>{\centering}m{1cm}||>{\centering}m{0.6cm}||c||>{\centering}m{1cm}|>{\centering}m{1cm}||m{0.8cm}?}\hlineB{5}
			Bench-mark points&$\sqrt s$ in TeV&Cut& Signal&Back-ground&Signi-ficance\\ \hline\hline
			\multirow{6}{*}{BP1}&	\multirow{2}{*}{0.2}&$|M_{lj}-M_{\phi}|\leq10$ GeV&5870.1&43725.0&26.4\\ 
			\cline{3-6}
			&&cut1+$(-0.2)\leq \cos \theta_{\ell j}\leq1$&5549.6&32989.8&28.3\\	\cline{2-6}
			&	\multirow{2}{*}{2}&$|M_{lj}-M_{\phi}|\leq10$ GeV&80.3&319.4&4.0\\ 	\cline{3-6}
			&&cut1+$(0.9)\leq \cos \theta_{\ell j}\leq1$&50.9&114.2&4.0\\	
			\cline{2-6}
			&	\multirow{2}{*}{3}&$|M_{lj}-M_{\phi}|\leq10$ GeV&33.6&219.8&2.1\\ 	\cline{3-6}
			&&cut1+$(0.9)\leq \cos \theta_{\ell j}\leq1$&19.4&44.2&2.4\\	
			\hline\hline
			
			\multirow{4}{*}{BP2}&	\multirow{2}{*}{2}&$|M_{lj}-M_{\phi}|\leq10$ GeV&27.0&2003.6&0.6\\ 
			\cline{3-6}
			&&cut1+$0\leq \cos \theta_{\ell j}\leq1$&20.8&129.1&1.7\\	\cline{2-6}
			&	\multirow{2}{*}{3}&$|M_{lj}-M_{\phi}|\leq10$ GeV&11.99&1660.7&0.3\\ 
			\cline{3-6}
			&&cut1+$0\leq \cos \theta_{\ell j}\leq1$&10.8&167.5&0.8\\ \hline\hline
			\multirow{4}{*}{BP3}&
			\multirow{2}{*}{2}&$|M_{lj}-M_{\phi}|\leq10$ GeV&19.4&1061.6&0.6\\ 
			\cline{3-6}
			&&cut1+$(-0.9)\leq \cos \theta_{\ell j}\leq1$&13.8&391.5&0.7\\	
			\cline{2-6}
			&\multirow{2}{*}{3}&$|M_{lj}-M_{\phi}|\leq10$ GeV&7.6&815.0&0.3\\ 
			\cline{3-6}
			&&cut1+$(-0.8)\leq \cos \theta_{\ell j}\leq1$&7.2&254.7&0.4\\	\hlineB{5}
		\end{tabular}
		\caption{Signal background analysis for  leptoquark  $(\widetilde {R}_2^{\nicefrac{+2}{3}})^c$  with luminosity 100 fb$^{-1}$ at $e$-$\gamma$ collider.}
		\label{tab:R2t_recons}
	\end{center}
\end{table}
The signal-background analysis for $(\widetilde {R}_2^{\nicefrac{+2}{3}})^c$ with luminosity of 100 fb$^{-1}$ has been rendered in table \ref{tab:R2t_recons}. For BP1, the cut on invariant mass of lepton-jet pair shows significances of $26.4\sigma$, $4.0\sigma$ and $2.1\sigma$ respectively, at three different values of centre of momentum energy;\hfill 
	\begin{figure}[H]
		\begin{center}	
			\includegraphics[scale=0.4]{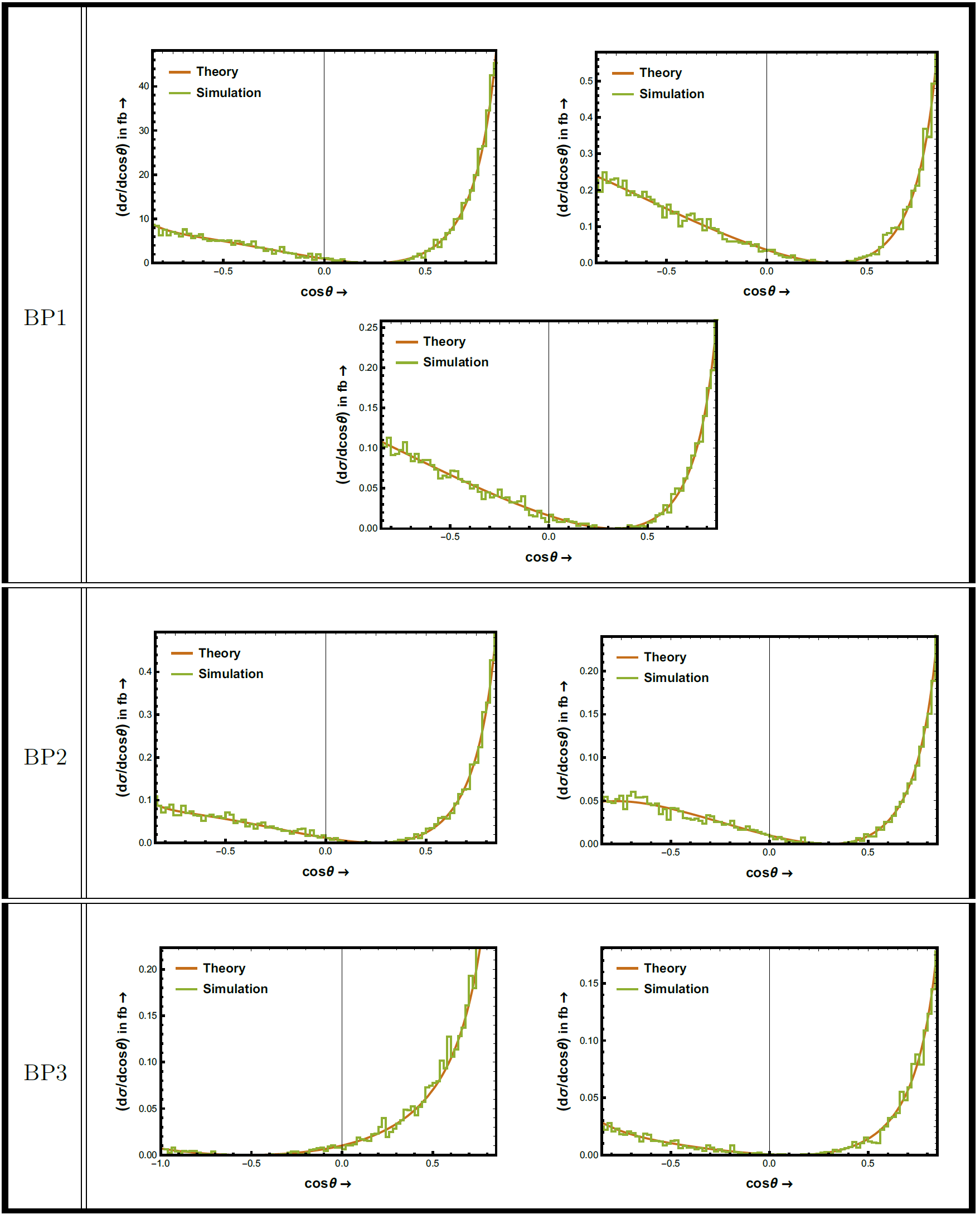}
			
		\end{center}
		\caption{Angular distribution for the production of  $(\widetilde {R}_2^{\nicefrac{+2}{3}})^c$ at various centre of momentum energies for different benchmark points, arranged in the order of table \ref{tab:R2t_recons}. The brown (smooth) curves indicate the theoretical expectations whereas the green (jagged) lines signify the PYTHIA simulated data.}
		\label{fig:R2t}
	\end{figure}
	
\noindent after applying the angular cuts, as described in case of $(S_1^{\nicefrac{+1}{}})^c$, the significances become $28.3\sigma$, $4.0\sigma$ and $2.4\sigma$ respectively. In case of BP2, only  significances of $0.6\sigma$ and $0.3\sigma$ are achieved by cut1 at 2 TeV and 3 TeV centre of momentum energies respectively, which increase to $1.7\sigma$ and $0.8\sigma$ respectively after implementation of the angular cut  $0\leq \cos \theta_{\ell j}\leq1$. For BP3 with 2 TeV energy, the significances reached by the two cuts are $0.6\sigma$ and $0.7\sigma$ and the same for 3 TeV energy are $0.3\sigma$ and $0.4\sigma$ respectively. It should be noticed that the significances are quite low in case of $(\widetilde {R}_2^{\nicefrac{+2}{3}})^c$ compared to $(S_1^{\nicefrac{+1}{}})^c$ especially with BP2 and BP3, and hence escalation in luminosity is essential for amelioration of the statistics.

Angular distributions for this case have been limned in fig. \ref{fig:R2t} where the brown (even) and green (uneven) lines signify the theoretical estimates and simulated data respectively. The plots are arranged in the same order as of table \ref{tab:R2t_recons}. It  can be observed that the distribution in every graph comes to zero at some point of phase space. The positions of zeros can be verified from the right column ( titled ``$Q_{q}=\nicefrac{-1}{3}$ or $Q_\phi=\nicefrac{-2}{3}$'') of table \ref{tab:zeros}.

\subsubsection{Leptoquark $( {R}_2^{\nicefrac{+5}{3}})^c$}

The PYTHIA analysis for leptoquark $( {R}_2^{\nicefrac{+5}{3}})^c$ has been presented in table \ref{tab:R2_recons}. The cut on $M_{\ell j}$ provides significances of $94.5\sigma$, $13.9\sigma$ and $7.8\sigma$ for the signal events at three centre of momentum energies in case of BP1 which change to $98.7\sigma$, $13.2\sigma$ and $8.1\sigma$ respectively after using suitable angular cuts on $\cos\theta_{\ell j}$. For BP2, signal events are produced with significances $14.4\sigma$ and $8.1\sigma$ at 2 TeV and 3 TeV centre of momentum energies respectively, any they get increased to $21.9\sigma$ and $14.8\sigma$ after applying the angular cut as $0\leq \cos \theta_{\ell j}\leq1$. 
For BP3 at $\sqrt s=$ 2 TeV, the significances become 
$11.4\sigma$ and $11.7\sigma$ after implementation of the two cuts and the same become
$5.9\sigma$ and $8.6\sigma$ respectively for $\sqrt s=$ 3 TeV.

\begin{figure}[H]
		\begin{center}	
			\includegraphics[scale=0.4]{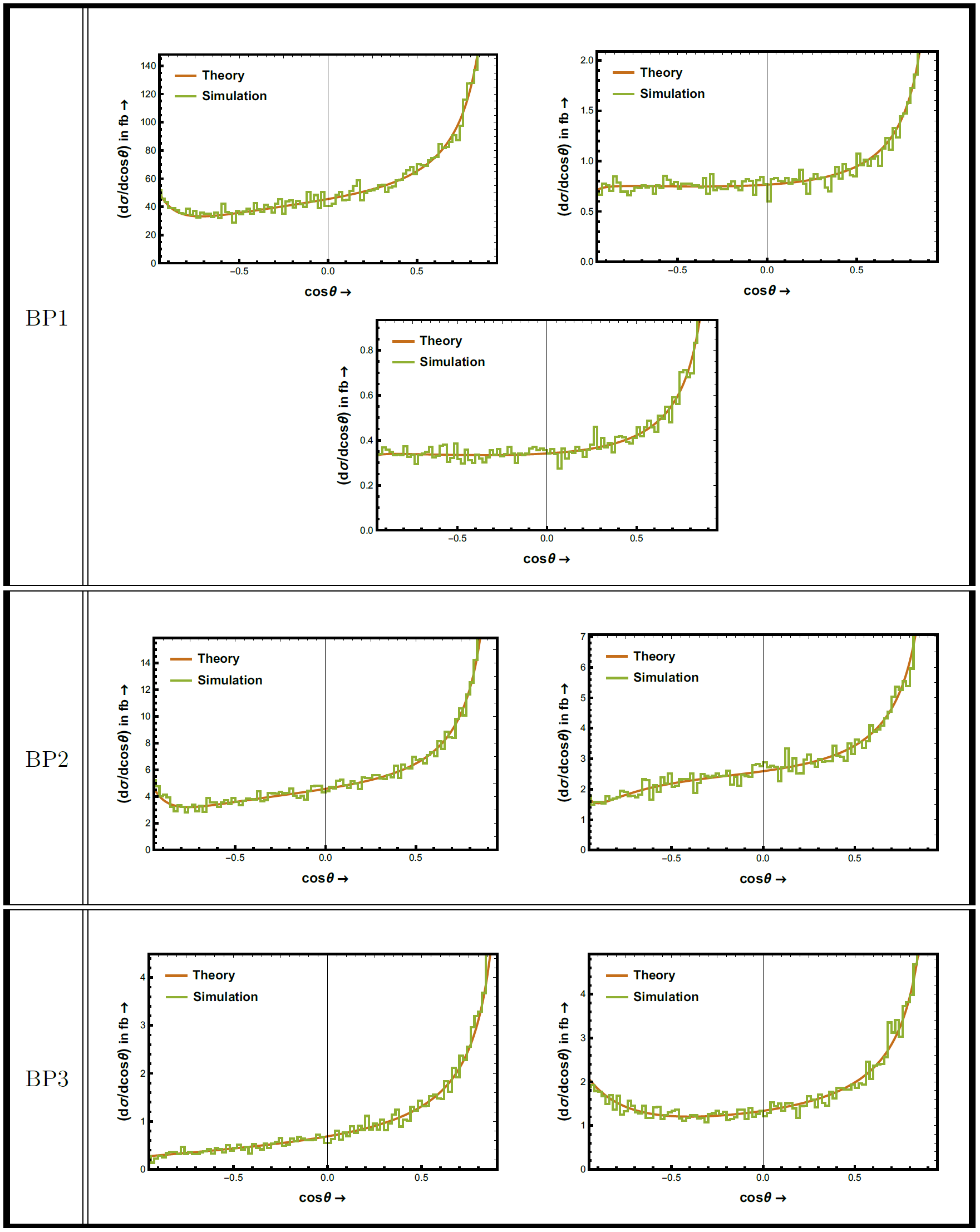}
		\end{center}
		\caption{Angular distribution for the production of  $( {R}_2^{\nicefrac{+5}{3}})^c$  at various centre of momentum energies for different benchmark points, arranged in the order of table \ref{tab:R2_recons}. The brown (smooth) curves indicate the theoretical expectations whereas the green (jagged) lines signify the PYTHIA simulated data.}
		\label{fig:R2}
	\end{figure}

The fig. \ref{fig:R2} describes the differential distribution with respect to the cosine of angle between leptoquark and initial state electron in this scenario. The plots are arranged in the order of table \ref{tab:R2_recons}. As the earlier cases the green (jagged) and brown (smooth) lines indicate the simulated data with 100 bins and the theoretical expectation for various benchmark points at different centre of momentum energy respectively. Unlike the other two cases, the angular distributions never vanish inside the physical region since this leptoquark does not satisfy Eq. \eqref{eq:condition}. 

\begin{table}[H]
	\begin{center}
		\renewcommand{\arraystretch}{1.4}
		\begin{tabular}{?>{\centering}m{1cm}||>{\centering}m{0.6cm}||c||>{\centering}m{1cm}|>{\centering}m{1cm}||m{0.8cm}?}\hlineB{5}
			Bench-mark points&$\sqrt s$ in TeV&Cut& Signal&Back-ground&Signi-ficance\\ \hline\hline
			
			\multirow{6}{*}{BP1}&	\multirow{2}{*}{0.2}&$|M_{lj}-M_{\phi}|\leq10$ GeV&24719.2&43725.0&94.5\\ 
			\cline{3-6}
			&&cut1+$(-0.2)\leq \cos \theta_{\ell j}\leq1$&23448.8&32989.8&98.7\\	\cline{2-6}
			&	\multirow{2}{*}{2}&$|M_{lj}-M_{\phi}|\leq10$ GeV&365.7&319.4&13.9\\ 	\cline{3-6}
			&&cut1+$(0.9)\leq \cos \theta_{\ell j}\leq1$&251.7&114.2&13.2\\	
			\cline{2-6}
			&	\multirow{2}{*}{3}&$|M_{lj}-M_{\phi}|\leq10$ GeV&148.9&219.8&7.8\\ 	\cline{3-6}
			&&cut1+$(0.9)\leq \cos \theta_{\ell j}\leq1$&96.2&44.2&8.1\\	
			\hline\hline
			
			\multirow{4}{*}{BP2}&	\multirow{2}{*}{2}&$|M_{lj}-M_{\phi}|\leq10$ GeV&757.4&2003.6&14.4\\ 
			\cline{3-6}
			&&cut1+$0\leq \cos \theta_{\ell j}\leq1$&585.5&129.1&21.9\\	\cline{2-6}
			&	\multirow{2}{*}{3}&$|M_{lj}-M_{\phi}|\leq10$ GeV&362.6&1660.7&8.1\\ 
			\cline{3-6}
			&&cut1+$0\leq \cos \theta_{\ell j}\leq1$&329.5&167.5&14.8\\ \hline\hline
			\multirow{4}{*}{BP3}&
			\multirow{2}{*}{2}&$|M_{lj}-M_{\phi}|\leq10$ GeV&440.9&1061.6&11.4\\ 
			\cline{3-6}
			&&cut1+$(-0.9)\leq \cos \theta_{\ell j}\leq1$&311.2&391.5&11.7\\	
			\cline{2-6}
			&	\multirow{2}{*}{3}&$|M_{lj}-M_{\phi}|\leq10$ GeV&187.8&815.0&5.9\\ 
			\cline{3-6}
			&&cut1+$(-0.8)\leq \cos \theta_{\ell j}\leq1$&180.0&254.7&8.6\\	\hlineB{5}
		\end{tabular}
		\caption{Signal background analysis for  leptoquark  $( {R}_2^{\nicefrac{+5}{3}})^c$   with luminosity 100 fb$^{-1}$ at $e$-$\gamma$ collider.}
		\label{tab:R2_recons}
	\end{center}
\end{table}

\subsubsection{Leptoquarks $(S_3^{\nicefrac{+4}{3}})^c$}

The signal-background analysis for $(\widetilde {S}_3^{\nicefrac{+4}{3}})^c$ with luminosity of $100\text{ }fb^{-1}$ has been rendered in table \ref{tab:S3_recons}. For BP1, the cut on invariant mass of lepton-jet pair shows significances of $36.1\sigma$, $6.2\sigma$ and $3.2\sigma$  at  centre of momentum energies to be 200 GeV, 2 TeV and 3 TeV respectively. The angular cuts modify these significances to become $38.7\sigma$, $6.8\sigma$ and $4.2\sigma$ respectively. In case of BP2, the significances  achieved by cut1 at 2 TeV and 3 TeV centre of momentum energies are $0.9\sigma$ and $0.5\sigma$ only, which increase to $2.5\sigma$ and $1.4\sigma$ respectively after implementation of the angular cut  $0\leq \cos \theta_{\ell j}\leq1$. For BP3 with 2 TeV energy, the significances reached by the two cuts are $0.6\sigma$ and $0.7\sigma$ respectively, which change to  $0.3\sigma$ and $0.7\sigma$ at $\sqrt s$ to be 3 TeV. In this case also the significances are quite low compared to $(S_1^{\nicefrac{+1}{3}})^c$ especially with BP2 and BP3. Increase in luminosity is needed for improvement of the statistics. 

\begin{table}[H]
	\begin{center}
		\renewcommand{\arraystretch}{1.4}
		\begin{tabular}{?>{\centering}m{1cm}||>{\centering}m{0.6cm}||c||>{\centering}m{1cm}|>{\centering}m{1cm}||m{0.8cm}?}\hlineB{5}
			Bench-mark points&$\sqrt s$ in TeV&Cut& Signal&Back-ground&Signi-ficance\\ \hline\hline
			\multirow{6}{*}{BP1}&	\multirow{2}{*}{0.2}&$|M_{lj}-M_{\phi}|\leq10$ GeV&8237.3&43725.0&36.1\\ 
			\cline{3-6}
			&&cut1+$(-0.2)\leq \cos \theta_{\ell j}\leq1$&7812.9&32989.8&38.7\\	\cline{2-6}
			&	\multirow{2}{*}{2}&$|M_{lj}-M_{\phi}|\leq10$ GeV&132.7&319.4&6.2\\ 	\cline{3-6}
			&&cut1+$(0.9)\leq \cos \theta_{\ell j}\leq1$&99.3&114.2&6.8\\	
			\cline{2-6}
			&	\multirow{2}{*}{3}&$|M_{lj}-M_{\phi}|\leq10$ GeV&53.6&219.8&3.2\\ 	\cline{3-6}
			&&cut1+$(0.9)\leq \cos \theta_{\ell j}\leq1$&38.0&44.2&4.2\\	
			\hline\hline
			
			\multirow{4}{*}{BP2}&	\multirow{2}{*}{2}&$|M_{lj}-M_{\phi}|\leq10$ GeV&40.4&2003.6&0.9\\ 
			\cline{3-6}
			&&cut1+$0\leq \cos \theta_{\ell j}\leq1$&31.4&129.1&2.5\\	\cline{2-6}
			&	\multirow{2}{*}{3}&$|M_{lj}-M_{\phi}|\leq10$ GeV&21.2&1660.7&0.5\\ 
			\cline{3-6}
			&&cut1+$0\leq \cos \theta_{\ell j}\leq1$&19.6&167.5&1.4\\ \hline\hline
			\multirow{4}{*}{BP3}&
			\multirow{2}{*}{2}&$|M_{lj}-M_{\phi}|\leq10$ GeV&20.4&1061.6&0.6\\ 
			\cline{3-6}
			&&cut1+$(-0.9)\leq \cos \theta_{\ell j}\leq1$&14.4&391.5&0.7\\	
			\cline{2-6}
			&	\multirow{2}{*}{3}&$|M_{lj}-M_{\phi}|\leq10$ GeV&9.7&815.0&0.3\\ 
			\cline{3-6}
			&&cut1+$(-0.8)\leq \cos \theta_{\ell j}\leq1$&9.3&254.7&0.6\\	\hlineB{5}
		\end{tabular}
		\caption{Signal background analysis for  leptoquark   $(S_3^{\nicefrac{+4}{3}})^c$  with luminosity 100 fb$^{-1}$ at $e$-$\gamma$ collider.}
		\label{tab:S3_recons}
	\end{center}
\end{table}

\begin{figure}[H]
		\begin{center}	
				\includegraphics[scale=0.4]{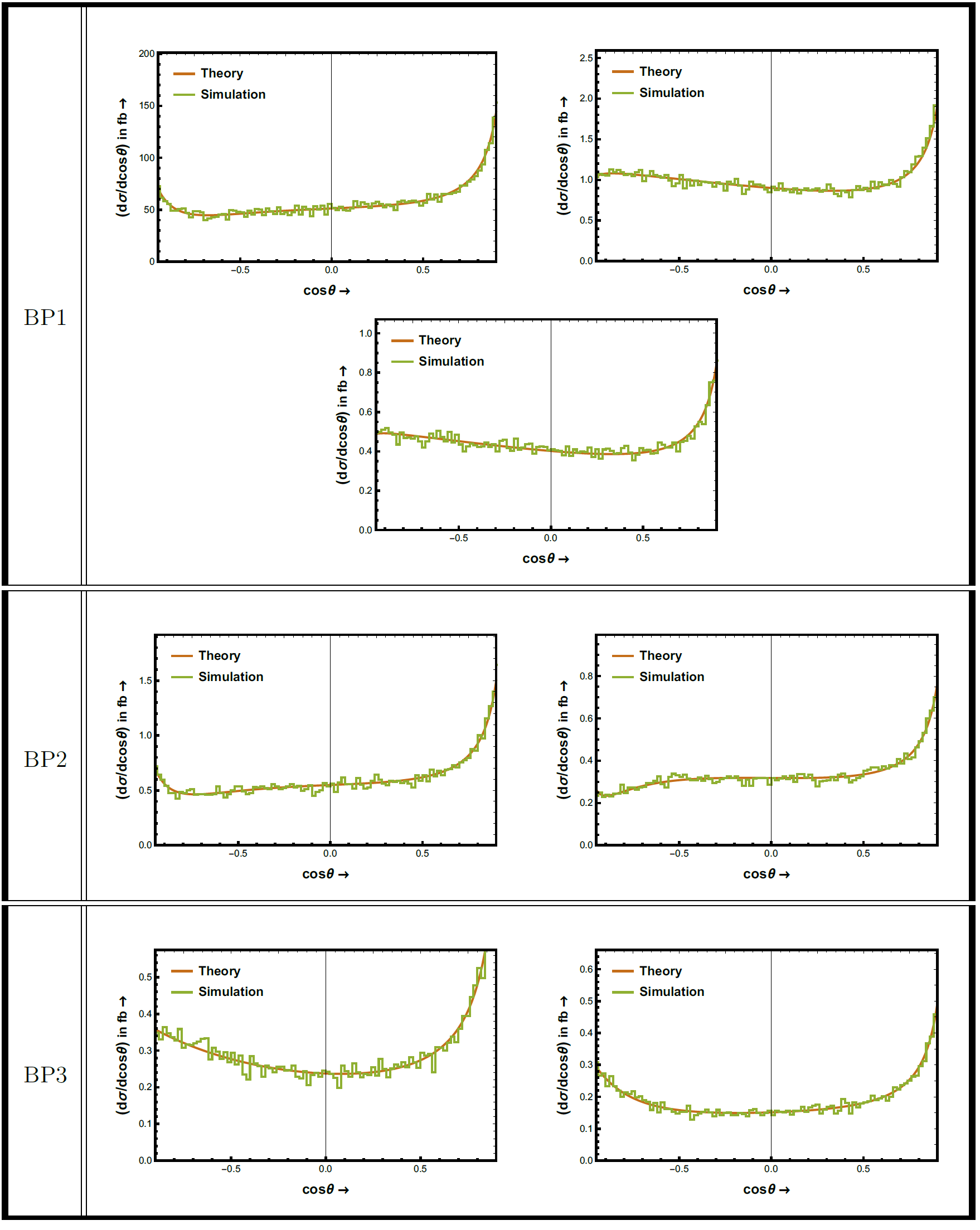}

		\end{center}
		\caption{Angular distribution for the production of  $(S_3^{\nicefrac{+4}{3}})^c$ at various centre of momentum energies for different benchmark points, arranged in the order of table \ref{tab:S3_recons}. The brown (smooth) curves indicate the theoretical expectations whereas the green (jagged) lines signify the PYTHIA simulated data.}
		\label{fig:S3}
	\end{figure}

Fig. \ref{fig:S3} shows the comparison between theoretical expectation and PYTHIA simulated data for for the production of  $(S_3^{\nicefrac{+4}{3}})^c$. The plots are arranged in the order of table \ref{tab:S3_recons}. As the earlier cases the green (uneven) and brown (even) lines indicate the simulated data with 100 bins and the theoretical expectation for various benchmark points at different centre of momentum energy respectively. In this case also no zero of differential distribution in any of the diagrams is found since its charge is smaller than $-1$ unit.

\subsection{Vector leptoquarks}
\label{sub:sig-bg-vector}

\subsubsection{Leptoquark $(U_{1\mu}^{\nicefrac{+2}{3}})^c$}

\begin{table}[H]
	\begin{center}
		\renewcommand{\arraystretch}{1.4}
			\begin{tabular}{?>{\centering}m{1cm}||>{\centering}m{0.6cm}||c||c|>{\centering}m{1cm}||m{0.8cm}?}\hlineB{5}
				Bench-mark points&$\sqrt s$ in TeV&Cut& Signal&Back-ground&Signi-ficance\\ \hline\hline
			\multirow{6}{*}{BP1}&	\multirow{2}{*}{0.2}&$|M_{lj}-M_{\phi}|\leq10$ GeV&10399.3&43725.0&44.7\\ 
			\cline{3-6}
			&&cut1+$(-0.2)\leq \cos \theta_{\ell j}\leq1$&9700.3&32989.8&46.9\\	\cline{2-6}
			&	\multirow{2}{*}{2}&$|M_{lj}-M_{\phi}|\leq10$ GeV&14666.5&319.4&119.8\\ 	\cline{3-6}
			&&cut1+$(0.9)\leq \cos \theta_{\ell j}\leq1$&9555.0&114.2&97.17\\	
			\cline{2-6}
			&	\multirow{2}{*}{3}&$|M_{lj}-M_{\phi}|\leq10$ GeV&14799.6&219.8&120.8\\ 	\cline{3-6}
			&&cut1+$(0.9)\leq \cos \theta_{\ell j}\leq1$&8736.1&44.2&93.2\\	
			\hline\hline
			
			\multirow{4}{*}{BP2}&	\multirow{2}{*}{2}&$|M_{lj}-M_{\phi}|\leq10$ GeV&443.3&2003.6&9.0\\ 
			\cline{3-6}
			&&cut1+$0\leq \cos \theta_{\ell j}\leq1$&337.5&129.1&15.6\\	\cline{2-6}
			&	\multirow{2}{*}{3}&$|M_{lj}-M_{\phi}|\leq10$ GeV&530.0&1660.7&11.3\\ 
			\cline{3-6}
			&&cut1+$0\leq \cos \theta_{\ell j}\leq1$&483.7&167.5&19.0\\ \hline\hline
			\multirow{4}{*}{BP3}&
			\multirow{2}{*}{2}&$|M_{lj}-M_{\phi}|\leq10$ GeV&144.4&1061.6&4.2\\ 
			\cline{3-6}
			&&cut1+$(-0.9)\leq \cos \theta_{\ell j}\leq1$&102.2&391.5&4.6\\	
			\cline{2-6}
			&	\multirow{2}{*}{3}&$|M_{lj}-M_{\phi}|\leq10$ GeV&63.9&815.0&2.2\\ 
			\cline{3-6}
			&&cut1+$(-0.8)\leq \cos \theta_{\ell j}\leq1$&60.7&254.7&3.4\\	\hlineB{5}
		\end{tabular}
		\caption{Signal background analysis for  $(U_{1\mu}^{\nicefrac{+2}{3}})^c$  with luminosity 100 fb$^{-1}$ at $e$-$\gamma$ collider.}
		\label{tab:U1_recons}
	\end{center}
\end{table}

In table \ref{tab:U1_recons}, we summarise the signal background analysis for the vector singlet leptoquark $(U_{1\mu}^{\nicefrac{+2}{3}})^c$. For BP1 at $\sqrt s=$ 200 GeV, the invariant mass cut of 10 GeV gives $44.7\sigma$ significance and further application of the angular cut of $(-0.2)\leq \cos \theta_{\ell j}\leq1$ changes it to $46.9\sigma$. For BP1 at centre of momentum energies to be 2 TeV and 3 TeV, the significances 
after the first cut are $119.8\sigma$ and $120.8\sigma$ respectively. In these cases, the signal events after the first cut are so large in number relative to the background events that the angular cut becomes obsolete. In case of BP2, the cut on $M_{\ell j}$ produce signal events with significances $9.0\sigma$ and $11.3\sigma$ for the two values of $\sqrt s$, which get enhanced to $15.6\sigma$ and $19.0\sigma$ respectively after constraining the angle $\theta_{\ell j}$ within the limit $0\leq \cos \theta_{\ell j}\leq1$. For BP3 at $\sqrt s=2$ TeV, the angular cut increases the significance to $4.6\sigma$ from $4.2\sigma$. Likewise, for BP3 at $\sqrt s=2$ TeV, the angular cut increases the significance to $3.4\sigma$ from $2.2\sigma$.

	\begin{figure}[H]
		\begin{center}	
			\includegraphics[scale=0.4]{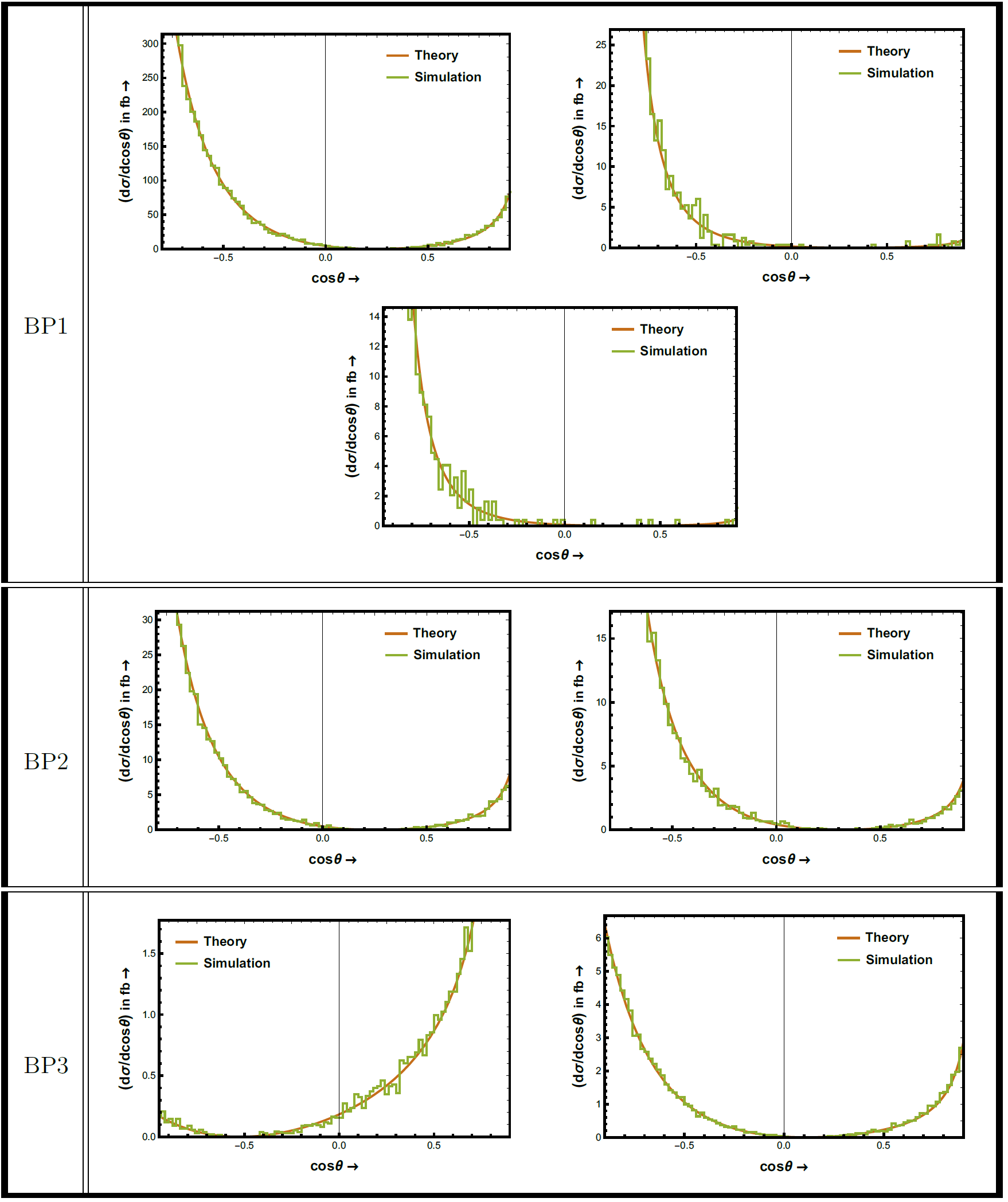}
		
		\end{center}
		\caption{Angular distribution for the production of  $(U_{1\mu}^{\nicefrac{+2}{3}})^c$ at various centre of momentum energies for different benchmark points, arranged in the order of table \ref{tab:U1_recons}. The brown (smooth) and green (jagged) lines indicate the theoretical expectations and the PYTHIA simulated data.}
		\label{fig:U1}
	\end{figure}

Angular distribution for this case has been limned in fig.  \ref{fig:U1} where the brown (smooth) and green (ragged) lines signify the theoretical estimates and simulated data respectively. The plots are arranged in terms of benchmark points and centre of momentum energy according to the table \ref{tab:U1_recons}. All the curves show zero certainly at some points of phase space which matches with the right column of table \ref{tab:zeros}.

\subsubsection{Leptoquark $(V_{2\mu}^{\nicefrac{+4}{3}})^c$}

\begin{table}[H]
	\begin{center}
		\renewcommand{\arraystretch}{1.4}
			\begin{tabular}{?>{\centering}m{1cm}||>{\centering}m{0.6cm}||c||c|>{\centering}m{1cm}||m{0.8cm}?}\hlineB{5}
				Bench-mark points&$\sqrt s$ in TeV&Cut& Signal&Back-ground&Signi-ficance\\ \hline\hline
			\multirow{6}{*}{BP1}&	\multirow{2}{*}{0.2}&$|M_{lj}-M_{\phi}|\leq10$ GeV&294306.3&43725.0&506.2\\ 
			\cline{3-6}
			&&cut1+$(-0.2)\leq \cos \theta_{\ell j}\leq1$&275902.1&32989.8&496.4\\	\cline{2-6}
			&	\multirow{2}{*}{2}&$|M_{lj}-M_{\phi}|\leq10$ GeV&399147.0&319.4&631.5\\ 	\cline{3-6}
			&&cut1+$(0.9)\leq \cos \theta_{\ell j}\leq1$&257096.9&114.2&506.9\\	
			\cline{2-6}
			&	\multirow{2}{*}{3}&$|M_{lj}-M_{\phi}|\leq10$ GeV&404429.7&219.8&635.8\\ 	\cline{3-6}
			&&cut1+$(0.9)\leq \cos \theta_{\ell j}\leq1$&238127.0&44.2&487.9\\	
			\hline\hline
			
			\multirow{4}{*}{BP2}&	\multirow{2}{*}{2}&$|M_{lj}-M_{\phi}|\leq10$ GeV&1560.1&2003.6&26.1\\ 
			\cline{3-6}
			&&cut1+$0\leq \cos \theta_{\ell j}\leq1$&1215.5&129.1&33.1\\	\cline{2-6}
			&	\multirow{2}{*}{3}&$|M_{lj}-M_{\phi}|\leq10$ GeV&1920.1&1660.7&32.1\\ 
			\cline{3-6}
			&&cut1+$0\leq \cos \theta_{\ell j}\leq1$&1754.7&167.5&40.0\\ \hline\hline
			\multirow{4}{*}{BP3}&
			\multirow{2}{*}{2}&$|M_{lj}-M_{\phi}|\leq10$ GeV&119.3&1061.6&3.5\\ 
			\cline{3-6}
			&&cut1+$(-0.9)\leq \cos \theta_{\ell j}\leq1$&85.5&391.5&3.9\\	
			\cline{2-6}
			&	\multirow{2}{*}{3}&$|M_{lj}-M_{\phi}|\leq10$ GeV&139.1&815.0&4.5\\ 
			\cline{3-6}
			&&cut1+$(-0.8)\leq \cos \theta_{\ell j}\leq1$&132.4&254.7&6.7\\	\hlineB{5}
		\end{tabular}
		\caption{Signal background analysis for  $(V_{2\mu}^{\nicefrac{+4}{3}})^c$  with luminosity 100 fb$^{-1}$ at $e$-$\gamma$ collider.}
		\label{tab:V2_recons}
	\end{center}
	
\end{table}

The signal background analysis for leptoquark $(V_{2\mu}^{\nicefrac{+4}{3}})^c$ has been shown in table \ref{tab:V2_recons}. For BP1, the significances of leptoquark production is very high ($506.2\sigma$, $631.5\sigma$ and $635.8\sigma$) at all the three values of $\sqrt s$ and angular cuts become almost obsolete. For BP2, the significances after first cut are $26.2\sigma$ and $32.1\sigma$  which get enhanced to $33.1\sigma$ and $40.0\sigma$ respectively after the second cut at the two different values of $\sqrt s$. For BP3 at 2 TeV centre of momentum energy the significances after the two cuts are $3.5\sigma$ and $3.9\sigma$ respectively. At 3 TeV centre of momentum energy for the same benchmark point, the significances after the two cuts become $4.5\sigma$ and $6.7\sigma$ respectively.
	
	\begin{figure}[H]
		\begin{center}	
			\includegraphics[scale=0.4]{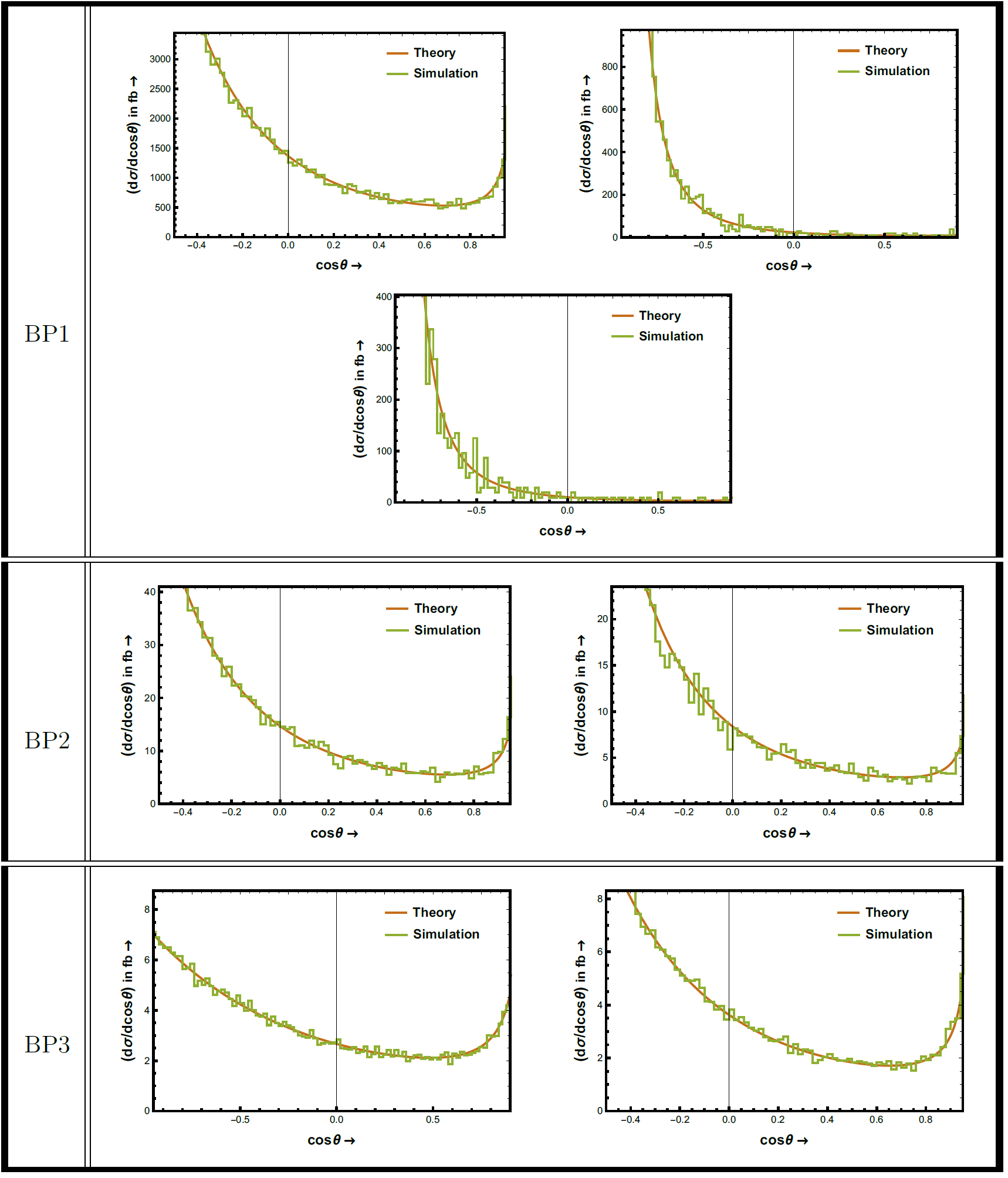}
		\end{center}
		\caption{Angular distribution for the production of  $(V_{2\mu}^{\nicefrac{+4}{3}})^c$ at various centre of momentum energies for different benchmark points, arranged in the order of table \ref{tab:V2_recons}. The brown (smooth) and  green (jagged) lines indicate the theoretical expectations  the PYTHIA simulated data.}
		\label{fig:V2}
	\end{figure}

In fig. \ref{fig:V2}, we show the angular distribution for the production of leptoquark $(V_{2\mu}^{\nicefrac{+4}{3}})^c$ associated with an antiquark $\bar d$ for all the three benchmark points at different centre of momentum energies as described in table \ref{tab:V2_recons}. As before, the brown (even) and green (uneven) lines signify the theoretical expectations and the PYTHIA simulated data respectively. In this case also, no zero in any of the plots is found.

\subsubsection{Leptoquark $(\widetilde V_{2\mu}^{\nicefrac{+1}{3}})^c$}

\begin{table}[H]
	\begin{center}
		\renewcommand{\arraystretch}{1.4}
			\begin{tabular}{?>{\centering}m{1cm}||>{\centering}m{0.6cm}||c||c|>{\centering}m{1cm}||m{0.8cm}?}\hlineB{5}
				Bench-mark points&$\sqrt s$ in TeV&Cut& Signal&Back-ground&Signi-ficance\\ \hline\hline
			\multirow{6}{*}{BP1}&	\multirow{2}{*}{0.2}&$|M_{lj}-M_{\phi}|\leq10$ GeV&102107.7&43725.0&267.4\\ 
			\cline{3-6}
			&&cut1+$(-0.2)\leq \cos \theta_{\ell j}\leq1$&96573.2&32989.8&268.3\\	\cline{2-6}
			&	\multirow{2}{*}{2}&$|M_{lj}-M_{\phi}|\leq10$ GeV&17380.0&319.4&130.6\\ 	\cline{3-6}
			&&cut1+$(0.9)\leq \cos \theta_{\ell j}\leq1$&11072.0&114.2&104.7\\	
			\cline{2-6}
			&	\multirow{2}{*}{3}&$|M_{lj}-M_{\phi}|\leq10$ GeV&16809.0&219.8&128.8\\ 	\cline{3-6}
			&&cut1+$(0.9)\leq \cos \theta_{\ell j}\leq1$&9738.8&44.2&98.5\\	
			\hline\hline
			
			\multirow{4}{*}{BP2}&	\multirow{2}{*}{2}&$|M_{lj}-M_{\phi}|\leq10$ GeV&236.5&2003.6&5.0\\ 
			\cline{3-6}
			&&cut1+$0\leq \cos \theta_{\ell j}\leq1$&179.6&129.1&10.2\\	\cline{2-6}
			&	\multirow{2}{*}{3}&$|M_{lj}-M_{\phi}|\leq10$ GeV&117.5&1660.7&2.8\\ 
			\cline{3-6}
			&&cut1+$0\leq \cos \theta_{\ell j}\leq1$&105.7&167.5&6.4\\ \hline\hline
			\multirow{4}{*}{BP3}&
			\multirow{2}{*}{2}&$|M_{lj}-M_{\phi}|\leq10$ GeV&154.1&1061.6&4.4\\ 
			\cline{3-6}
			&&cut1+$(-0.9)\leq \cos \theta_{\ell j}\leq1$&109.6&391.5&4.9\\	
			\cline{2-6}
			&	\multirow{2}{*}{3}&$|M_{lj}-M_{\phi}|\leq10$ GeV&62.5&815.0&2.1\\ 
			\cline{3-6}
			&&cut1+$(-0.8)\leq \cos \theta_{\ell j}\leq1$&60.1&254.7&3.4\\	\hlineB{5}
		\end{tabular}
		\caption{Signal background analysis for  leptoquark  $(\widetilde V_{2\mu}^{\nicefrac{+1}{3}})^c$  with luminosity 100 fb$^{-1}$ at $e$-$\gamma$ collider.}
		\label{tab:V2t_recons}
	\end{center}
	
\end{table}

Table \ref{tab:V2t_recons} summarises the reconstruction of leptoquark $(\widetilde V_{2\mu}^{\nicefrac{+1}{3}})^c$ at 100 fb$^{-1}$ luminosity. In this case also the significance for production of the leptoquark is quite high after the first cut for BP1 ($267.4\sigma$, $130.6\sigma$ and $128.8\sigma$ respectively) and hence the second cuts become unimportant. For BP2, the significances after the invariant mass cut are $5.0\sigma$ and $2.8\sigma$ which get improved to $10.2\sigma$ and $6.4\sigma$ respectively after the angular cut for 2 TeV and 3 TeV centre of momentum energies respectively. For BP3 at $\sqrt s=$ 2 TeV, the significance goes to $4.9\sigma$ from $4.4\sigma$ after implementing the angular cut of $(-0.8)\leq \cos \theta_{\ell j}\leq1$. For $\sqrt s=$ 3 TeV the corresponding change in significance is from $2.1\sigma$ to $4.2\sigma$.

	\begin{figure}[H]
		\begin{center}	
				\includegraphics[scale=0.4]{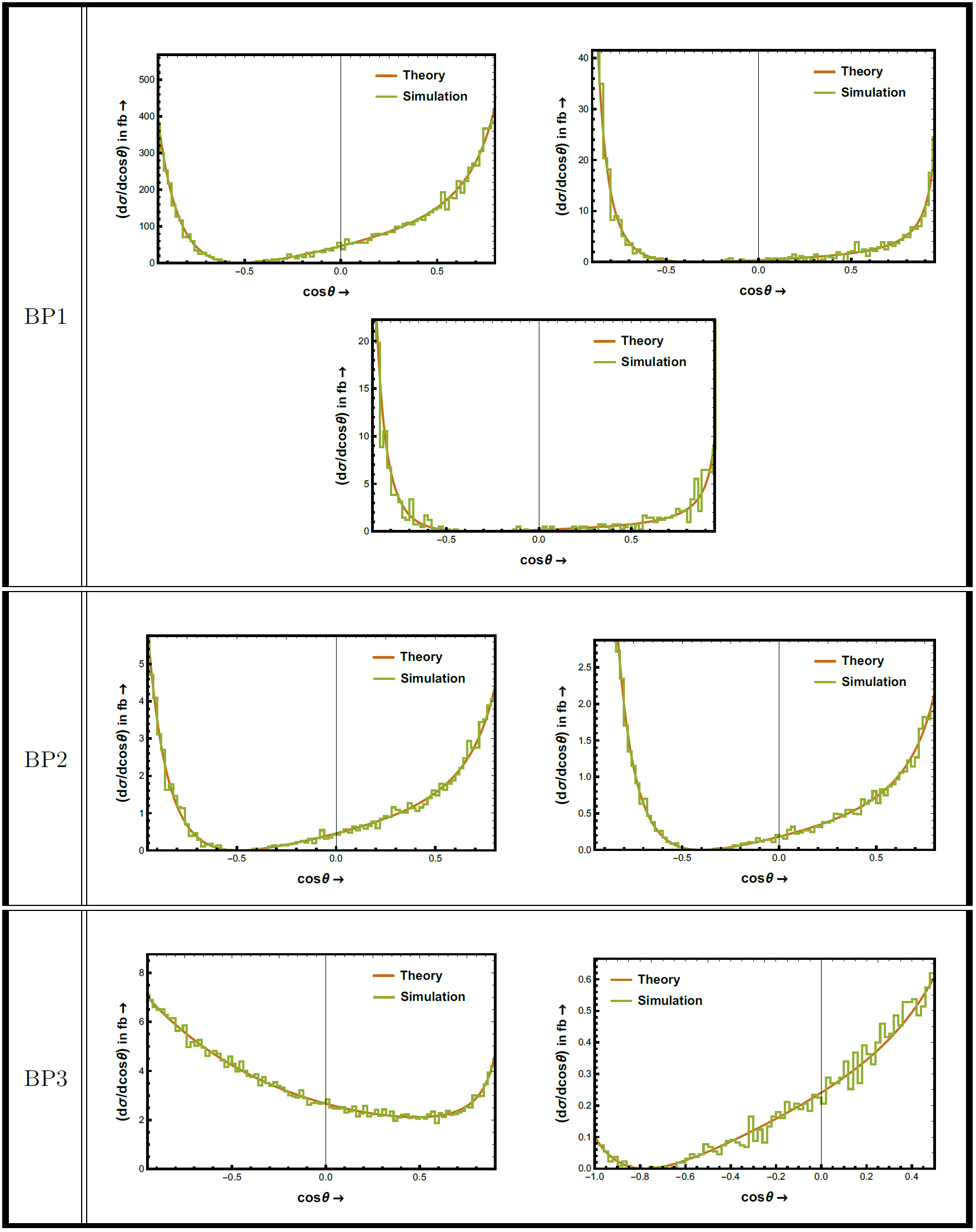}
		\end{center}
		\caption{Angular distribution for the production of  $(\widetilde V_{2\mu}^{\nicefrac{+1}{3}})^c$ at various centre of momentum energies for different benchmark points, arranged in the order of table \ref{tab:V2t_recons}. The brown (smooth) curves indicate the theoretical expectations whereas the green (jagged) lines signify the PYTHIA simulated data.}
		\label{fig:V2t}
	\end{figure}

In fig. \ref{fig:V2t}, we show the differential distribution for the production of this leptoquark. We ordered the graphs in the same way as in table \ref{tab:V2t_recons}. The brown (smooth) and green (coarse) lines signify the theoretical estimates and the simulated data respectively. As expected the distributions at different centre of momentum energies for various benchmark points go to zero at different points of phase space except the plot at the left panel in third row. The positions of zeros can be verified from the left column ( titled ``$Q_{\bar q}=\nicefrac{-2}{3}$ or $Q_\phi=\nicefrac{-1}{3}$'') of table \ref{tab:zeros}.

\subsubsection{Leptoquark $(U_{3\mu}^{\nicefrac{+5}{3}})^c$}

\begin{table}[H]
	\begin{center}
		\renewcommand{\arraystretch}{1.4}
			\begin{tabular}{?>{\centering}m{1cm}||>{\centering}m{0.6cm}||c||c|>{\centering}m{1cm}||m{0.8cm}?}\hlineB{5}
				Bench-mark points&$\sqrt s$ in TeV&Cut& Signal&Back-ground&Signi-ficance\\ \hline\hline
			\multirow{6}{*}{BP1}&	\multirow{2}{*}{0.2}&$|M_{lj}-M_{\phi}|\leq10$ GeV&402140.1&43725.0&602.2\\ 
			\cline{3-6}
			&&cut1+$(-0.2)\leq \cos \theta_{\ell j}\leq1$&376284.9&32989.8&588.2\\	\cline{2-6}
			&	\multirow{2}{*}{2}&$|M_{lj}-M_{\phi}|\leq10$ GeV&421151.4&319.4&648.7\\ 	\cline{3-6}
			&&cut1+$(0.9)\leq \cos \theta_{\ell j}\leq1$&268692.3&114.2&518.2\\	
			\cline{2-6}
			&	\multirow{2}{*}{3}&$|M_{lj}-M_{\phi}|\leq10$ GeV&421146.5&219.8&648.8\\ 	\cline{3-6}
			&&cut1+$(0.9)\leq \cos \theta_{\ell j}\leq1$&247085.8&44.2&497.0\\	
			\hline\hline
			
			\multirow{4}{*}{BP2}&	\multirow{2}{*}{2}&$|M_{lj}-M_{\phi}|\leq10$ GeV&1038.7&2003.6&18.8\\ 
			\cline{3-6}
			&&cut1+$0\leq \cos \theta_{\ell j}\leq1$&814.4&129.1&26.5\\	\cline{2-6}
			&	\multirow{2}{*}{3}&$|M_{lj}-M_{\phi}|\leq10$ GeV&1110.4&1660.7&21.1\\ 
			\cline{3-6}
			&&cut1+$0\leq \cos \theta_{\ell j}\leq1$&1014.0&167.5&29.5\\ \hline\hline
			\multirow{4}{*}{BP3}&
			\multirow{2}{*}{2}&$|M_{lj}-M_{\phi}|\leq10$ GeV&162.4&1061.6&4.6\\ 
			\cline{3-6}
			&&cut1+$(-0.9)\leq \cos \theta_{\ell j}\leq1$&115.5&391.5&5.1\\	
			\cline{2-6}
			&	\multirow{2}{*}{3}&$|M_{lj}-M_{\phi}|\leq10$ GeV&119.3&815.0&3.9\\ 
			\cline{3-6}
			&&cut1+$(-0.8)\leq \cos \theta_{\ell j}\leq1$&113.9&254.7&5.9\\	\hlineB{5}
		\end{tabular}
		\caption{Signal background analysis for  leptoquark $(U_{3\mu}^{\nicefrac{+5}{3}})^c$  with luminosity 100 fb$^{-1}$ at $e$-$\gamma$ collider.}
		\label{tab:U3_recons}
	\end{center}
\end{table}

We present the PYTHIA analysis for leptoquark $(U_{3\mu}^{\nicefrac{+5}{3}})^c$ in table \ref{tab:U3_recons} with a luminosity of 100 fb$^{-1}$. By putting a cut on the invariant mass of lepton jet pair, we get the signals with very high significances ($602.2\sigma$, $648.7$ and $648.8\sigma$) in case of BP1 for all the three centre of momentum energies. The angular  cut in this  case  becomes  inessential. The leptoquark for \hfill
\begin{figure}[H]
	\begin{center}	
		\includegraphics[scale=0.4]{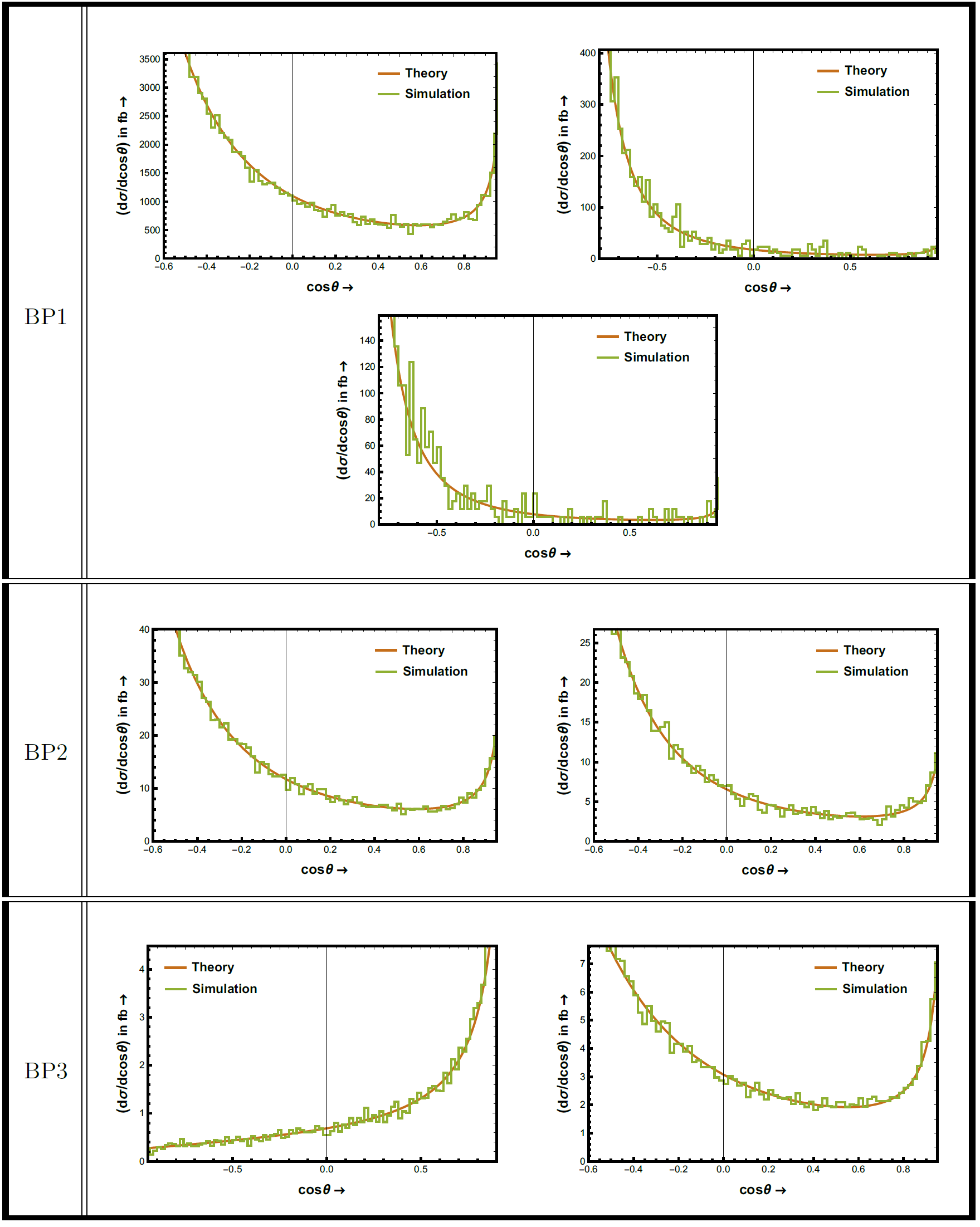}			
	\end{center}
	\caption{Angular distribution for the production of  $(U_{3\mu}^{\nicefrac{+5}{3}})^c$ at various centre of momentum energies for different benchmark points, arranged in the order of table \ref{tab:U3_recons}. The brown (smooth) curves indicate the theoretical expectations whereas the green (jagged) lines signify the PYTHIA simulated data.}
	\label{fig:U3}
\end{figure}
\noindent BP2 can be reconstructed with the significances $18.8\sigma$ and $21.1\sigma$ at $\sqrt s$ to be 2 TeV and 3 TeV respectively. Using the angular cut, the significances can be upgraded to $26.5\sigma$ and $29.5\sigma$ respectively.  For BP3 at 2 TeV, the cut of 10 GeV on $M_{\ell j}$ around the mass of leptoquark provides  $4.6\sigma$ significance for the signal events whereas the angular cut of $(-0.9)\leq \cos \theta_{\ell j}\leq1$ enhances it to $5.1\sigma$. For same benchmark point at 3 TeV, significance for the signal events goes to $5.9\sigma$ from  $3.9\sigma$ after applying the angular cut of $(-0.8)\leq \cos \theta_{\ell j}\leq1$.

In fig. \ref{fig:U3}, we show the angular distribution for the production of leptoquark $(U_{3\mu}^{\nicefrac{+5}{3}})^c$ associated with a $u$-quark for all the three bench mark points at different centre of momentum energies. The brown (even) and green (uneven) lines signify the theoretical expectations and the PYTHIA simulated data respectively. In this case also, no zero in any of the distributions is found.

\section{Effects of non-monochromatic photons}
\label{sec:non-mono}

All of our simulations until this point were performed with monochromatic photon beams. However, the experimental collider technology developed so far cannot deal with monochromatic photons in the initial state. Rather the photons used in modern colliders have some energy distribution. The two ways for the production of these high energetic photons, we are going to discuss, are: laser backscattering and quasi-real photons emitted by fast charged leptons. For the second process one can use protons instead of leptons too but it would make the environment messier due to presence of strongly interacting particle in initial state and therefore, we would stick to leptons only for our purpose of discussion. 

In case of laser backscattering, a laser beam interacts with high energy electrons or positrons and thus highly photons produced in the backward direction due to Compton scattering. This facility will be available in linear $e^+ e^-$ colliders like CLIC \cite{clic}, ILC \cite{ilc} etc. This option for photon is there in ClacHEP, but not in MadGraph \cite{mad}. The distribution of photon in this case is given by \cite{EPhC3b}:
\begin{equation}
\frac{1}{\sigma_c} \frac{d\sigma_c^{}}{dy} =f(x,y)=
 \frac{2\,\sigma_0^{}}{x\,\sigma_c^{}}\Big[1-y+\frac{1}{1-y}-\frac{4y}{x(1-y)}+\frac{4y^2}{x^2(1-y)^2}\Big] \quad\text{for } 0<y<y_{max}\,,
\end{equation} 
where, $y$ is the fraction of energy for backscattered photon relative to the energy of parent charged lepton, the maximum value of $y$ is $y_{max}$ which can also be written as $\big(\frac{x}{1+x}\big)$, the constant $\sigma_0=(\pi e^4/m_e^2)$ with $e$ and $m_e^{}$ being electric charge and mass of positron respectively and the total cross-section for Compton scattering is given by:
\begin{equation}
\sigma_c^{}=\frac{2\sigma_0^{}}{x}\Big[\Big(1-\frac{4}{x}-\frac{8}{x^2}\Big)\ln(1+x)+\frac{1}{2}+\frac{8}{x}-\frac{1}{2(1+x)^2}\Big]\,.
\end{equation} 
If laser and positron with energies $\omega_0$ and $E$ collide at a small angle $\alpha_0$ for the backscattering, then the quantity $x$ is defined as $x=(4E\,\omega_0/m_e^2)\cos^2(\alpha_0/2)$. However, the value of $x$ is taken as 4.82 in CalcHEP.

On the other hand, any fast moving charged particle can be considered as an electromagnetic radiation field by equivalent photon approximation (EPA) \cite{EPA1,EPA2,EPA3}. This radiation can be interpreted as a flux of quasi-real photons with some energy distribution. Following Williams-Weizs\"acker approximation, this distribution can be taken as:
\begin{equation}
f(y,q_{max}^2)=\frac{\alpha}{\pi}\bigg[\Big(\frac{1}{y}-1+\frac{y}{2}\Big)\ln\Big(\frac{q_{max}^2}{q_{min}^2}\Big)+\Big(1-\frac{1}{y}\Big)\Big(1-\frac{q_{min}^2}{q_{max}^2}\Big)\bigg]\quad \text{for } 0<y<1\,,
\end{equation}
where, $y$ is the fraction of energy for the quasi-real photon with respect to that of the positron, $\alpha$ is electromagnetic coupling constant, the minimum value for $q^2$ is $q_{min}^2$ which can be expressed as $\Big(\frac{m_e^2\, y^2}{1-y}\Big)$ and the maximum value for $q^2$ is $q_{max}^2$ that signifies the region for photon virtuality. It should kept in mind that the four momentum of quasi-real photon is denoted by $q^\mu$ in lab frame. This scheme is available in both MadGraph and ClacHEP.

To visualise the effects of non-monochromatic photons on the zeros of angular distribution for the  production of leptoquark associated with a quark (or anti-quark) at $e$-$\gamma$ collider, we pick four different scenarios, one from each of the leptoquark models having the zero inside physical region. Two of them are taken from lower energy and mass region and rest two are taken from higher energy and mass region. These four scenarios are as follows: a) BP1 for $(S_1^{\nicefrac{+1}{3}})^c$ at $\sqrt{s}=0.2$ TeV, b) BP2 for $(\widetilde R_2^{\nicefrac{+2}{3}})^c$ at $\sqrt{s}=2$ TeV c) BP3 for $(U_{1\mu}^{\nicefrac{+2}{3}})^c$ at $\sqrt{s}=2$ TeV and d) BP1 for $(\widetilde V_{2\mu}^{\nicefrac{+1}{3}})^c$ at $\sqrt{s}=0.2$ TeV. The signal background analysis and the cross-section for  production of leptoquarks in these four cases with laser backscattering and EPA has been presented in tables \ref{tab:las_syn_cross} and \ref{tab:las_syn} respectively. It can easily be seen from these tables that the production cross-section,signal event and the significance get enhanced to a great extent by laser backscattering compared to monochromatic photon beams whereas EPA scheme diminishes them notably. The increments in significances under laser backscattering for the above four scenarios are 83\%, 47\%, 11\% and 40\%, respectively, whereas under EPA, the significances reduce by 27\%, 65\%, 91\% and 51\%, respectively, for those cases. The production cross-section and signal event for laser backscattering are slightly lower than that of monochromatic case in only BP3 scenario for $(U_{1\mu}^{\nicefrac{+2}{3}})^c$ at $\sqrt{s}=2$ TeV. This occurs because of phase space suppression for low energy photons due to heavy mass of leptoquark in BP3 scenario.
\begin{table}[H]
	\begin{center}
		\renewcommand{\arraystretch}{1.4}
		\begin{tabular}{?>{\centering}m{2cm}||>{\centering}m{2cm}||>{\centering}m{2cm}||>{\centering}m{2cm}||m{2cm}?}
			\hlineB{5}
			\multirow{3}{*}{Photon}&\multicolumn{4}{c?}{Cross-section in fb}\\ \cline{2-5}
			&\hspace*{-3mm}\multirow{1}{*}{ $(S_1^{\nicefrac{+1}{3}})^c$, BP1} \hspace*{-3mm}\multirow{1}{*}{$\sqrt{s}=0.2$ TeV} &\hspace*{-3mm}\multirow{1}{*}{$(\widetilde R_2^{\nicefrac{+2}{3}})^c$, BP2} \hspace*{-3mm}\multirow{1}{*}{$\sqrt{s}=2$ TeV} & \hspace*{-3mm}\multirow{1}{*}{$(U_{1\mu}^{\nicefrac{+2}{3}})^c$, BP3 } \hspace*{-3mm}\multirow{1}{*}{$\sqrt{s}=2$ TeV}  & \hspace*{-3mm}\multirow{1}{*}{$(\widetilde V_{2\mu}^{\nicefrac{+1}{3}})^c$, BP1} \hspace*{-3mm}\multirow{1}{*}{$\sqrt{s}=0.2$ TeV}  \\ \hline\hline
			\multirow{1}{*}{Laser back-} \multirow{1}{*}{scattering}&688.20&4.87&11.11&\hspace*{4mm}3337.54\\ \hline
			EPA&101.42&0.81&0.40&\hspace*{5mm}486.94\\ \hline
			\hspace*{-3mm}\multirow{1}{*}{Monochromatic}&430.24&2.89&14.84&\hspace*{4mm}2127.02\\ \hlineB{5}
		\end{tabular}
		\caption{Cross-section for production of leptoquarks in the chosen four scenarios with laser backscattering, equivalent photon approximation and monochromatic photon at $e$-$\gamma$ collider.}
		\label{tab:las_syn_cross}
	\end{center}
\end{table}

\begin{table}[H]
	\begin{center}
		\renewcommand{\arraystretch}{1.4}
		\begin{tabular}{?>{\centering}m{0.7cm}|>{\centering}m{0.4cm}|>{\centering}m{0.2cm}||>{\centering}p{1.4cm}||>{\centering}m{1.2cm}|c|>{\centering}m{1cm}||m{0.8cm}?}\hlineB{5}
		\hspace*{-2mm}Model&\hspace*{-3.5mm}Bench-\hspace*{-3.5mm}mark \hspace*{-3.5mm}points&\hspace*{-2mm}$\sqrt s$ \hspace*{-2mm}in \hspace*{-2mm}TeV&Photon&Cut& Signal&Back-ground&\hspace*{-2mm}Signi- \hspace*{-2mm}\multirow{1}{*}{ficance}\\ \hlineB{5}
			\multirow{6}{*}{\hspace*{-3mm}$(S_1^{\nicefrac{+1}{3}})^c$}&\multirow{6}{*}{\hspace*{-3mm} BP1}&	\multirow{6}{*}{\hspace*{-2mm}0.2}&\hspace*{-3mm}\multirow{1}{*}{Laser Back-}&cut1&18518.7&36417.9&79.0\\ 
			\cline{5-8}
			&&&\hspace*{-1.5mm}\multirow{1}{*}{scattering}&\hspace*{-3mm}cut1+cut2& 17499.5&18173.8&92.65\\	\cline{4-8}
			&&&\multirow{2}{*}{EPA}&cut1& 2863.6 &1964.4&41.2\\ 	\cline{5-8}
			&&&&\hspace*{-3mm}cut1+cut2&2051.6&1038.2&36.9\\	
			\cline{4-8}
			&&&\hspace*{-1mm}mono-&cut1&11133.6&43725.0&47.5\\ 	\cline{5-8}
			&&&\hspace*{-2mm}\multirow{1}{*}{chromatic}&\hspace*{-3mm}cut1+cut2&10537.8&32989.8&50.5\\	
\hline\hline
	\multirow{6}{*}{\hspace*{-3mm}$(\widetilde R_2^{\nicefrac{+2}{3}})^c$}&\multirow{6}{*}{\hspace*{-3mm} BP2}&	\multirow{6}{*}{\hspace*{-1mm}2}&\hspace*{-3mm}\multirow{1}{*}{Laser Back-}&cut1&62.7&2078.4&1.4\\ 
	\cline{5-8}
	&&&\hspace*{-1.5mm}\multirow{1}{*}{scattering}&\hspace*{-3mm}cut1+cut2&        48.1&326.1&2.5\\	\cline{4-8}
	&&&\multirow{2}{*}{EPA}&cut1&11.7&390.2&0.6\\ 	\cline{5-8}
	&&&&\hspace*{-3mm}cut1+cut2& 7.0&230.8&0.5\\	
	\cline{4-8}
	&&&\hspace*{-1mm}mono-&cut1&27.0&2003.6&0.6\\ 	\cline{5-8}
	&&&\hspace*{-2mm}\multirow{1}{*}{chromatic}&\hspace*{-3mm}cut1+cut2&20.8&129.1&1.7\\	\hline\hline
	
	\multirow{6}{*}{\hspace*{-3mm}$(U_{1\mu}^{\nicefrac{+2}{3}})^c$}&\multirow{6}{*}{\hspace*{-3mm} BP3}&	\multirow{6}{*}{\hspace*{-2mm} 2}&\hspace*{-3mm}\multirow{1}{*}{Laser Back-}&cut1&131.8&446.3&5.5\\ 
	\cline{5-8}
	&&&\hspace*{-1.5mm}\multirow{1}{*}{scattering}&\hspace*{-3mm}cut1+cut2&93.4&246.0&5.1\\	\cline{4-8}
	&&&\multirow{2}{*}{EPA}&cut1&4.5&86.1&0.5\\ 	\cline{5-8}
	&&&&\hspace*{-3mm}cut1+cut2&3.5&79.7&0.4\\	
	\cline{4-8}
	&&&\hspace*{-1mm}mono-&cut1&144.4&1061.6&4.2\\ 	\cline{5-8}
	&&&\hspace*{-2mm}\multirow{1}{*}{chromatic}&\hspace*{-3mm}cut1+cut2&102.2&391.5&4.6\\	\hline\hline
	
	\multirow{6}{*}{\hspace*{-3mm}$(\widetilde V_{2\mu}^{\nicefrac{+1}{3}})^c$}&\multirow{6}{*}{\hspace*{-3mm} BP1}&	\multirow{6}{*}{\hspace*{-2mm}0.2}&\hspace*{-3mm}\multirow{1}{*}{Laser Back-}&cut1&168135.3&36417.9&371.8\\ 
	\cline{5-8}
	&&&\hspace*{-1.5mm}\multirow{1}{*}{scattering}&\hspace*{-3mm}cut1+cut2&157708.8& 18173.8&376.0\\	\cline{4-8}
	&&&\multirow{2}{*}{EPA}&cut1&24697.2&1964.4&151.3\\ 	\cline{5-8}
	&&&&\hspace*{-3mm}cut1+cut2&17881.0&1038.2&130.0\\	
	\cline{4-8}
	&&&\hspace*{-1mm}mono-&cut1&102107.7&43725.0&267.4\\ 	\cline{5-8}
	&&&\hspace*{-2mm}\multirow{1}{*}{chromatic}&\hspace*{-3mm}cut1+cut2&96573.2&32989.8&268.3\\	
	\hlineB{5}
	
	\end{tabular}
		\caption{Signal background analysis for chosen four scenarios with laser backscattering, equivalent photon approximation and monochromatic photon at $e$-$\gamma$ collider of luminosity 100 fb$^{-1}$. The term ``cut1'' indicates the invariant mass cut $|M_{lj}-M_{\phi}|\leq10$ GeV whereas the term ``cut2'' denotes the angular cut corresponding to particular benchmark point and centre of momentum energy as shown in other tables for signal background analysis.}
		\label{tab:las_syn}
	\end{center}
\end{table}

\begin{figure}[H]
	\includegraphics[width=7.2cm,height=5.5cm]{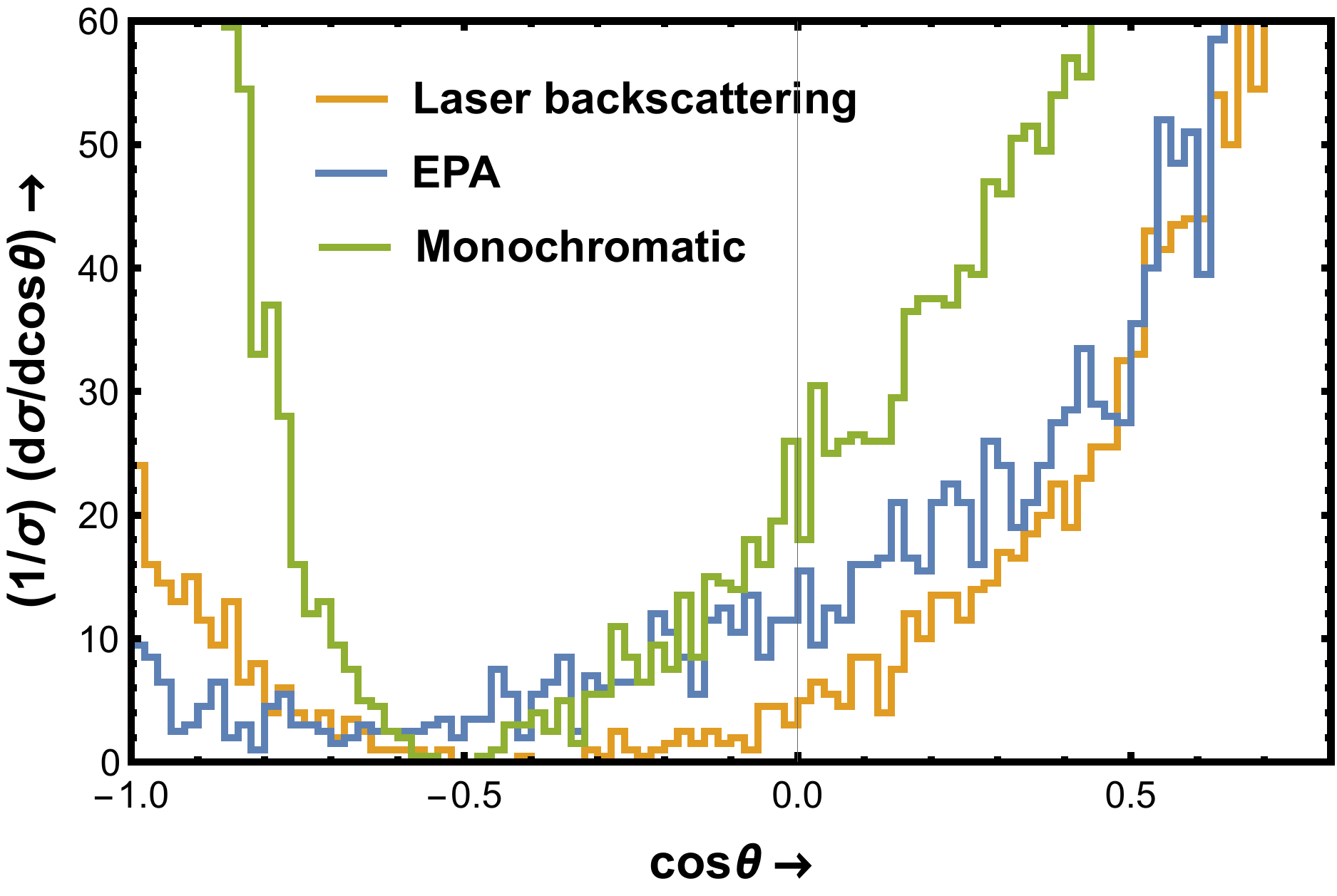}\hfill\includegraphics[width=7.2cm,height=5.4cm]{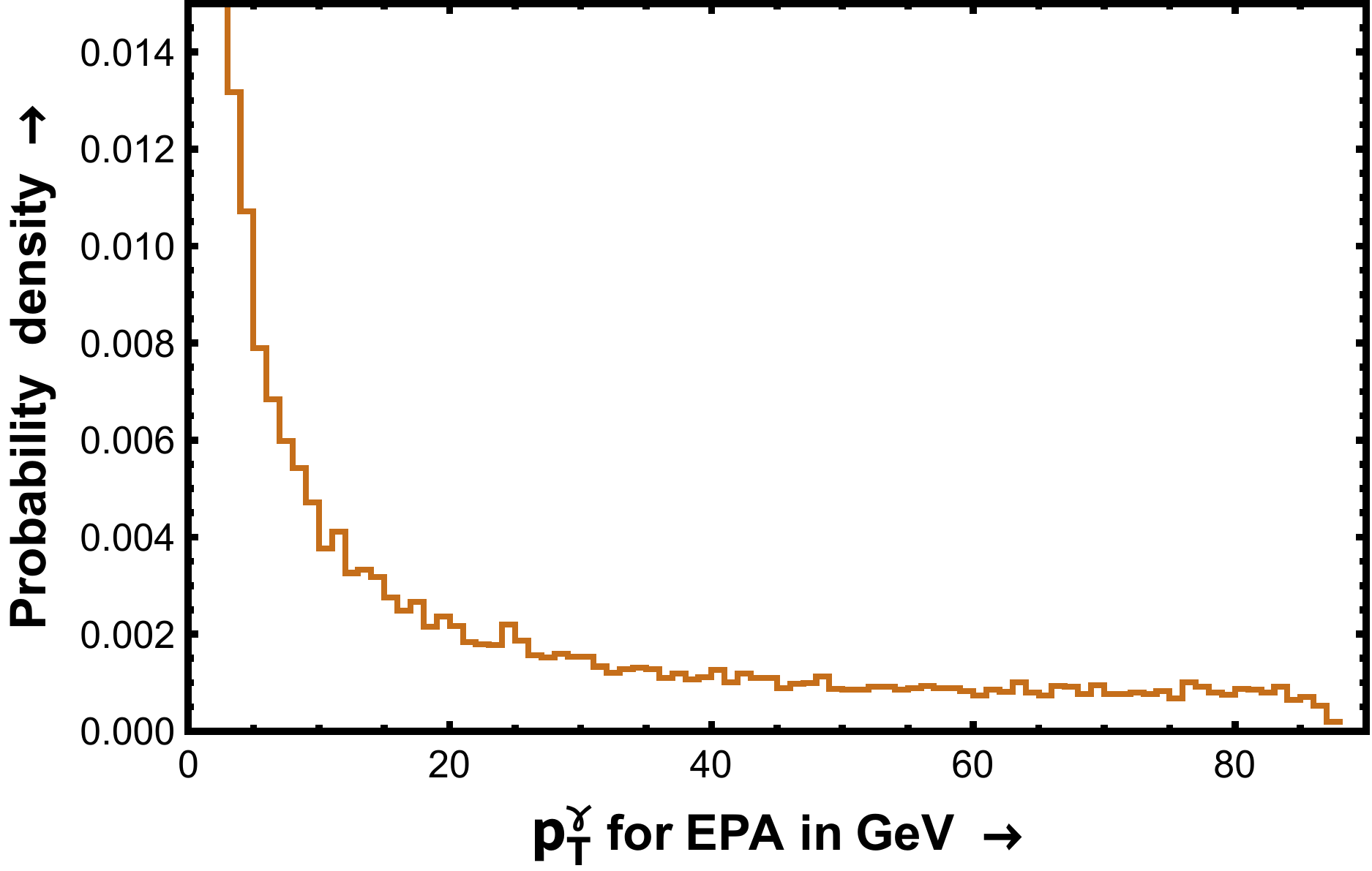}
	\caption{The comparison among laser backscattering, EPA and monochromatic photons (represented by orange, blue and green line respectively) in terms of weighted differential distribution ($\frac{1}{\sigma}\cdot\frac{d\sigma}{dcos\theta}$) for the production of $(\widetilde V_{2\mu}^{\nicefrac{+1}{3}})^c$ in BP1 scenario at $\sqrt{s}=0.2$ TeV is shown in the left panel. The distribution for transverse momentum of photon from 100 GeV positron under EPA scheme is shown in right panel.}
	\label{fig:comp}
\end{figure}

The weighted differential distributions ($\frac{1}{\sigma}\cdot\frac{d\sigma}{dcos\theta}$) for the production of $(\widetilde V_{2\mu}^{\nicefrac{+1}{3}})^c$ in BP1 scenario at $\sqrt{s}=0.2$ TeV for laser backscattering, EPA and monochromatic photons, which are represented by orange, blue and green line respectively, are shown in the left panel of fig. \ref{fig:comp}. As expected, the three distributions do not coincide. Still laser backscattering looks optimistic since it preserves the zero of angular distribution (though slightly deviated from the monochromatic case). The slight shift of zero in this case is due to variation in $\sqrt{s}$ at each collision causing from the distribution of photon energy. But in the EPA scheme, the zero gets smeared off. The reason behind this smearing effect lies in the distribution for transverse momentum of photons $(p_T^\gamma)$ which is depicted in  right panel of fig. \ref{fig:comp}. Although most of the photons in EPA have small $p_T^\gamma$, there is a non-zero possibility for them to acquire very high $p_T^\gamma$ too. As can be seen from the right panel of fig. \ref{fig:comp}, the highest $p_T^\gamma$ achieved by photons from 100 GeV positron under EPA is around 90 GeV. Due to this non-zero $p_T^\gamma$ the direction of photons changes in case of each collision and hence $e$ and $\gamma$ no longer move collinearly in opposite directions. 

Another important point to mention is that here we have presented the angular distribution for all the cases in terms of angle between electron and leptoquark; however, one can use the distribution in terms of angle between photon and leptoquark too. In the case of monochromatic photons, the system lies in centre of momentum frame and the two angles mentioned above are supplementary to each other; hence the distributions with respect to them are equivalent apart from a negative sign. For laser backscattering, the system no longer lies in centre of momentum frame due to varying energy of photons; however the above-mentioned two angles still remain supplementary to each other since all the photons are found with zero $p_T^\gamma$ only and therefore, the two distributions look quite similar discarding the negative sign. But in EPA scenario, neither the system lies in centre of momentum frame nor the angle between photon-leptoquark  and electron-leptoquark remain supplementary as well; hence the angular distributions with respect to these two angles disagree conspicuously. Talking in terms of zeros of angular distributions, the situations are worse while considering angle between photon and leptoquark under EPA. Similar kind of things happen for other leptoquark models with different benchmark point and centre of momentum energy also.

\section{Conclusion}
\label{sec:concl}

In conclusion, we have studied zeros of single photon tree-level amplitude at the $e$-$\gamma$ collider producing a leptoquark associated with a quark (or antiquark). Unlike other colliders, we find that the position of zeros of single photon tree-level amplitude in this case does depend on the centre of momentum energy as well as the mass and charge of leptoquark.  The cosine of angle between leptoquark and initial state electron, at which zero of the angular distribution happens, approaches $\nicefrac{\pm1}{3}$ asymptotically depending on the charge of the produced leptoquark for very high value of $\sqrt s$ with respect to the mass of leptoquark. No zero in the differential distribution can be found for leptoquarks having charges smaller than -1 unit. In a PYTHIA based analysis we look for both light and heavy leptoquarks at both low and high energy scales. Light leptoquarks having small couplings to quarks and leptons of all generation is not completely ruled out by Tevatron. In our simulation, we reconstruct the leptoquark from the lepton-jet pair and then study the differential distribution against the cosine of the angle between it and the initial state electron which matches with the theoretical expectation. We have also studied the consequences of using non-monochromatic photons for the production of leptoquarks at electron-photon colliders. The production cross-section and significance increase notably under laser backscattering and decrease terribly under equivalent photon approximation. It turns out that non-zero transverse momentum of photons smears off the zeros of angular distributions completely in equivalent photon approximation whereas laser backscattering preserves them (though slightly deviated from the monochromatic case) since all the photons here move only in the direction  opposite to electron. It seems that laser backscattering is very promising for investigating the production of leptoquarks at $e$-$\gamma$ collider by means of zeros of differential distribution.

\section{Acknowledgement}

The authors thank Abhay Deshpande, Yoshitaka Kuno, Saurabh Sandilya and Vishal Bhardwaj for some useful discussions. The authors also thank Alexander Pukhov and Olivier Mattelaer for private communications. PB and AK also thank SERB India, grant no: CRG/2018/004971 for the financial support.

\end{document}